\RequirePackage{lineno}
\documentclass[pdflatex,aps,prd,nofootinbib,showpacs,superscriptaddress]{revtex4-1}
\pdfoutput=1
\usepackage{amsmath,amssymb,color,fancyhdr,hyperref,lastpage,lineno} 
\usepackage{graphicx}
\graphicspath{{Figures/}}

\usepackage[latin1]{inputenc}
\usepackage[english]{babel}

\def \simgt{\,\rlap{\lower 7.5 pt\hbox{$\mathchar \sim$}}\raise 3 pt \hbox{$>$}\,}
\def \simlt{\,\rlap{\lower 7.5 pt\hbox{$\mathchar \sim$}}\raise 3 pt \hbox{$<$}\,}

\def\lsim{\raise0.3ex\hbox{$<$\kern-0.75em\raise-1.1ex\hbox{$\sim$}}}
\def\gsim{\raise0.3ex\hbox{$>$\kern-0.75em\raise-1.1ex\hbox{$\sim$}}}
\newcommand{\hmu}{\hat{\mu}}

\newcommand{\be}{\begin{eqnarray}}
\newcommand{\ee}{\end{eqnarray}}

\newcommand{\ord}{\mathcal{O}}

\makeatletter
\ifx\@bibitemshut\undefined\let\@bibitemShut=\@empty\fi
\makeatother

\begin{document}
\title{The QCD Equation of State to \boldmath$\ord(\mu_B^6)$ from Lattice QCD}

\author{A. Bazavov}
\affiliation{Department of Computational Mathematics, Science and Engineering 
and Department of Physics and Astronomy, Michigan State University, 
East Lansing, MI 48824, USA}
\author{H.-T. Ding}
\affiliation{ Key Laboratory of Quark \& Lepton Physics (MOE) and Institute of
Particle Physics, Central China Normal University, Wuhan 430079, China}
\author{P. Hegde\footnote{prasad@chep.iisc.ernet.in}}
\affiliation{Center for High Energy Physics, Indian Institute of Science,
Bangalore 560012, India}
\author{O. Kaczmarek}
\affiliation{ Key Laboratory of Quark \& Lepton Physics (MOE) and Institute of
Particle Physics, Central China Normal University, Wuhan 430079, China}
\affiliation{Fakult\"at f\"ur Physik, Universit\"at Bielefeld, D-33615 Bielefeld,
Germany}
\author{F. Karsch}
\affiliation{Fakult\"at f\"ur Physik, Universit\"at Bielefeld, D-33615 Bielefeld,
Germany}
\affiliation{Physics Department, Brookhaven National Laboratory, Upton, NY 11973, USA}
\author{E. Laermann}
\affiliation{Fakult\"at f\"ur Physik, Universit\"at Bielefeld, D-33615 Bielefeld,
Germany}
\author{Y. Maezawa}
\affiliation{Yukawa Institute for Theoretical Physics, Kyoto University,  
Kyoto 606-8317, Japan}
\author{Swagato Mukherjee}
\affiliation{Physics Department, Brookhaven National Laboratory, Upton, NY 11973, USA}
\author{H. Ohno}
\affiliation{Physics Department, Brookhaven National Laboratory, Upton, NY 11973, USA}
\affiliation{Center for Computational Sciences, University of Tsukuba, Tsukuba,
Ibaraki 305-8577, Japan}
\author{P. Petreczky}
\affiliation{Physics Department, Brookhaven National Laboratory, Upton, NY 11973, USA}
\author{H. Sandmeyer}
\affiliation{Fakult\"at f\"ur Physik, Universit\"at Bielefeld, D-33615 Bielefeld,
Germany}
\author{P. Steinbrecher}
\affiliation{Fakult\"at f\"ur Physik, Universit\"at Bielefeld, D-33615 Bielefeld,
Germany}
\affiliation{Physics Department, Brookhaven National Laboratory, Upton, NY 11973, USA}
\author{C. Schmidt}
\affiliation{Fakult\"at f\"ur Physik, Universit\"at Bielefeld, D-33615 Bielefeld,
Germany}
\author{S. Sharma}
\affiliation{Physics Department, Brookhaven National Laboratory, Upton, NY 11973, USA}
\author{W. Soeldner}
\affiliation{Institut f\"ur Theoretische Physik, Universit\"at Regensburg, D-93040
Regensburg, Germany}
\author{M. Wagner}
\affiliation{NVIDIA GmbH, D-52146 W\"urselen, Germany}

\begin{abstract}
We calculated 
the QCD equation of state using Taylor expansions that include
contributions from up to sixth order in the baryon, strangeness and 
electric charge chemical potentials. Calculations have been performed with 
the Highly Improved Staggered Quark action in the temperature range
$T\in [135~{\rm MeV}, 330~{\rm MeV}]$ using up to four different sets
of lattice cut-offs corresponding to lattices of size $N_\sigma^3\times N_\tau$
with aspect ratio $N_\sigma/N_\tau=4$ and $N_\tau =6-16$.
The strange quark mass is tuned to its
physical value and we use two strange to light quark mass ratios
$m_s/m_l=20$ and $27$, which in the continuum limit correspond to a pion 
mass of about $160$~MeV and $140$~MeV respectively.
Sixth-order results for Taylor expansion coefficients are used 
to estimate truncation errors of the fourth-order expansion. We show that
truncation errors are small for baryon chemical potentials less then
twice the temperature ($\mu_B\le 2T$). The
fourth-order equation of state thus is suitable for the modeling of dense 
matter created in heavy ion collisions with center-of-mass energies down 
to $\sqrt{s_{NN}}\sim 12$ GeV. We provide a parametrization of basic
thermodynamic quantities that can be readily used in hydrodynamic simulation
codes. 
The results on up to sixth order expansion coefficients
of bulk thermodynamics are used 
for the calculation of lines of constant pressure,
energy and entropy densities in the $T$-$\mu_B$ plane
and are compared with the crossover line for the QCD chiral transition
as well as with experimental results
on freeze-out parameters in heavy ion collisions.
These coefficients also provide estimates for the location of a 
possible critical point. We argue
that results on sixth order expansion coefficients disfavor the existence 
of a critical point in the QCD phase diagram for $\mu_B/T\le 2$ and
$T/T_c(\mu_B=0) > 0.9$.

\vspace{0.2in}
\begin{center}
\bf{\today}
\end{center}
\end{abstract}

\vspace{0.2in}

\maketitle
\section{Introduction}
\label{sec:intro}
The temperature and density dependence of bulk thermodynamic quantities,
commonly summarized as the equation of state (EoS), provide the most basic 
characterization of equilibrium properties of strong-interaction matter.
Its analysis within the framework 
of lattice regularized Quantum Chromodynamics (QCD) has been refined ever since the 
early calculations performed in pure $SU(N)$ gauge theories 
\cite{Engels:1980ty}. Quite recently, the continuum extrapolated 
results for the EoS of QCD with physical light and strange quark masses have
been calculated
\cite{Borsanyi:2013bia,Bazavov:2014pvz}. Bulk thermodynamic observables such as pressure 
($P$), energy density ($\epsilon$) and entropy density ($s$) as well as second order 
quantities such as the specific heat ($C_V$) and velocity of sound ($c_s$) have now
been obtained at vanishing chemical potentials for the three quark flavors 
$(\mu_u, \mu_d, \mu_s)$. In accordance with the analysis of the chiral transition
temperature, $T_c\simeq (154\pm 9)$~MeV \cite{Bazavov:2011nk}, bulk 
thermodynamic observables change smoothly in the transition region. 
At low temperature they are
found to be in quite good agreement with hadron resonance gas (HRG) model calculations,
although some systematic deviations have been observed, which may be 
attributed to the existence of additional resonances which are not taken into 
account in HRG model calculations based on well established resonances listed 
in the particle data tables \cite{Majumder:2010ik,Bazavov:2014xya}.

The EoS at vanishing chemical potentials does already provide important input into the 
modelling of the hydrodynamic evolution of hot and dense matter created in heavy ion 
collisions. While this is appropriate for the thermal conditions met in these 
collisions
at the LHC and the highest RHIC beam energies, knowledge of the EoS at non-vanishing
baryon ($\mu_B$), strangeness ($\mu_S$) and electric charge ($\mu_Q$) chemical potentials is 
indispensable for the hydrodynamic modelling of the conditions met in the beam energy scan (BES)
at RHIC. Due to the well-known sign problem for lattice QCD formulations at non-zero chemical
potential a direct calculation of the EoS at non-zero $(\mu_B,\ \mu_Q,\ \mu_S)$ is 
unfortunately not yet possible. 
At least for small values of the chemical potentials this can be circumvented
by using a Taylor expansion of the thermodynamic potential \cite{Gavai:2001fr, Allton:2002zi}.
In this way some results for EoS at non-zero baryon chemical potential have been
obtained on coarse lattices \cite{Allton:2002zi,Gavai:2003mf,Allton:2003vx}. 
These calculations have even been extended to 
sixth order in the baryon chemical potential \cite{Allton:2005gk,Ejiri:2005uv}.
First continuum extrapolated results for the EoS using second order Taylor 
expansion
coefficients have been obtained within the stout discretization scheme for 
staggered fermions \cite{Borsanyi:2012cr} and simulations at imaginary 
chemical potential have been used to arrive at a sixth order result for the
QCD EoS \cite{Gunther:2016vcp} and up to eighth order for some 
generalized susceptibilities \cite{DElia:2016jqh} through analytic continuation. 

Results for higher order expansion coefficients are clearly needed if one wants to 
cover the range of chemical potentials, $0\le \mu_B/T \lsim\ 3$ that is expected to be 
explored with the BES at RHIC by varying the beam energies in the range 
$7.7~{\rm GeV}\le \sqrt{s_{NN}} \le 200~{\rm GeV}$. Of course,  the Taylor 
expansions  will break down, should the elusive critical point in the QCD 
phase diagram 
\cite{CEP,Stephanov} 
turn out to be present in this range of baryon chemical potentials. 
The convergence of the series thus needs to be monitored carefully.

This paper is organized as follows. In the next section we briefly discuss 
Taylor series for a HRG model in Boltzmann approximation. This helps to
argue for the significance of sixth order Taylor expansions. In 
Section ~\ref{sec:outline}, we present the basic framework of Taylor series
expansions, introduce expansions in the presence of global constraints and
discuss some details of our calculations and the ensembles used. In 
Section~\ref{sec:eos0} we discuss the $6^{th}$ order Taylor expansion of
QCD thermodynamics in the simplified case of vanishing strangeness and 
electric charge chemical potentials. Section~\ref{sec:eosSN} is devoted to 
the corresponding discussion of strangeness neutral systems $n_S=0$ with 
fixed net electric charge ($n_Q$) to net baryon-number ($n_B$) ratio, which is 
of relevance  for the description of hot and dense matter formed in heavy ion 
collisions where typically $n_Q/n_B\simeq 0.4$.
We discuss the relevance of a non-vanishing electric charge chemical 
potential by considering electric charge neutral ($n_Q/n_B=0$) as well as
isospin symmetric ($n_Q/n_B=1/2$) systems.
At the end of this section we present a parametrization of the equation
of state that can easily be used as input for the modeling of the thermal
conditions met in heavy ion collisions. In Section~\ref{sec:LCP} we present results
on lines of constant pressure, energy density and entropy density and compare
their dependence on $\mu_B$ with empirical results for the freeze-out
conditions observed in heavy ion collisions. We comment on the radius 
of convergence of the Taylor series for the pressure and resulting 
constraints for the location of a possible critical point in 
Section~\ref{sec:radius}.  Finally we present our 
conclusions in Section ~\ref{sec:conclusion}. Details on (A) the statistics
and simulation parameters, (B) explicit expressions for the expansions
of electric charge and baryon number chemical potentials, 
and (C) explicit
expressions for the expansion parameters of the lines of constant
physics are given in three Appendices A-C.

\section{Taylor expansions and the low and high temperature limits of strong interaction matter}
\label{sec:HRG}

The main aim of this work is to supply a EoS of strong-interaction matter 
using up to sixth order Taylor expansions for bulk thermodynamic observables. 
As we will see later at present results on sixth order expansion coefficients
in the Taylor series will mainly help to constrain 
truncation errors in the fourth order expansion rather than providing
accurate results on the sixth order contribution to thermodynamic 
quantities. 
We will argue that  our analysis provides reliable results for the EoS for 
baryon chemical potentials up to $\mu_B/T \simeq 2$ at temperatures below
$T\simeq 160$~MeV and for an even larger range 
in $\mu_B/T$ at higher temperatures.
  
Before turning to a discussion of lattice QCD results on the EoS, it may be 
useful to analyze truncation effects
in the hadron resonance gas (HRG) model, which seems to provide a good approximation for thermodynamics in
the low temperature, hadronic regime. For simplicity let us consider the case of vanishing electric
charge and strangeness chemical potentials, $\mu_Q=\mu_S=0$. 
At temperatures close to the transition temperature $T_c\simeq 154$~MeV and
for baryon chemical potentials less than a few times the transition 
temperature,
the baryon sector of a HRG is well described in the Boltzmann
approximation. In a HRG model calculation based on non-interacting hadrons
the pressure may then be  written as
\begin{eqnarray}
P(T,\mu_B)&=&P_M(T)+P_B(T,\hmu_B) \nonumber \\
&=&P_M(T)+P_B(T,0) + P_B(T,0) \left( \cosh (\hmu_B) -1\right) \; ,
\label{PHRg}
\end{eqnarray}
where we introduced the notation $\hmu_B \equiv \mu_B/T$ and $P_{M}(T)$
($P_B(T,\hmu_B)$) 
denote the meson (baryon) contributions to the pressure.
A similar relation holds for the energy density,
\begin{eqnarray}
\epsilon(T,\mu_B)&=& \epsilon_M(T)+\epsilon_B(T,\hmu_B) 
\nonumber \\
&=& \epsilon_M(T)+\epsilon_B(T,0)+ \epsilon_B(T,0) 
\left( \cosh (\hmu_B) -1\right) \; ,
\label{EHRG}
\end{eqnarray}
with $\epsilon_{M/B} \equiv T^2 \left( \partial (P_{M/B}/T)/\partial T\right)_{\hmu_B}$. 
The $\mu_B$-dependent contribution 
thus is simple and can easily be represented by a Taylor series. Truncating 
this expansion at $(2n)$-th order we obtain
\begin{eqnarray}
(\Delta (P/T^4))_{2n} \equiv \frac{(P_B(T,\mu_B)- P_B(T,0))_{2n}}{T^4} &=& 
\sum_{k=1}^{n} \frac{\chi_{2k}^{B,HRG}(T)}{(2k)!}\hmu_B^{2k} 
\simeq \frac{P_B(T,0)}{T^4} \sum_{k=1}^{n} \frac{1}{(2k)!}\hmu_B^{2k}  \; ,
\end{eqnarray}
where in the last equality we made use of the fact that in HRG models
constructed from non-interacting, point-like hadrons, all expansion 
coefficients are identical when using a Boltzmann approximation for the 
baryon sector, i.e. all baryon number susceptibilities are identical,
$\chi_{2k}^{B,HRG}=P_B(T,0)$. The ratios of these susceptibilities 
are unity,   
$\chi_{2k}^{B,HRG}/\chi_{2(k-1)}^{B,HRG} =\chi_{2k}^{B,HRG}/\chi_{2}^{B,HRG} =1$.
Similarly one finds for the net baryon-number density,
\begin{equation}
\frac{n_B}{T^3}= \frac{P_B(T,0)}{T^4}\sinh{\hmu_B}=
\sum_{k=1}^{\infty} \frac{\chi_{2k}^{B,HRG}(T)}{(2k-1)!}\hmu_B^{2k-1} 
\simeq \frac{P_B(T,0)}{T^4} \sum_{k=1}^{\infty} \frac{1}{(2k-1)!}\hmu_B^{2k-1}  \; .
\end{equation}
Higher order
corrections are thus more important in the net baryon-number density than
in the expansions of the pressure or energy density. For instance,
the contribution to $\mu_B n_B/T^4$ at  
${\cal O}(\hmu_B^{2k})$ 
is a factor $2k$ larger than the corresponding ${\cal O}(\hmu_B^{2k})$
expansion coefficient of the pressure.

\begin{figure}[t]
\begin{center}
\includegraphics[width=0.50\textwidth]{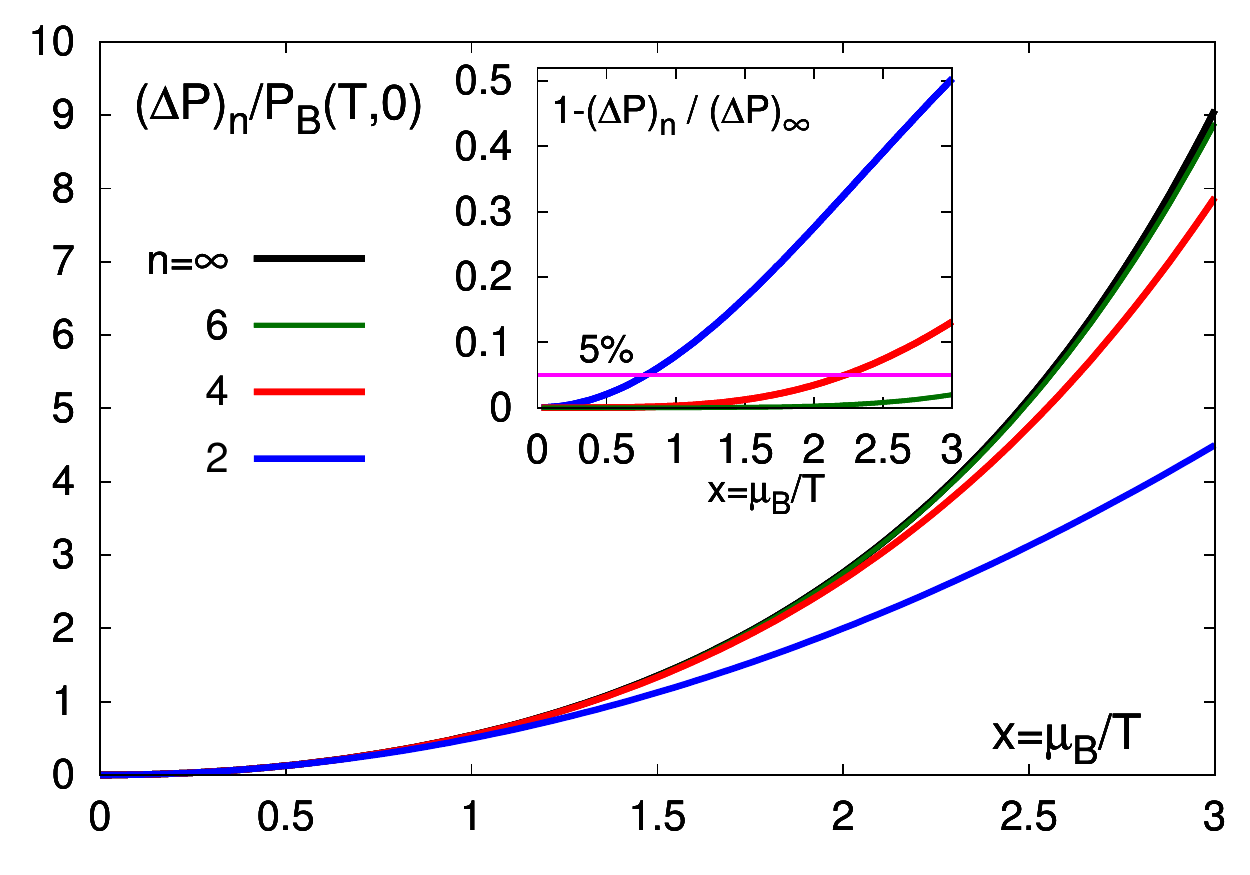}
\end{center}
\caption{$n$-th order Taylor series, $(\Delta P)_n$ for
$(\Delta P)_\infty=P_B(T,0)(\cosh(x)-1)$
compared to the exact result. The insertion shows the relative
error due to truncation of the Taylor series after $n$-th order.
Note that the sixth order result is hardly visible behind the
exact result.
}
\label{fig:cosh}
\end{figure}

In Fig.~\ref{fig:cosh} we show results from a Taylor series expansion
of the $\mu_B$-dependent part of the pressure in a HRG model truncated after 
leading order (LO),  next-to-leading order (NLO) and next-to-next-to-leading 
(NNLO) order. 
These truncated expansions are compared to the exact result, i.e.
$\left( \Delta P \right)_\infty (T)=P_B(T) (\cosh(\hmu_B)-1)$. The insertion 
shows the deviation of the $n$-th order truncated Taylor series 
($\left( \Delta P \right)_n(T)$) from
the exact result ($\left( \Delta P \right)_\infty(T)$).
As can be seen already the fourth order Taylor series provides a good
approximation for the pressure (and energy as well as entropy density)
of a HRG for all $\mu_B \le 2 T$. At $\mu_B=2T$ the fourth order Taylor
series for the $\mu_B$-dependent contribution to the pressure deviates
by less than 5\% from the exact result. These deviations are, of course,
even smaller in the total pressure which in the temperature 
range of interest is dominated by the meson contribution. Even at
$T = 170$~MeV, which certainly is already above the range of 
applicability of HRG models, the baryonic contribution to the 
pressure (energy density) amounts only to about 20\% (30\%). 
A 5\% truncation error 
in the $\mu_B$-dependent contribution to the pressure or energy density 
thus amounts to less than a 2\% effect in the total pressure or 
energy density.
Similar estimates hold for the more general case of 
non-vanishing $\mu_Q$ and $\mu_S$. 

Of course, the good convergence properties of the Taylor series for
the pressure in HRG models also reflect that the radius of convergence of 
this series is infinite. If there exists a critical point in the QCD 
phase diagram one cannot expect to find that the Taylor series is that
well behaved. Still the HRG result provides
a benchmark also for the QCD case. If the radius of convergence of 
the Taylor series for the QCD pressure is finite and, in particular, smaller
than $\mu_B \simeq 3 T$, one should find large deviations in the generalized 
susceptibilities from the corresponding HRG results. Ratios
of susceptibilities have to grow  asymptotically like,
$\chi_{2k}^{B,QCD}/\chi_{2(k-1)}^{B,QCD}\sim k^2$ in order to yield
a finite radius of convergence for a Taylor expansion. We will 
come back to a discussion of this asymptotic behavior after having 
discussed our sixth order calculation of Taylor expansion coefficients.

Let us briefly mention also the high temperature limit. At large
values of the temperature, the pressure approaches that of a massless
ideal gas of quarks and gluons. In this limit the pressure is just
a second order polynomial in $\hmu^2_f$,
\begin{eqnarray}
\frac{P_{ideal}}{T^4}  =  \frac{8 \pi^2}{45} + 
 \sum_{f=u,d,s} \left[\frac{7 \pi^2}{60} +
\frac{1}{2}  \left(\frac{\mu_f}{T}\right)^2 
+ \frac{1}{4 \pi^2} \left(\frac{\mu_f}{T}\right)^4 
\right] \quad ,
\label{free}
\end{eqnarray}
In this limit a fourth order Taylor expansion thus provides the exact results
for the basic bulk thermodynamic observables. This also is correct
in leading order perturbation theory, i.e.  at ${\cal O} (g^2)$ \cite{Vuorinen:2003fs}.

\section{Outline of the Calculation}
\label{sec:outline}
\subsection{Taylor series in baryon number, electric charge and strangeness
chemical potentials}
Our goal is the calculation of Taylor expansion coefficients for basic
bulk thermodynamic observables of strong-interaction matter
in terms of chemical potentials $\mu_X$ for conserved charges 
($X=B,\ Q,\ S)$. We start with the expansion of the pressure, $P$, in terms 
of the dimensionless ratios
$\hat{\mu}_X\equiv \mu_X/T$, which are the logarithms of fugacities,
\begin{equation}
\frac{P}{T^4} = \frac{1}{VT^3}\ln\mathcal{Z}(T,V,\hmu_u,\hmu_d,\hmu_s) = \sum_{i,j,k=0}^\infty%
\frac{\chi_{ijk}^{BQS}}{i!j!\,k!} \hmu_B^i \hmu_Q^j \hmu_S^k \; ,
\label{Pdefinition}
\end{equation}
with $\chi_{000}^{BQS}\equiv P(T,0)/T^4$.
The chemical potentials for conserved charges are related to the quark 
chemical potentials 
$(\mu_u, \mu_d, \mu_s)$,
\begin{eqnarray}
\mu_u&=&\frac{1}{3}\mu_B + \frac{2}{3}\mu_Q \; , \nonumber \\
\mu_d&=&\frac{1}{3}\mu_B - \frac{1}{3}\mu_Q \; ,\nonumber \\
\mu_s&=&\frac{1}{3}\mu_B - \frac{1}{3}\mu_Q - \mu_S \; .
\label{potential}
\end{eqnarray}
The expansion coefficients $\chi_{ijk}^{BQS}$, i.e. the so-called 
generalized susceptibilities, can be calculated at vanishing chemical 
potential\footnote{We often suppress the argument ($T$) of the generalized 
susceptibilities. We also suppress superscripts and subscripts of 
$\chi_{ijk}^{BQS}$ whenever one of the subscripts vanishes, e.g. 
$\chi_{i0k}^{BQS}\equiv \chi_{ik}^{BS}$.},
\begin{equation}
\chi_{ijk}^{BQS}\equiv \chi_{ijk}^{BQS}(T) =\left. 
\frac{\partial P(T,\hmu)/T^4}{\partial\hmu_B^i \partial\hmu_Q^j \partial\hmu_S^k}\right|_{\hmu=0} \; .
\label{suscept}
\end{equation}
From Eq.~\ref{Pdefinition} it is straightforward to obtain the Taylor series 
for the number densities,
\begin{equation}
\frac{n_X}{T^3} = \frac{\partial P/T^4}{\partial \hmu_X} \; ,\; X=B,\ Q,\ S \;.
\label{number}
\end{equation}
This only requires knowledge of the expansion coefficients entering
the series for $P/T^4$. 
The energy ($\epsilon$) and entropy ($s$) densities, on the other hand, also
require derivatives of the generalized susceptibilities with respect 
to temperature, which are the expansion coefficients of the trace 
anomaly, 
\begin{equation}
\Delta(T,\hmu_B,\hmu_Q,\hmu_S)\equiv \frac{\epsilon - 3P}{T^4} =
T \frac{\partial P/T^4}{\partial T} 
=\sum_{i,j,k=0}^\infty%
\frac{\Xi_{ijk}^{BQS}}{i!j!\,k!}%
\hmu_B^i \hmu_Q^j \hmu_S^k  \; ,
\label{e3p}
\end{equation}
with $i+j+k$ even and
\begin{equation}
\Xi_{ijk}^{BQS}(T) = T \frac{{\rm d} \chi_{ijk}^{BQS}(T)}{{\rm d} T}\; .
\label{chiT}
\end{equation}
With this one finds for the Taylor expansions
of the energy and entropy densities,
\begin{eqnarray}
\frac{\epsilon}{T^4} &=& 
\sum_{i,j,k=0}^\infty
\frac{ \Xi_{ijk}^{BQS}  +3\chi_{ijk}^{BQS}}{i!\,j!\,k!}
\hmu_B^i \hmu_Q^j \hmu_S^k 
\; ,
\label{energy}
\\
\frac{s}{T^3} &=& 
\frac{\epsilon +p -\mu_B n_B - \mu_Q n_Q -\mu_S n_S}{T^4}
\nonumber \\
&=& \sum_{i,j,k=0}^\infty
\frac{\Xi_{ijk}^{BQS}  +(4-i-j-k)\chi_{ijk}^{BQS}}{i!j!\,k!}
\hmu_B^i \hmu_Q^j \hmu_S^k \; .
\label{entropy}
\end{eqnarray}

\subsection{Constrained series expansions}
\label{sec:constraint}

In our calculations we generated all generalized susceptibilities
up to $6^{th}$ order, which are needed to set up the general Taylor series 
in terms 
of the three conserved charge chemical potentials as discussed in the previous 
subsection. In the following we will, however, consider only thermodynamic
systems, in which the 
electric charge and strangeness chemical potentials are fixed by additional 
constraints and become functions of the baryon chemical potential and 
temperature. 
We only consider constraints that can be fulfilled order by order in the 
Taylor series
expansion. That is, for the construction of the $6^{th}$ order Taylor series 
of the pressure in terms of $\hmu_B$ we need to know the expansion of 
$\hmu_Q(T,\mu_B)$ and $\hmu_S(T,\mu_B)$ up to
fifth order in $\hmu_B$,
\begin{eqnarray}
\hmu_Q(T,\mu_B) &=& q_1(T)\hmu_B + q_3(T) \hmu_B^3+q_5(T) \hmu_B^5+ \dots \; , \nonumber \\
\hmu_S(T,\mu_B) &=& s_1(T)\hmu_B + s_3(T) \hmu_B^3+s_5(T) \hmu_B^5 +\dots \; . 
\label{qs}
\end{eqnarray}
The above parametrization includes the cases of vanishing electric charge
and strangeness chemical potentials, $\mu_Q=\mu_S=0$, which we are going 
to discuss in the next section as well as the strangeness neutral case
with fixed electric charge to baryon-number ratio, which we will 
analyze in Section~\ref{sec:eosSN}.

Implementing the constraints specified in Eq.~\ref{qs} in the Taylor
series for the pressure and net conserved-charge number densities one obtains 
series in terms of the baryon chemical potential only, 
\begin{eqnarray}
\frac{P(T,\mu_B)}{T^4} -\frac{P(T,0)}{T^4} &=& 
\sum_{k=1}^{\infty} P_{2k}(T) \hmu_B^{2k} \; ,
\label{Pn} \\
\frac{n_X}{T^3} &=& \sum_{k=1}^{\infty} N_{2k-1}^X \hmu_B^{2k-1} \;\; ,\;\;
X=B,\ Q,\ S \;\; .
\label{nX}
\end{eqnarray}
Using
\begin{equation}
\hmu_B \frac{{\rm d} P/T^4}{{\rm d}\hmu_B}  = \hmu_B \frac{n_B}{T^3} +
\hmu_B \frac{{\rm d} \hmu_Q}{{\rm d}\hmu_B} \frac{n_Q}{T^3} +
\hmu_B \frac{{\rm d} \hmu_S}{{\rm d}\hmu_B} \frac{n_S}{T^3}\; ,
\end{equation} 
and the series expansions of $\hmu_Q$ and $\hmu_S$ given in Eq.~\ref{qs}
one easily finds the relation between the expansion coefficients for 
the pressure and number densities,
\begin{equation}
P_{2n} = \frac{1}{2n}\left( N_{2n-1}^B+ 
\sum_{k=1}^{n} (2k-1) \left( s_{2k-1} N^S_{2n-2k+1} +
q_{2k-1} N^Q_{2n-2k+1}\right)
\right) \; .
\label{PNX}
\end{equation}

When imposing constraints on the electric charge and strangeness
chemical potentials, these generally become temperature dependent
functions as indicated in Eq.~\ref{qs}. The temperature derivative
of $P/T^4$ at fixed $\hmu_B$ in the constraint case and the partial
derivative of $P/T^4$ at fixed $(\hmu_B,\hmu_Q, \hmu_S)$, which defines
the trace anomaly $\Delta$ (Eq.~\ref{e3p}), thus are related through
\begin{eqnarray}
T\frac{{\rm d}P/T^4}{{\rm d}T} &=&
\Delta + T\hmu_Q' \frac{n_Q}{T^3}+T \hmu_S' \frac{n_S}{T^3} \; ,
\label{total} 
\end{eqnarray}
where the (total) temperature derivative ${\rm d}/{\rm d}T$ is taken at 
fixed $\hmu_B$
and $\hmu_X'={\rm d}\hmu_X/{\rm d}T$. With this we obtain the Taylor series
for the trace anomaly,
\begin{equation}
\Delta(T,\hmu_B) =
\frac{\epsilon -3P}{T^4}= \left(\frac{\epsilon -3P}{T^4}\right)_{\hmu_B=0}
+\sum_{n=1}^\infty (T P_{2n}'(T) - h_{2n}(T)) \hmu_B^{2n} \; ,
\label{tracea}
\end{equation}
with 
\begin{equation}
h_{2n} =
\sum_{k=1}^{n} \left( s'_{2k-1} N^S_{2n-2k+1} +
q'_{2k-1} N^Q_{2n-2k+1}\right) \; .
\end{equation}
We also introduce 
\begin{equation}
t_{2n} =
\sum_{k=1}^{n} \left( s_{2k-1} N^S_{2n-2k+1} +
q_{2k-1} N^Q_{2n-2k+1}\right) \; .
\end{equation}
With this the Taylor series expansion of the energy 
and entropy densities for constraint cases, in which $\hmu_Q$ and
$\hmu_S$ satisfy Eq.~\ref{qs}, becomes
\begin{eqnarray}
\frac{\epsilon(T,\mu_B)}{T^4} - \frac{\epsilon(T,0)}{T^4} &=& 
\sum_{n=1}^{\infty} \epsilon_{2n}(T) \hmu_B^{2n}\; ,
\label{energyc} \\
\frac{s(T,\mu_B)}{T^3} - \frac{s(T,0)}{T^3} &=& \sum_{n=1}^{\infty} \sigma_{2n}(T) \hmu_B^{2n}\; .
\label{entropyc}
\end{eqnarray}
with $\epsilon_{2n}(T) =3 P_{2n}(T)+T P_{2n}'(T) - h_{2n}(T)$
and $\sigma_{2k}(t) = 
4 P_{2n}(T)+TP_{2n}'(T) -N_{2n-1}^B(T)- h_{2n}(T)-t_{2n}(T)$.

\subsection{Numerical calculation of generalized susceptibilities up to
\boldmath${\cal O}(\mu^6)$}

The generalized susceptibilities $\chi_{ijk}^{BQS}$ have been calculated on 
gauge field configurations generated for (2+1)-flavor QCD using the 
Highly Improved Staggered Quark (HISQ) action \cite{Follana:2006rc} 
and the tree-level improved Symanzik gauge action. 

All calculations are performed using a strange quark mass $m_s$ tuned to its 
physical value. We performed calculations with two different light to strange
quark mass ratios, $m_l/m_s= 1/27$ and $1/20$. The former corresponds to
a pseudo-scalar Goldstone mass, which in the continuum limit yields a
pion mass $m_\pi\simeq 140$~MeV, the latter leads to a pion mass
$m_\pi\simeq 160$~MeV. 
These parameters are fixed using the line of constant
physics determined by HotQCD from the $f_K$ scale.
Using $f_K= 155.7(9)/\sqrt{2}$~MeV allows to determine the lattice spacing 
$a(\beta)$ at a given value of the gauge coupling $\beta$ and the 
corresponding set of quark masses ($m_l,m_s)$, which in turn fixes the 
temperature on a lattice with temporal extent 
$N_\tau$, i.e. $T=(N_\tau a)^{-1}$. More details on the scale determination
are given in \cite{Bazavov:2011nk}.

All calculations have been performed on lattices of size $N_\sigma^3 N_\tau$
with an aspect ratio $N_\sigma/N_\tau = 4$. We perform calculations
in the temperature interval $T\in [135~{\rm MeV},330~{\rm MeV}]$
using lattices with temporal extent $N_\tau=6,\ 8,\ 12$ and $16$, which 
corresponds to four different values of the lattice spacings at fixed 
temperature.
At temperatures $T\le 175$~MeV all calculations have been performed with the 
lighter, physical quark mass ratio $m_l/m_s= 1/27$. 
In the high temperature 
region quark mass effects
are small and we based our calculations on existing data sets for
$m_l/m_s= 1/20$, which have previously been
generated by the HotQCD collaboration and used for the calculation
of second order susceptibilities \cite{Bazavov:2012jq}. These data sets have 
been extended for the calculation of higher order susceptibilities. 
Gauge field configurations are stored after every $10^{th}$ molecular
dynamics trajectory of unit length. 

All calculations of $4^{th}$ and $6^{th}$ order expansion coefficients
have been performed on lattices with temporal extent $N_\tau=6$ and 8.
In these cases we gathered a large amount of statistics. 
At low temperatures we have generated up to 
$1.2$ million trajectories for $N_\tau=6$ and up to $1.8$ million
trajectories for $N_\tau=8$. At high temperature less than a tenth of 
this statistics turned out to be sufficient.
The $2^{nd}$ order expansion coefficients have been calculated on lattices
with four different temporal extends, $N_\tau=6,\ 8,\ 12,\ 16$. At fixed
temperature this corresponds to four different values of the lattice 
cut-off, which we used to extract continuum extrapolated results for the
second order expansion coefficients. 
We also extrapolated results for the higher order expansion coefficients
to the continuum limit. However, having at hand results from only
two lattice spacings for these expansion coefficients we consider these
extrapolations as estimates of the results in the continuum limit.

On each configuration the traces of all operators needed to construct
up to sixth order Taylor expansion coefficients have been calculated 
stochastically. 
For the calculation of $2^{nd}$ and $4^{th}$ order expansion 
coefficients we follow the standard approach of introducing a non-zero 
chemical potential in the QCD Lagrangian as an 
exponential prefactor for time-like gauge field variables 
\cite{Hasenfratz:1983ba},
i.e. the chemical potential $\mu_f$ for quark flavor $f$ is introduced through 
a factor ${\rm e}^{ \mu_fa}$ (${\rm e}^{- \mu_fa}$) 
on time-like links directed in the forward
(backward) direction. This insures that all observables calculated
are free of ultra-violet divergences. 
For the calculation of all $6^{th}$ 
order expansion coefficients we use the so-called linear-$\mu$ approach 
\cite{Gavai:2011uk,Gavai:2014lia}. This becomes possible as no 
ultra-violet divergences appear in $6^{th}$ order cumulants and above.
In the linear-$\mu$ formulation the number of operators that contribute
to cumulants is drastically reduced and their structure is simplified.
All operators appearing in the exponential formulation, that involve second
or higher order derivatives of the fermion matrix \cite{Allton:2005gk}, 
vanish. The remaining operators are identical in both formulations. One 
thus only has to calculate traces of observables that are of the form,
\begin{equation}
{\rm Tr} M_f^{-1} M_f' M_f^{-1} M_f'.... M_f^{-1} M_f' \; ,
\nonumber
\label{traces}
\end{equation}
where $M_f$ is the staggered fermion matrix for light ($f=l$) or 
strange ($f=s$) quarks, respectively, and  
$M_f'$ denotes its derivative with respect to the flavor chemical 
potential $\hmu_f$.  
The final error on these traces depends on
the noise due to the use of stochastic estimators for the inversion of the
fermion matrices $M_f$, as well as on
the gauge noise resulting from a finite set of gauge configurations that
get analyzed. We analyzed the signal to noise ratio for all traces
of operators that we calculate and identified the operator $D_1=M_f^{-1}M'_f$,
as being particularly sensitive to the stochastic noise contribution.
This operator has been measured using 2000 random noise vectors.
For the calculation of traces of all other operators we used
500 random noise vectors. We checked that this suffices to reduce
the stochastic noise well below the gauge noise.
The simulation parameters 
and the statistics accumulated in this calculation are summarized in
the tables of Appendix~\ref{app:statistics}.

All fits and continuum extrapolations shown in the following are
based on spline interpolations with coefficients that are allowed
to depend quadratically on the inverse temporal lattice size. 
Our fitting ansatz and the strategy followed to arrive at continuum
extrapolated results are described in detail in Ref.~\cite{Bazavov:2014pvz}.
For the current analysis we found it sufficient to use spline interpolations
with quartic polynomials and 3 knots whose location is allowed to vary 
in the fit range.

\section{Equation of state for \boldmath $\mu_Q=\mu_S=0$}
\label{sec:eos0}

Let us first discuss the Taylor expansion for bulk thermodynamic
observables in the case of vanishing electric charge and strangeness
chemical potentials. This greatly simplifies the discussion and yet
incorporates all the features of the more general case. Also the discussion
of truncation errors presented in this section carries over to the
more general situation.

\subsection{Pressure and net baryon-number density}

For $\mu_Q=\mu_S=0$ the Taylor expansion coefficients $P_{2n}$ and 
$N_{2n-1}^B$, introduced in Eqs.~\ref{Pn} and \ref{nX}, are simply 
related by 
\begin{equation}
P_{2n} = \frac{1}{2n} N_{2n-1}^B = \frac{1}{(2n)!} \chi_{2n}^B \; .
\label{Pncoef}
\end{equation}
The series for the pressure and net baryon-number density simplify to,
\begin{eqnarray}
\frac{P(T,\mu_B)-P(T,0)}{T^4} &=&  
\sum_{n=1}^\infty \frac{\chi_{2n}^{B}(T)}{(2n)!}
\left(\frac{\mu_B}{T}\right)^{2n}  
= \frac{1}{2} \chi_2^B(T) \hmu_B^2 \left( 1+ \frac{1}{12}
\frac{\chi_{4}^B(T)}{\chi_2^B(T)} \hmu_B^2
+ \frac{1}{360}\frac{\chi_{6}^B(T)}{\chi_2^B(T)} \hmu_B^4+\; ... \right)
\; ,
\label{PmuB}\\
\frac{n_B}{T^3} &= & 
\sum_{n=1}^\infty \frac{\chi_{2n}^{B}(T)}{(2n-1)!}
\hmu_B^{2n-1} 
= \chi_2^B(T) \hmu_B \left( 1+ \frac{1}{6}
\frac{\chi_{4}^B(T)}{\chi_2^B(T)} \hmu_B^2
+ \frac{1}{120}\frac{\chi_{6}^B(T)}{\chi_2^B(T)} \hmu_B^4+\; ... \right)
\; .
\label{nmuB}
\end{eqnarray} 

\begin{figure}[t]
\includegraphics[scale=0.7]{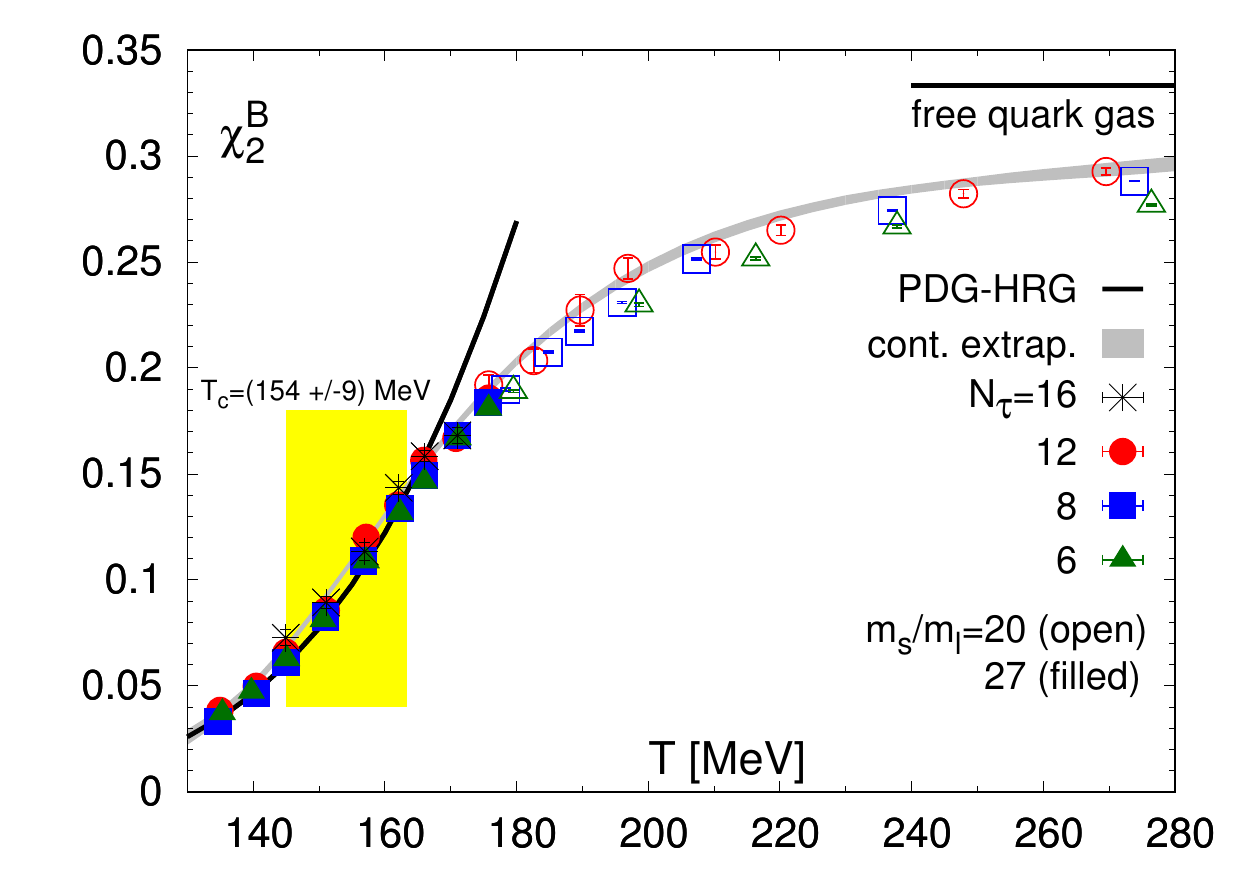}
\includegraphics[scale=0.7]{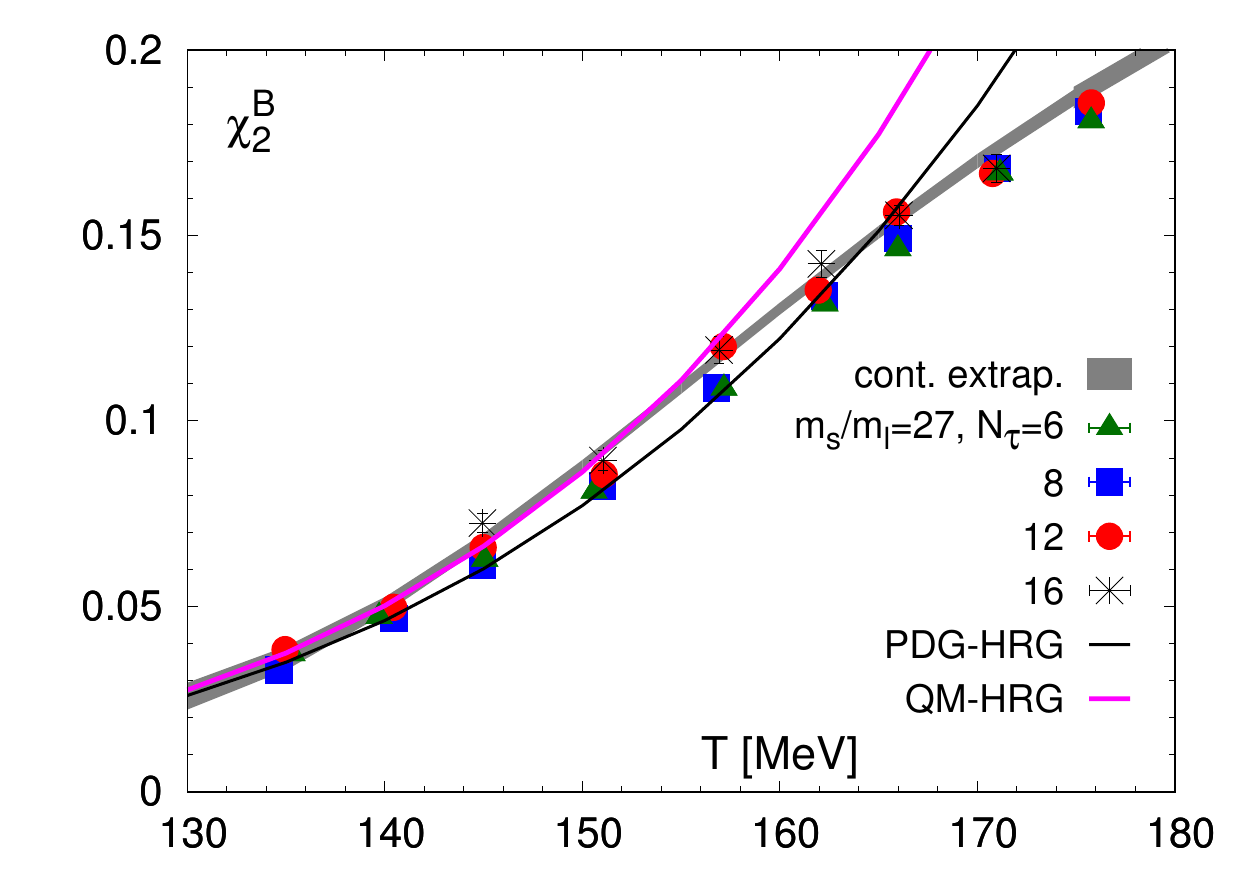}
\caption{The leading order (${\cal O}(\mu_B^2)$) correction to the pressure
calculated at zero baryon chemical potential. The left hand figure shows the leading
order correction in a large temperature range. The right hand part of the 
figure shows an enlarged view into the low temperature region. In addition to
the continuum extrapolation of the lattice QCD results we also show results
from HRG model calculations based on all hadron resonances listed
by the particle data group (PDG-HRG) and obtained in quark model
calculations (QM-PDG).
}
\label{fig:Bcumulants}
\end{figure}

In Eqs.~\ref{PmuB} and \ref{nmuB} we have factored out the leading order (LO)
$\mu_B$-dependent part in the series for the pressure as well as the net 
baryon-number density. This helps to develop a feeling for the importance 
of higher order
contributions and, in particular, the approach to the HRG limit at
low temperatures. Note that all ratios $\chi_{2n}^B/\chi_2^B$ are unity
in a HRG and, in the infinite temperature, ideal quark gas limit, 
$\chi_{4}^B/\chi_2^B=2/(3\pi^2)\simeq 0.068$ is the only
non-vanishing higher order expansion coefficient. From Eqs.~\ref{PmuB} and 
\ref{nmuB} it is evident that contributions from higher order expansion 
coefficients become more important in the number density than in the pressure. 
Relative to the LO result, the contributions of the NLO and NNLO expansion 
coefficients for $n_B/T^3$ are a factor two and three larger respectively 
than for the corresponding expansion coefficients in the pressure
series.

\begin{figure}[t]
\includegraphics[scale=0.68]{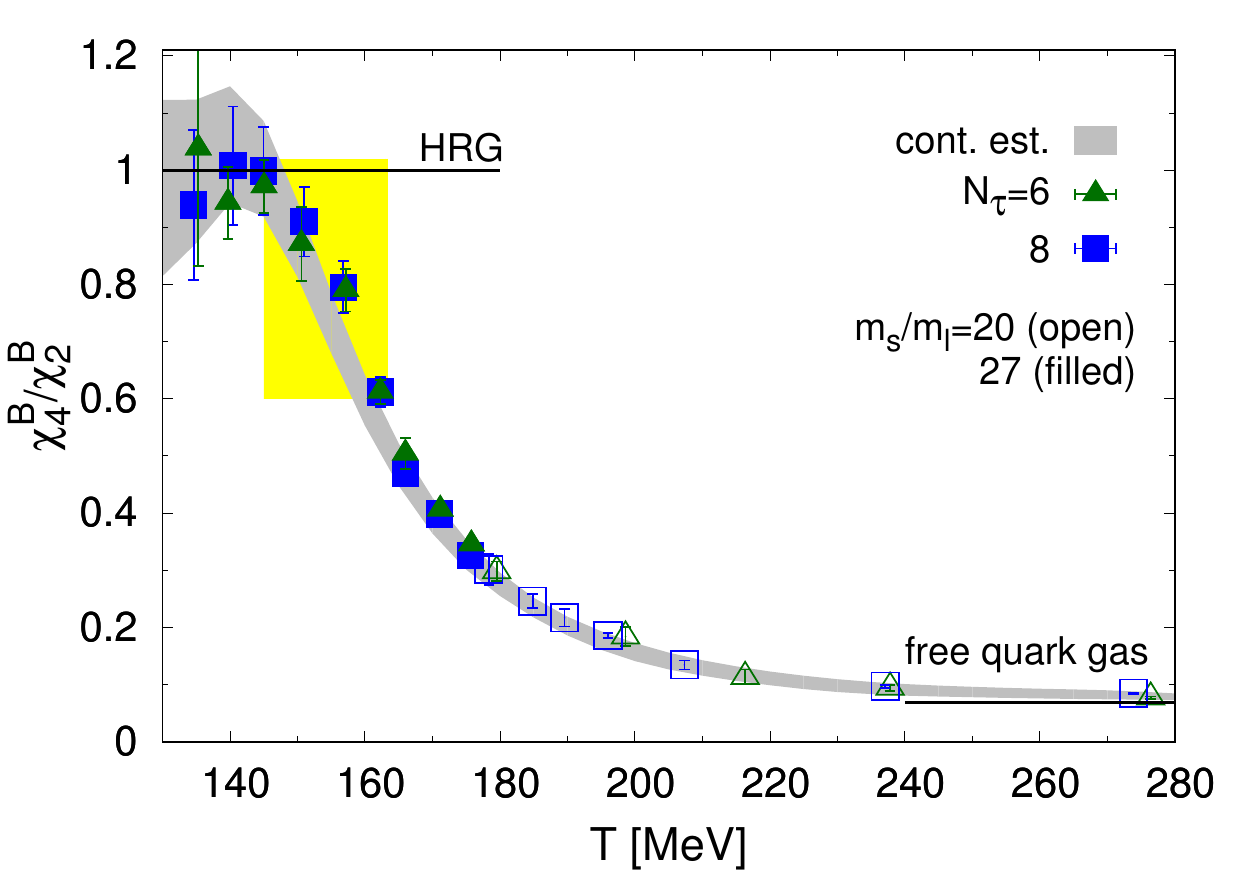}
\includegraphics[scale=0.68]{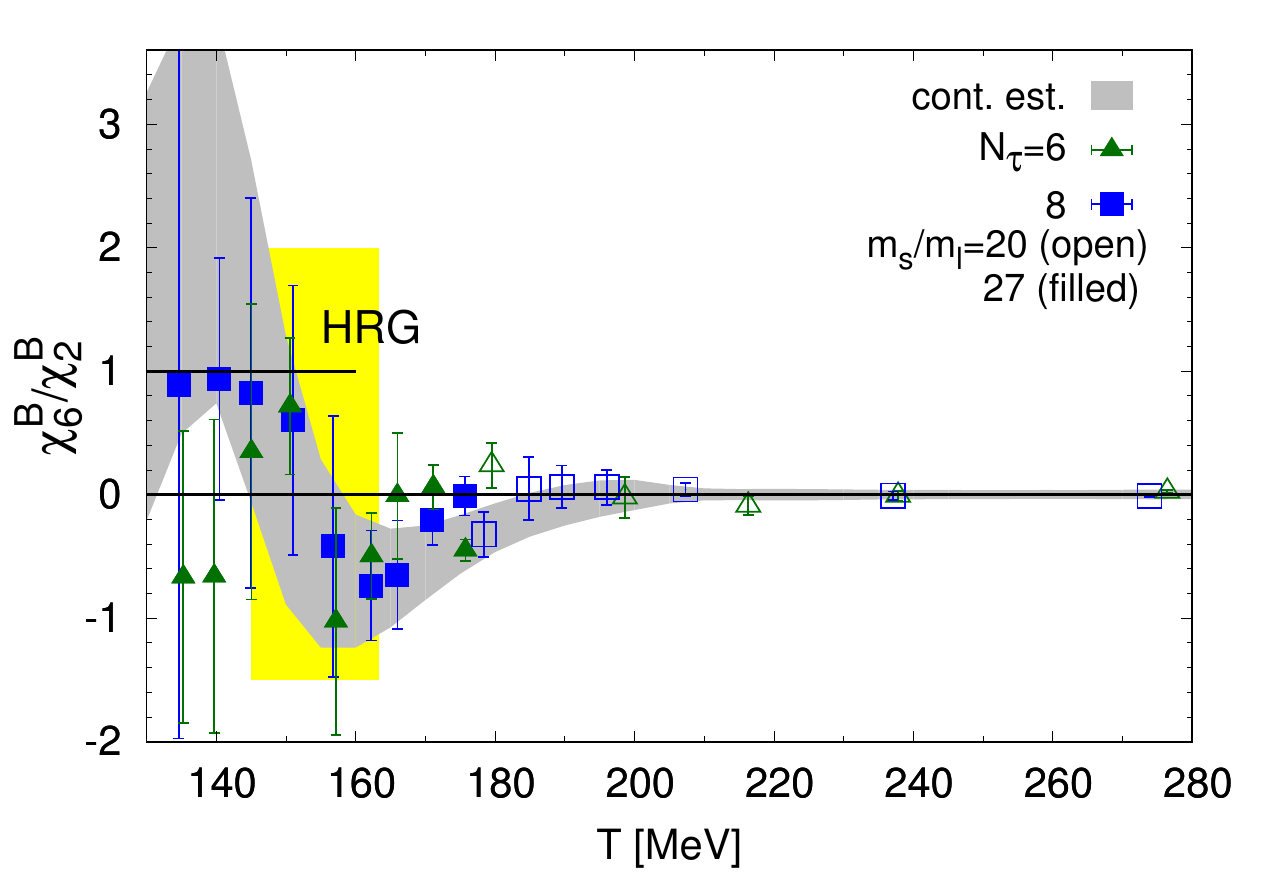}
\caption{{\it Left:}
The ratio of fourth and second order cumulants of net-baryon
number fluctuations ($\chi_4^B/\chi_2^B$) versus temperature.
{\it Right:} same as the left hand side, but for the ratio of
sixth and second order cumulants of net-baryon
number fluctuations ($\chi_6^B/\chi_2^B$). The boxes indicate the
transition region, $T_c=(154\pm 9)$~MeV. Grey bands show  
continuum estimate.
}
\label{fig:chi6B}
\end{figure}

We show the leading order coefficient $\chi_2^B(T)$  in 
Fig.~\ref{fig:Bcumulants} and the NLO ($\chi_4^B$)
and NNLO ($\chi_6^B$) coefficients divided by $\chi_2^B(T)$
in Fig.~\ref{fig:chi6B}.
The left hand part of Fig.~\ref{fig:Bcumulants} shows the leading order 
contribution $\chi_2^B$ in the entire temperature interval used in the current
analysis. For the LO expansion coefficients we also used data from 
simulations on $48^3\times 12$ lattices. Here we used existing
data for $m_l/m_s=1/20$ \cite{Bazavov:2014pvz} and generated new
ensembles for $m_l/m_s=1/27$ at nine temperature values below 
$T=175$~MeV. Furthermore, we used data on $64^3\times 16$  lattices
at a corresponding set of low temperature values. These data are taken
from an ongoing calculation of higher order susceptibilities 
performed by the HotQCD Collaboration\footnote{We thank the HotQCD 
Collaboration for providing access to the second order quark number
susceptibilities.}.
This allowed us to update
the continuum extrapolation for $\chi_2^B$ given in \cite{Bazavov:2012jq}.
The new continuum extrapolation shown in Fig.~\ref{fig:Bcumulants} is
consistent with our earlier results, but has significantly
smaller errors in the low temperature region. In the right hand part of 
this figure we compare the continuum extrapolated lattice QCD data for
$\chi_2^B$ with HRG model calculations. It is obvious that the continuum
extrapolated QCD results overshoot results obtained from a 
conventional, non-interacting HRG model calculations with resonances taken
from the particle data tables (PDG-HRG) and treated as point-like excitations.
We therefore compare the QCD results also with a HRG model that includes
additional strange baryons,which are not listed in the PDG but are predicted
in quark models and lattice QCD calculations. We successfully used such
an extended HRG model (QM-HRG) in previous calculations 
\cite{Majumder:2010ik,Bazavov:2014xya}. As can be seen in 
Fig.~\ref{fig:Bcumulants}~(left), continuum extrapolated results for 
$\chi_2^B$ agree well with QM-HRG calculations.

As can be seen
in the left hand part of Fig.~\ref{fig:chi6B}, the ratio 
$\chi_{4}^B/\chi_2^B$ approaches unity with decreasing temperature, but
is small at high temperatures where the leading order correction is large. 
The relative contribution of the NLO 
correction thus is largest in the hadronic phase,
where $\chi_{4}^B/\chi_2^B\simeq 1$. 
For temperatures $T\lsim 155$~MeV we find $\chi_{4}^B/\chi_2^B\le 0.8$.
The relative contribution 
of the NLO correction to the $\mu_B$-dependent part of the pressure 
(number density) in the crossover region and below thus is about 
8\% (16\%) at $\mu_B/T=1$ and rises 
to about 33\% (66\%) at $\mu_B/T=2$. At 
temperatures larger than $180$~MeV the relative contribution of the
NLO correction to pressure and number density at $\mu_B/T=2$  is less 
than 8\%  and 16\%, respectively.

The relative contribution of the ${\cal O}(\hmu_B^6)$ correction, 
$\chi_6^B/\chi_2^B$, is shown in the right hand part of Fig.~\ref{fig:chi6B}.  
The ideal gas limit for this ratio vanishes. Obviously the
ratio is already small for all temperatures $T>180$~MeV, i.e.
$\chi_6^B/\chi_2^B\le 0.5$. Consequently, for $\hmu_B=2$ 
the correction to the leading order result is less than 
2.2\% for the $\mu_B$-dependent part of the pressure and less than
7\% for the net baryon-number density. 
At lower temperatures the statistical errors on current
results for $\chi_6^B/\chi_2^B$ are still large. However, a crude estimate
for the magnitude of this ratio at all temperatures larger
than $130$~MeV suggests, $\left| \chi_6^B/\chi_2^B\right|\le 3$.
In the low temperature, hadronic regime and for $\hmu_B=2$ 
the ${\cal O}(\hmu_B^6)$ corrections to the $\mu_B$-dependent part of the 
pressure can be about 13\%. However, in the total
pressure, which also receives large contributions from the meson sector,
this will result only in an error of less than 3\%. In the calculation
of the net baryon-number density, on the other hand, the current uncertainty 
on ${\cal O}(\hmu_B^6)$ expansion coefficients
results in errors of about 40\% at temperatures below $T\simeq 155$~MeV.
In fact, as discussed already in section~\ref{sec:HRG}, higher order
corrections are larger in the Taylor expansion
of the number density. From Eq.~\ref{Pncoef} it follows for the
ratio of NLO and LO expansion coefficients, $N_5^B/N_1^B= 3 P_6/P_2$.
Clearly better statistics is needed
in the low temperature range to control higher order corrections to $n_B/T^3$.

\begin{figure}[htb]
\hspace{-0.9cm}\includegraphics[scale=0.81]{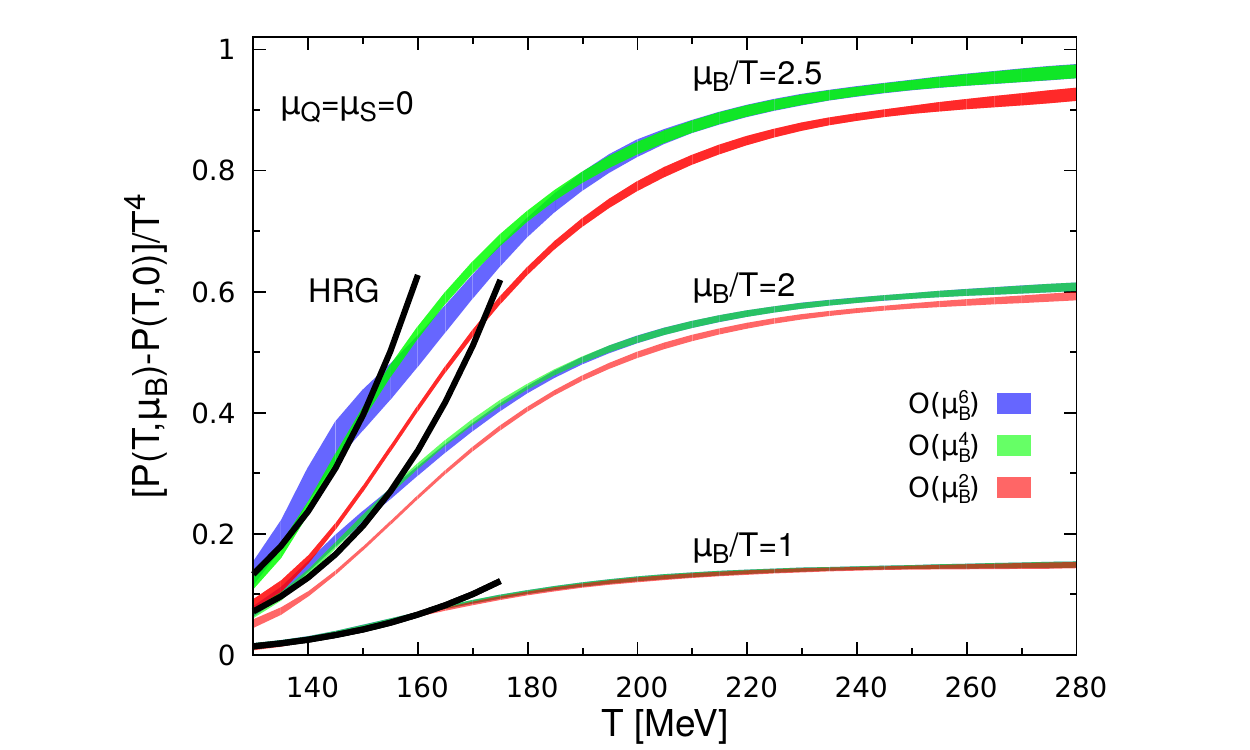}\hspace{-1.9cm}
\includegraphics[scale=0.81]{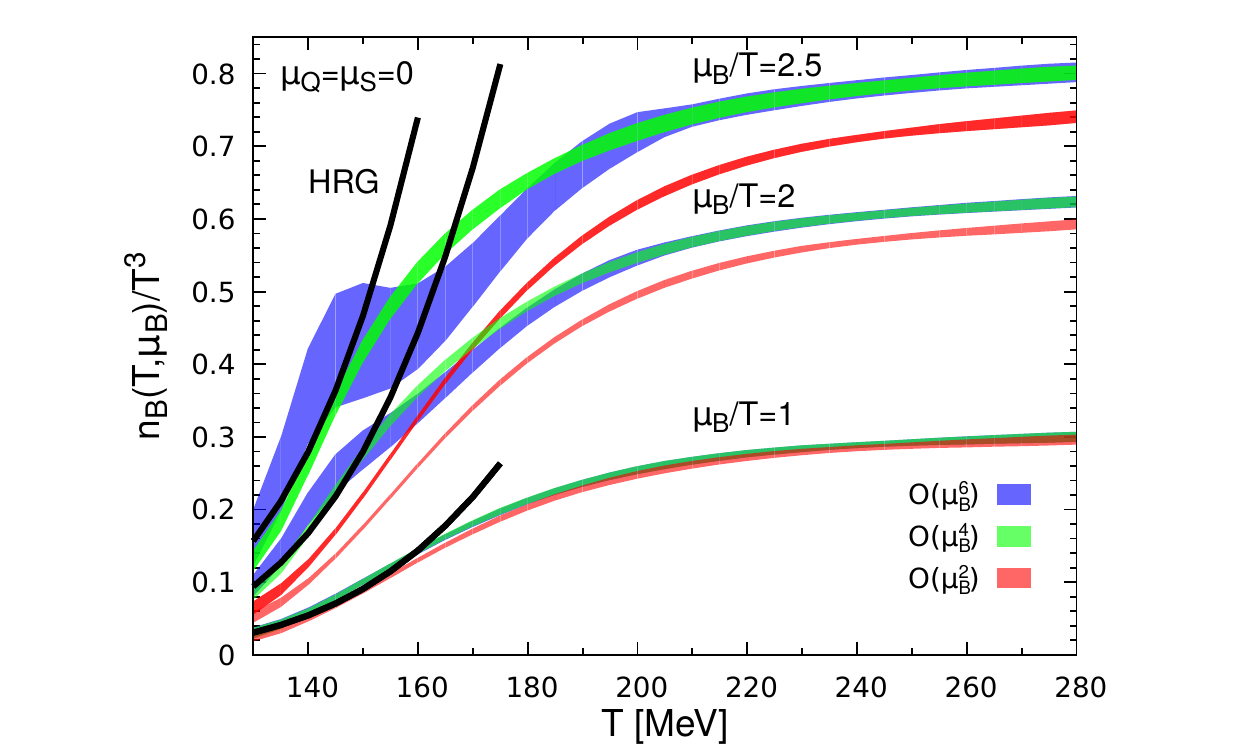}
\caption{The $\mu_B$-dependent contribution to
the pressure (left) and the baryon-number density (right)
in the case of vanishing electric charge and strangeness chemicals
potential for several values of the baryon chemical potential
in units of temperature. The different bands show results including
Taylor series results upto the order indicated. 
}
\label{fig:pressureB}
\end{figure}

In Fig.~\ref{fig:pressureB} we show results for
the $\mu_B$-dependent part of the pressure (left) and the net 
baryon-number density (right) calculated from Taylor series up to and including
LO, NLO and NNLO contributions, respectively.
This suggests that up to $\mu_B\simeq 2 T$ results for the pressure at low 
temperature are well described by a Taylor series truncated at NNLO, while
at higher temperature NNLO corrections are small even at $\mu_B\simeq 3 T$.
This also is the case for $n_B/T^3$, although the NNLO correction is large
at low temperatures and, at present, does not allow for a detailed quantitative
analysis of the baryon-number density in this temperature range. 

It also is obvious that the Taylor series for the pressure and $n_B/T^3$ in
the temperature range up to $T\simeq 180$~MeV are sensitive to the negative 
contributions of the $6^{th}$ order expansion coefficient. 
The occurrence of a dip in the sixth order expansion coefficient of 
the pressure has been expected to show up on the basis of general 
scaling arguments  for higher order derivatives of the QCD pressure
in the vicinity of the chiral phase transition \cite{Friman:2011pf}. It may,
however, also reflect the influence of a singularity on the imaginary
chemical potential axis \cite{Bonati:2016pwz} (Roberge-Weiss critical point
\cite{Roberge}) on Taylor series
of bulk thermodynamic observables in QCD.
Even with improved statistics it thus is expected that the wiggles, 
that start to show up in the expansion of pressure and net baryon-number
density above $\mu_B/T\simeq 2$ (see Fig.~\ref{fig:pressureB}) and reflect
the change of sign in the sixth order expansion coefficient, will persist.
Getting the magnitude of 
the dip in $\chi_6^B/\chi_2^B$ at $T\simeq 160$~MeV under control in 
future calculations thus is of importance for the understanding of this
non-perturbative regime of the QCD equation of state in the high 
temperature phase close to the transition region. This also indicates that
higher order corrections need to be calculated in order to control the
equation of state in this temperature regime.

\subsection{Net strangeness and net  electric charge densities} 

\begin{figure}[htb]
\hspace{-0.9cm}\includegraphics[scale=0.81]{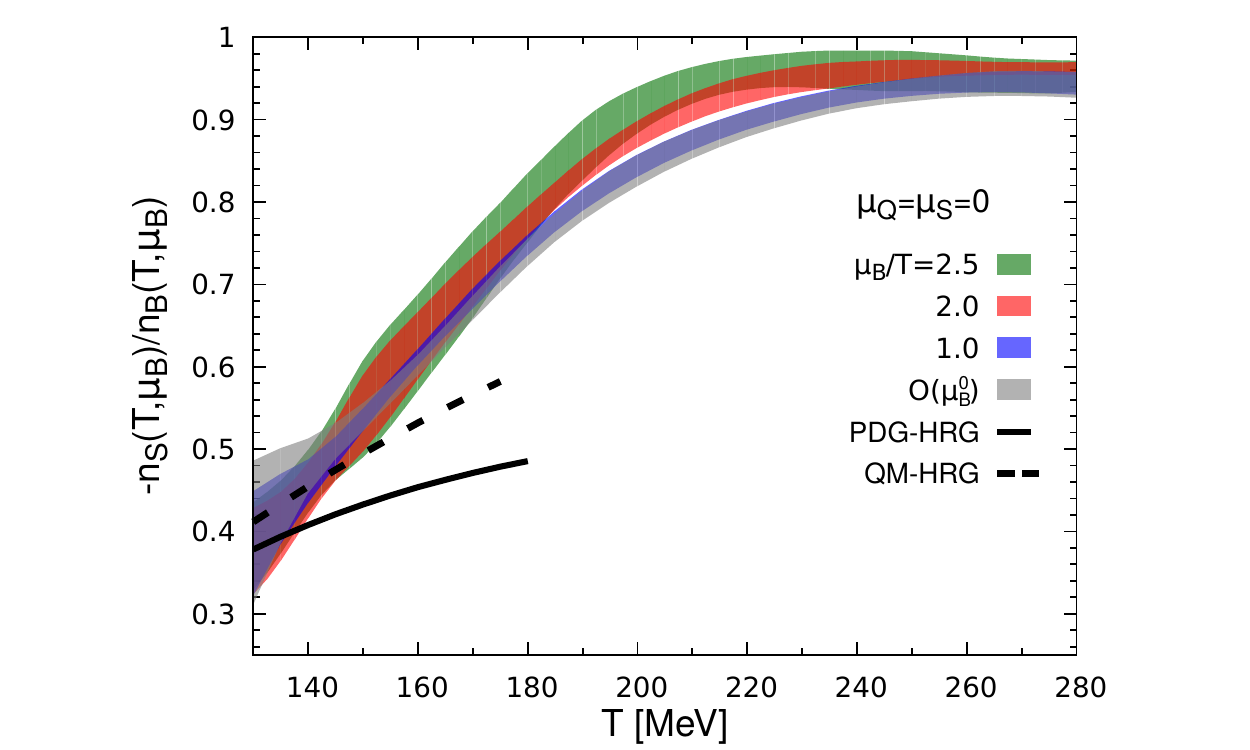}\hspace{-1.9cm}
\includegraphics[scale=0.81]{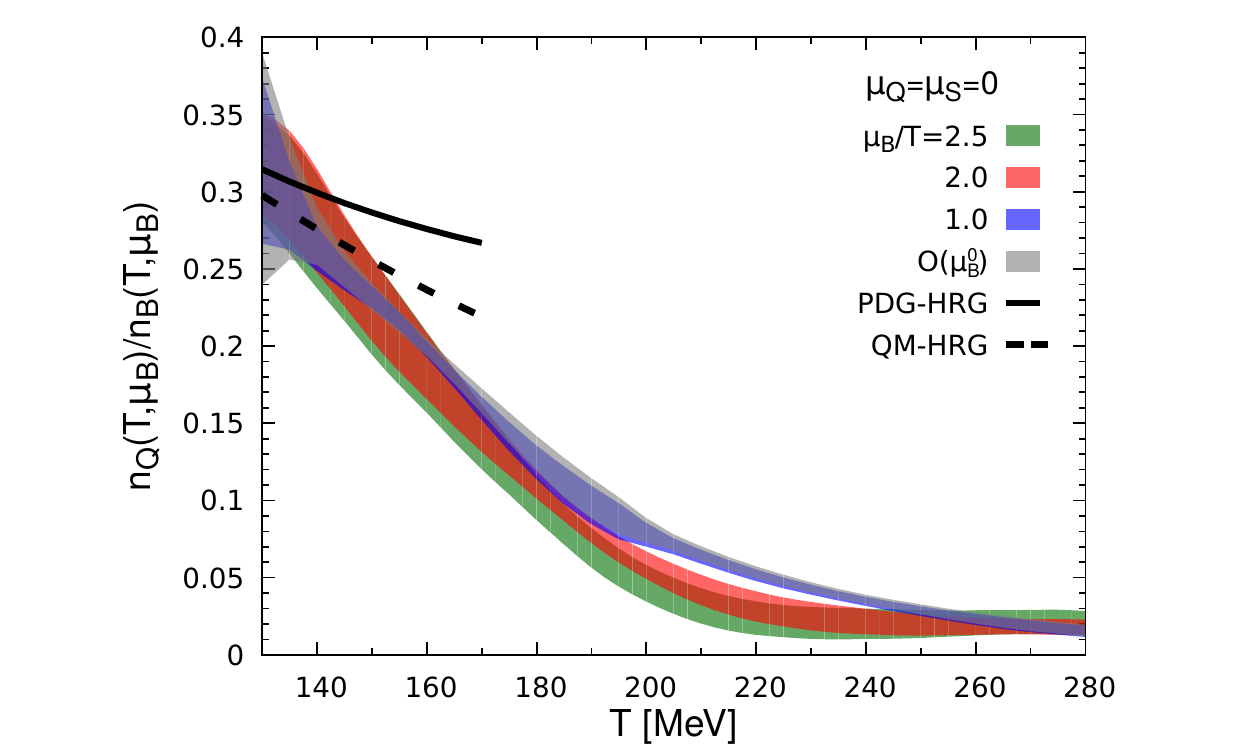}
\caption{The ratio of net strangeness and net baryon-number densities
(left) and the ratio of net electric charge and net baryon-number densities
(right).  At low temperatures results from hadron resonance gas 
calculations at $\mu_B=0$ are shown (see text). 
}
\label{fig:BXratios}
\end{figure}

For vanishing strangeness and electric charge chemical potentials the 
corresponding net strangeness ($n_S$) and net electric charge ($n_Q$) 
densities are 
nonetheless non-zero because the carriers of these quantum numbers also
carry baryon number. The ratios of number densities are given by
\begin{equation}
\frac{n_X}{n_B} = 
\frac{\chi_{11}^{BX} +\frac{1}{6}\chi_{31}^{BX} \hmu_B^2+ \frac{1}{120}\chi_{51}^{BX} \hmu_B^4}{\chi_{2}^{B} +\frac{1}{6}\chi_{4}^{B} \hmu_B^2+ \frac{1}{120}\chi_{6}^{B} \hmu_B^4}
\;\; ,\,\, X=Q,\ S\;\; .
\end{equation}
In a hadron resonance gas the ratios $n_S/n_B$ and
$n_Q/n_B$ are independent of the baryon chemical potential and, irrespective
of the value of $\hmu_B$, these ratios approach $-1$ and $0$, respectively,
in the $T\rightarrow \infty$ limit.
One thus may expect that these ratios only show a mild 
dependence on $\hmu_B$, which indeed is apparent from the results of the 
NNLO expansions shown in Fig.~\ref{fig:BXratios}. 

For $\mu_Q=\mu_S=0$ non-vanishing electric charge and strangeness densities
only arise due to a non-zero baryon-chemical potential. In the low temperature
HRG phase $n_Q$ and $n_S$ thus only receive contributions from charged baryons
or strange baryons, respectively. The ratios $n_Q/n_B$ and $n_S/n_B$ thus
are sensitive to the particle content in a hadron resonance  gas and a 
comparison with PDG-HRG and QM-HRG is particularly sensitive to the 
differences in the baryon content in these two models.
It is apparent from Fig.~\ref{fig:BXratios} that at low temperatures
the QM-HRG model provides a better description of the lattice QCD results 
than the PDG-HRG model.

\subsection{The energy and entropy densities} 

In order to calculate the energy and entropy densities, defined in 
Eqs.~\ref{energyc} and \ref{entropyc}, we need to extract the temperature 
derivative of the expansion coefficients of the pressure. We use as a starting 
point the representation of the pressure given in Eq.~\ref{PmuB} and  
calculate the temperature derivatives of  $\chi_n^B$ from the splines
used to fit this observable. With this we construct the 
expansion coefficients $\epsilon_n^B(T)$ and $\sigma_n^B$ defined in 
Eqs.~\ref{energy} and \ref{entropy},

\begin{eqnarray}
\Delta\left(\epsilon/T^4\right) &=& 
\frac{\epsilon(T,\mu_B)-\epsilon(T,0)}{T^4} = 
\sum_{k=1}^3 \epsilon_{2k} \hmu_B^{2k}=
\sum_{k=1}^3 \left(  TP'_{2k}+3P_{2k}\right) \hmu_B^{2k}
\; ,
\label{emuB} \\
\Delta\left(s/T^3\right) &=& 
\frac{s(T,\mu_B)-s(T,0)}{T^3} = 
\sum_{k=1}^3 \sigma_{2k} \hmu_B^{2k}=
\sum_{k=1}^3 (\epsilon_{2k}-(2k-1) P_{2k}) \hmu_B^{2k}
\; .
\label{smuB}
\end{eqnarray}

\begin{figure}[t]
\includegraphics[scale=0.7]{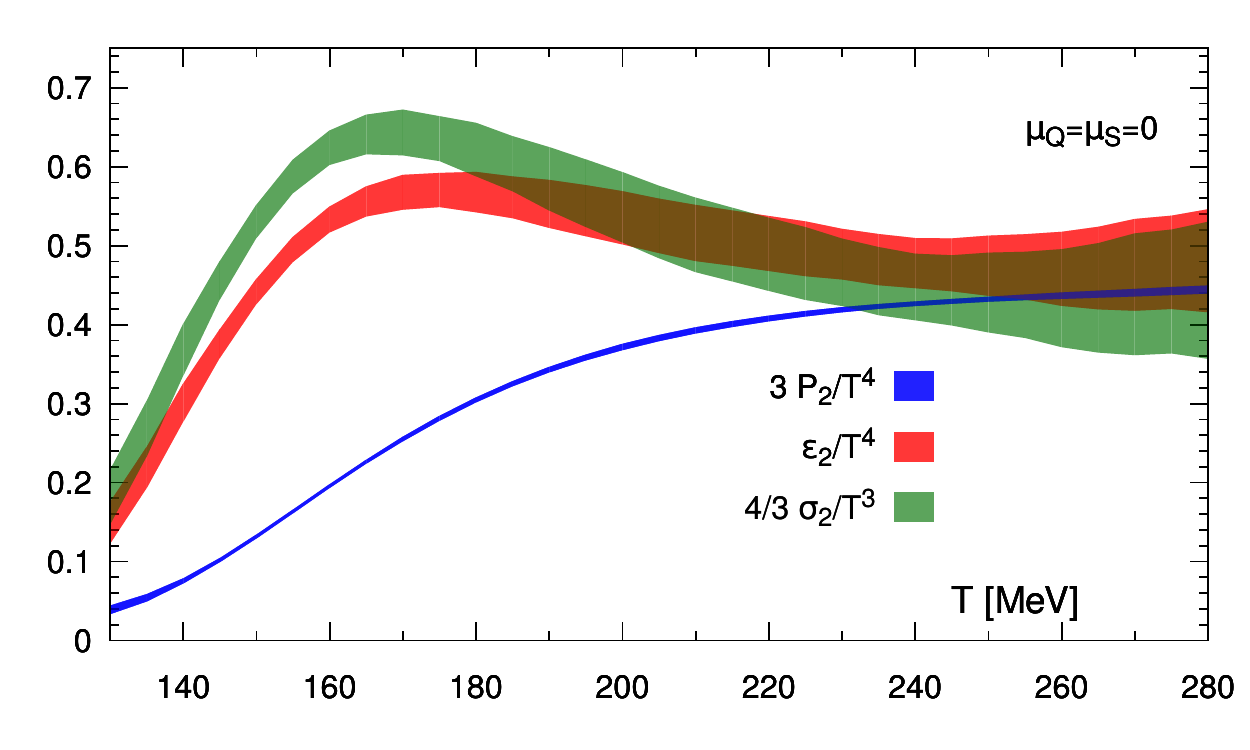}
\includegraphics[scale=0.7]{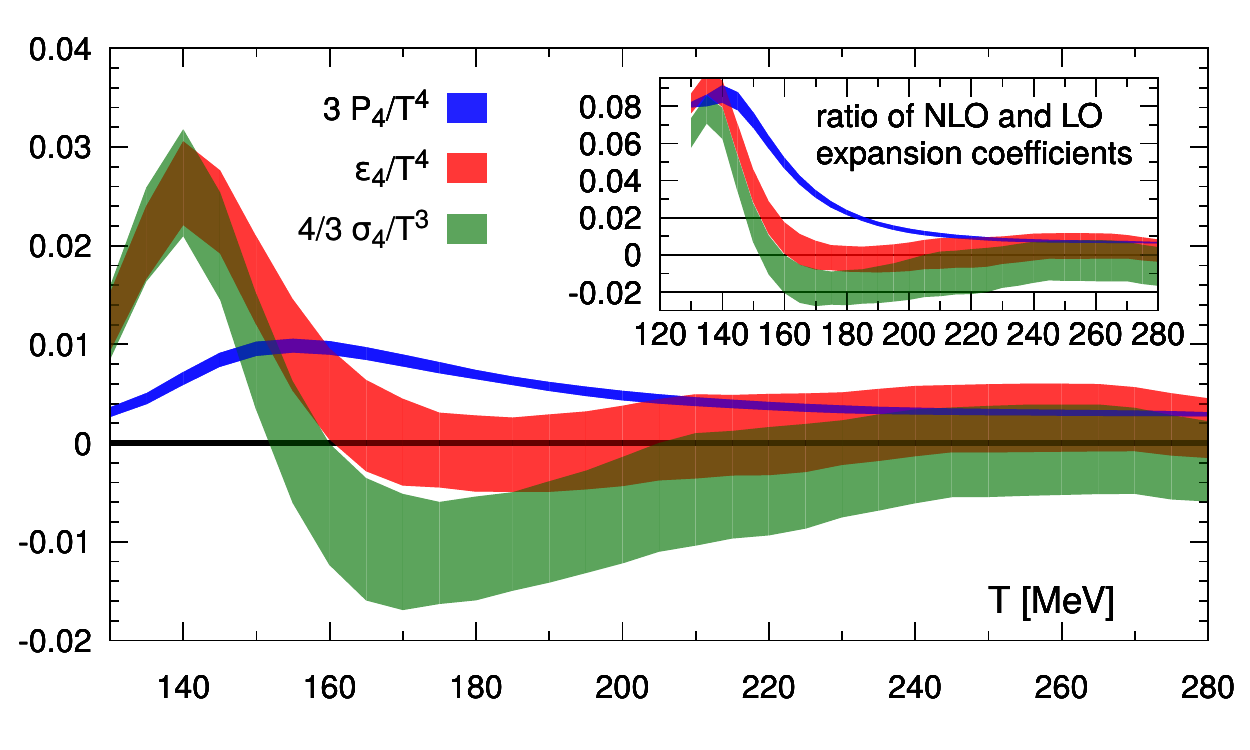}
\caption{Leading order (left) and next-to-leading order (right)
expansion coefficients for the $\mu_B$-dependent part of pressure,
the energy and entropy densities in the case $\mu_Q=\mu_S=0$. The
inset in the right hand figure shows the ratios of NLO and LO
expansion coefficients $P_4/P_2$, $\epsilon_4/\epsilon_2$ and
$\sigma_4/\sigma_2$. Note that the expansion coefficients for the
net baryon-number density are directly proportional to those of the
pressure series, i.e. $N^B_1=2 P_2$ and $N^B_3=4 P_4$.
}
\label{fig:ecumulants}
\end{figure}

We show the LO and NLO expansion coefficients for energy and entropy 
densities together with the expansion coefficient for the pressure
in Fig.~\ref{fig:ecumulants}. Because of Eq.~\ref{Pncoef} the expansion
coefficients of the net baryon-number density are simply proportional 
to those of the pressure.

Clearly the temperature dependence of the expansion 
coefficients of the energy and entropy densities shows more structure
than in the case of the pressure. 
Qualitatively this can be understood in terms
of the pseudo-critical behavior of bulk thermodynamic observables. 
Once thermodynamic quantities are dominated by contributions from the 
singular part of the free energy, which is expected to happen in the
transition region, they become functions of $(T-T_c)+\kappa \hmu_B^2$. 
The temperature
derivative of the expansion coefficient $P_2$, which gives $\epsilon_2$, thus
will show properties similar to those of $P_4$.
The LO correction $\epsilon_2^B/T^4$ 
has a mild peak, which results from the strongly peaked $T$-derivative of 
$\chi_2^B$ which is qualitatively similar to $\chi_4^B$,
and the NLO correction is negative in a small temperature interval above $T_c$, 
which arises from the negative $T$-derivative of $\chi_4^B$ at high temperature,
which resembles the negative part of $\chi_6^B$ at high temperature.

\begin{figure}[t]
\hspace{-0.9cm}\includegraphics[scale=0.81]{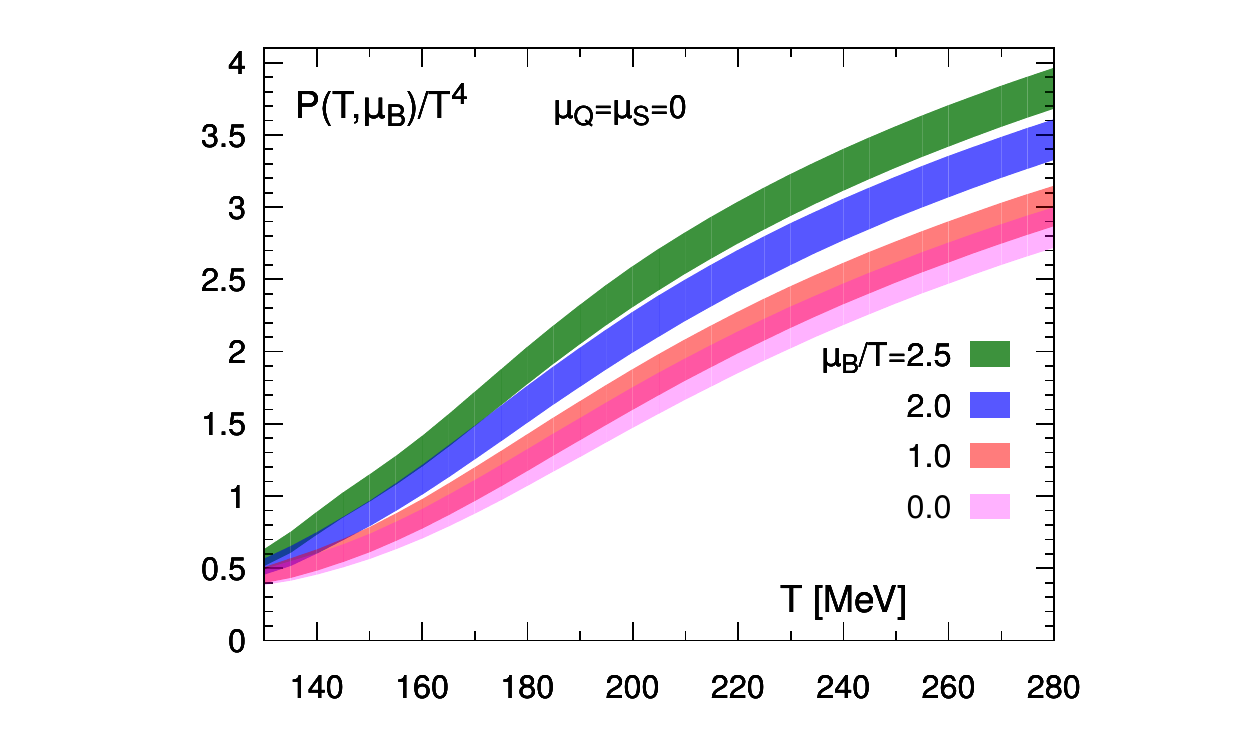}\hspace{-1.9cm}
\includegraphics[scale=0.81]{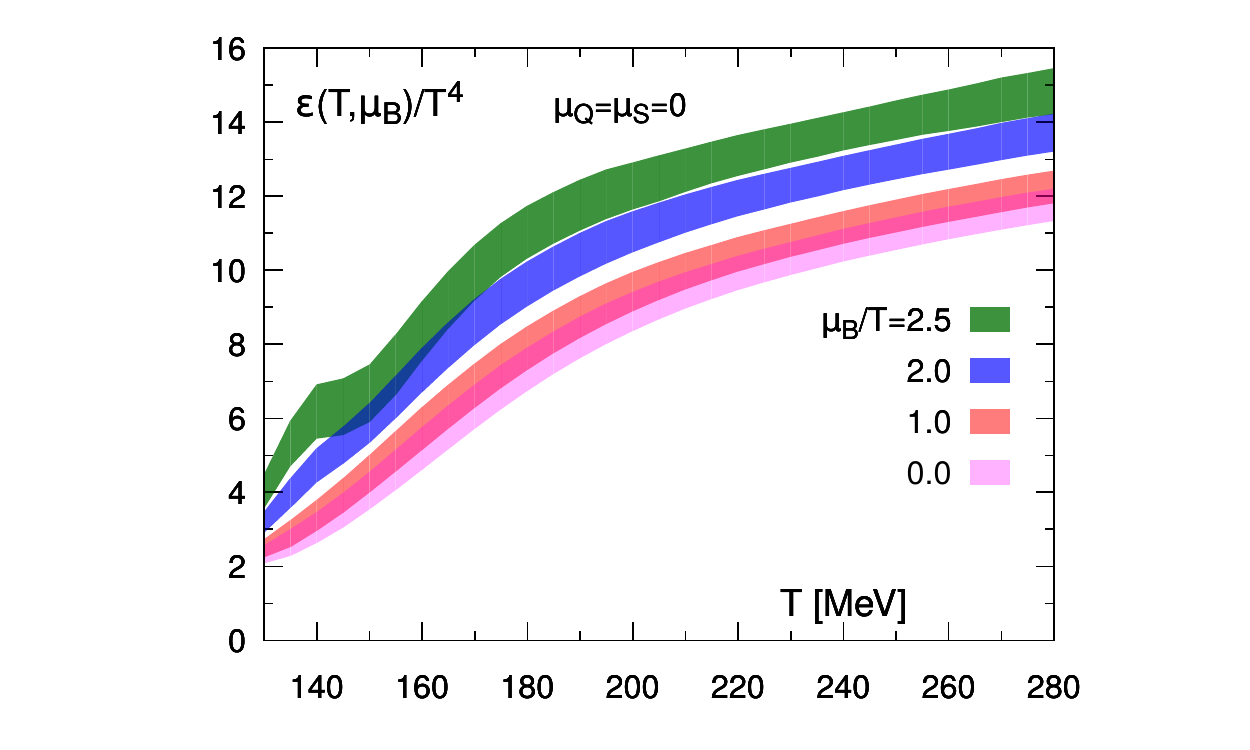}
\caption{(Left) The total pressure in (2+1)-flavor QCD
in ${\cal O}(\hmu_B^6)$ for several values of
$\mu_B/T$.
(Right) The total energy density in (2+1)-flavor QCD
in ${\cal O}(\hmu_B^6)$ for several values of
$\mu_B/T$.
The results for $\hmu_B=0$ are taken from
Ref.~\cite{Bazavov:2014pvz}.
}
\label{fig:energyB}
\end{figure}

Although the 
temperature dependence of $\epsilon_n$ and $\sigma_n$ differs from that of 
the pressure coefficient, $P_n$, the conclusions drawn for the relative 
strength of the expansion coefficients are identical in all cases. As can be
seen from the inset in Fig.~\ref{fig:ecumulants}~(right) the relative 
contribution of the NLO expansion coefficients never exceeds 10\%.
In particular, at temperatures larger than $180$~MeV the magnitude of the 
NLO expansion coefficients never exceeds 2\% of the LO expansion coefficients.
Again this leads to the conclusion that at $\mu_B/T=2$ and temperatures
above $180$~MeV the NLO correction contributes less than 8\% of the leading 
correction to $\mu_B$-dependent part of the energy and entropy densities. 
For $T\lsim 155$~MeV, however, the NLO contribution can rise to about 30\%.
A similar conclusion holds for the ${\cal O}(\hmu_B^6)$ corrections, although
it requires higher statistics to better quantify the magnitude of this
contribution. 
In Fig.~\ref{fig:energyB} we show results for the total pressure and 
total energy density. For $P/T^4$ and $\epsilon/T^4$ at $\mu_B=0$ we used 
the results obtained by the HotQCD Collaboration \cite{Bazavov:2014pvz}
and added to it the results from the ${\cal O}(\hmu_B^6)$ expansions 
presented above. This figure also makes it clear that despite of the
large error of higher order expansion coefficients, which we have discused
above, the error on the total pressure and energy density still is dominated
by errors on their values at $\mu_B=0$.

\section{Equation of state in strangeness neutral systems}
\label{sec:eosSN}

\subsection{Taylor expansion of pressure, baryon-number, energy
and entropy densities}
We now want to discuss the equation of state for strangeness neutral
systems with a fixed ratio of electric charge to baryon-number density,
i.e. we impose the constraints \cite{Bazavov:2012vg}
\begin{equation} 
n_S=0\;\;\; ,\;\;\; \frac{n_Q}{n_B}=r \; .
\label{neutral}
\end{equation}
These constraints can be realized through suitable choices of 
the electric charge and strangeness chemical potentials.
This thus is a particular case of the constraint expansion
discussed in Subsection~\ref{sec:constraint}. The expansion 
coefficients $q_n$, $s_n$, $n=1,\ 3, \ 5$ needed to satisfy these
constraints are given in Appendix~\ref{app:constraint}. 
For $r=0.4$ the constrained EoS obtained in this way is usually
considered to be most appropriate for applications to heavy ion
collisions. We will, however, in the following also comment
on other choices of $r$, including the case of isospin symmetric
systems ($r=1/2$) and electric charge neutral matter ($r=0$).

\begin{figure}[t]
\begin{center}
\includegraphics[width=78mm]{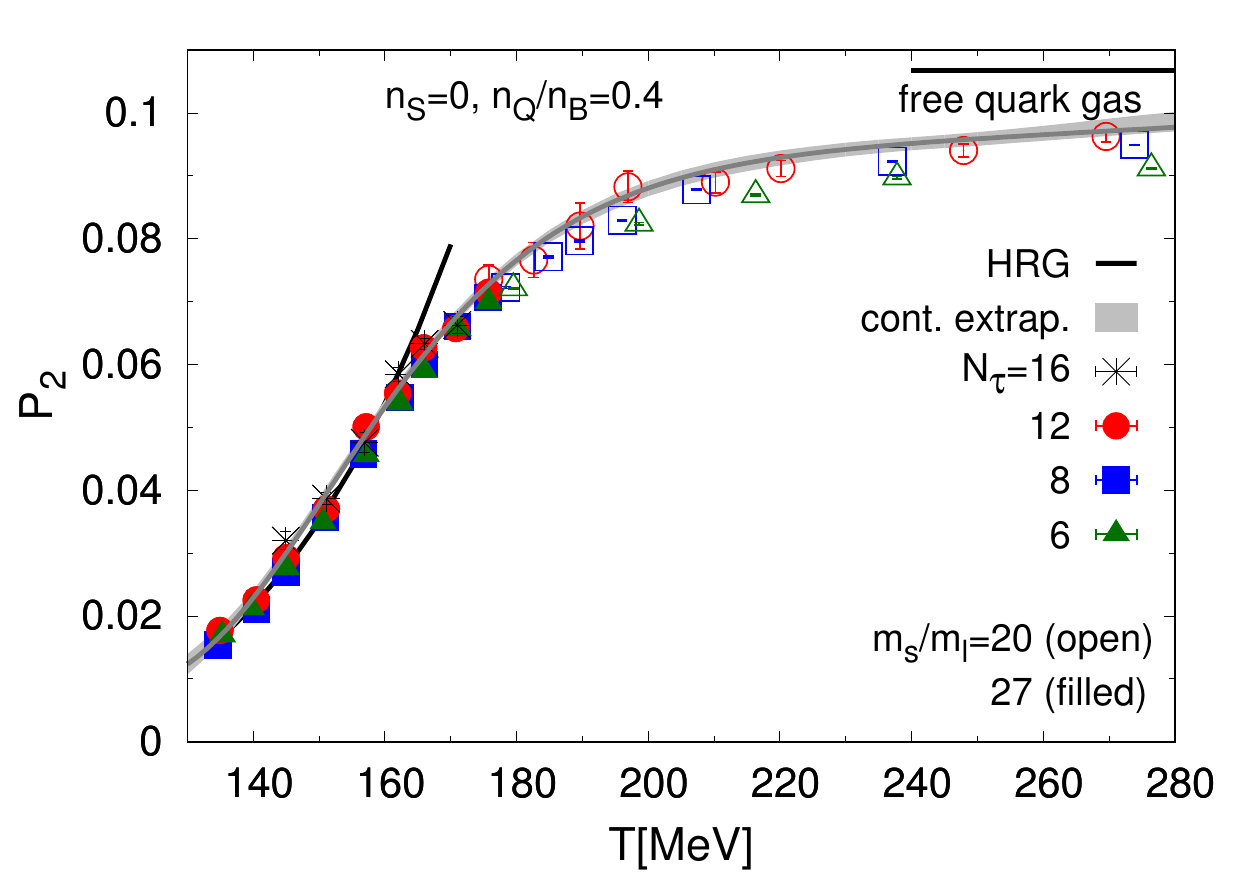}
\includegraphics[width=78mm]{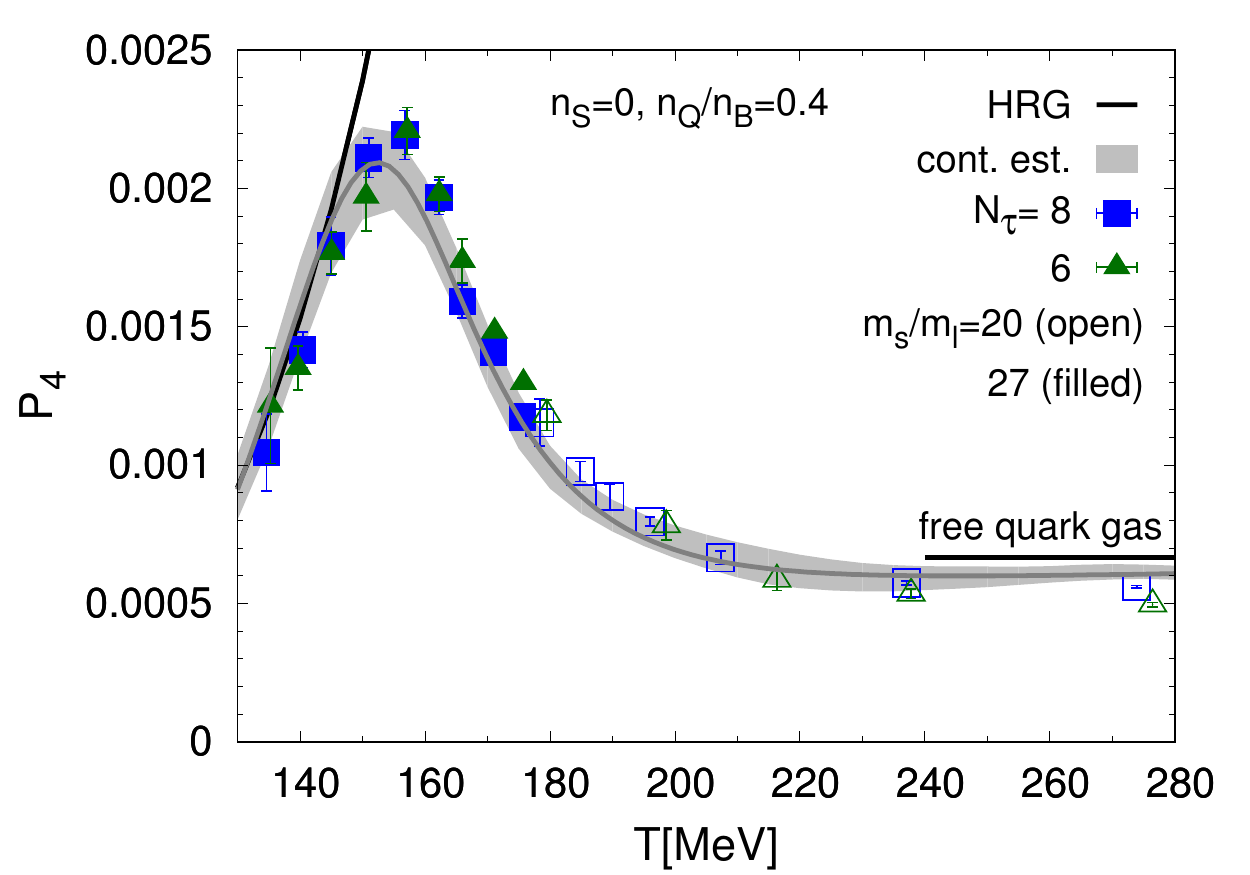}
\includegraphics[width=78mm]{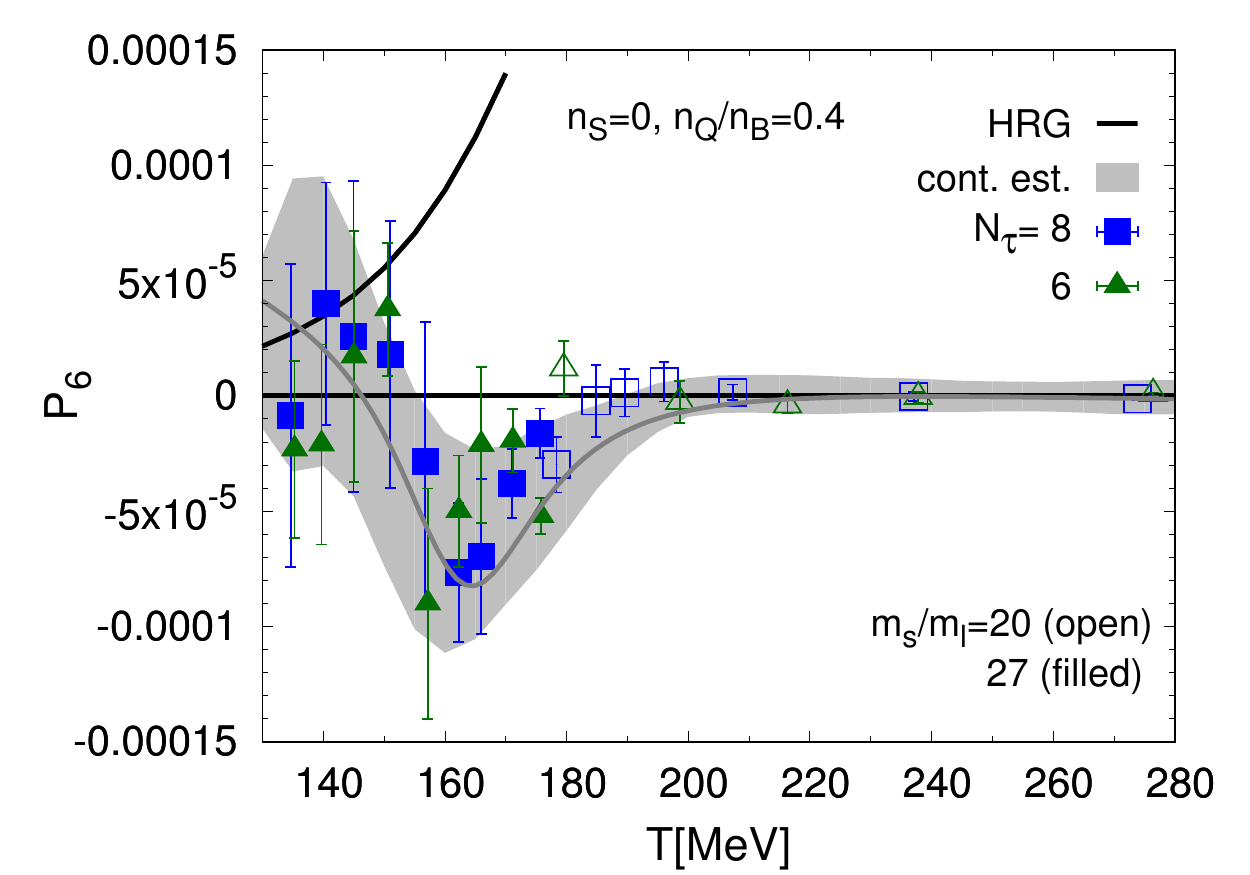}
\includegraphics[width=78mm]{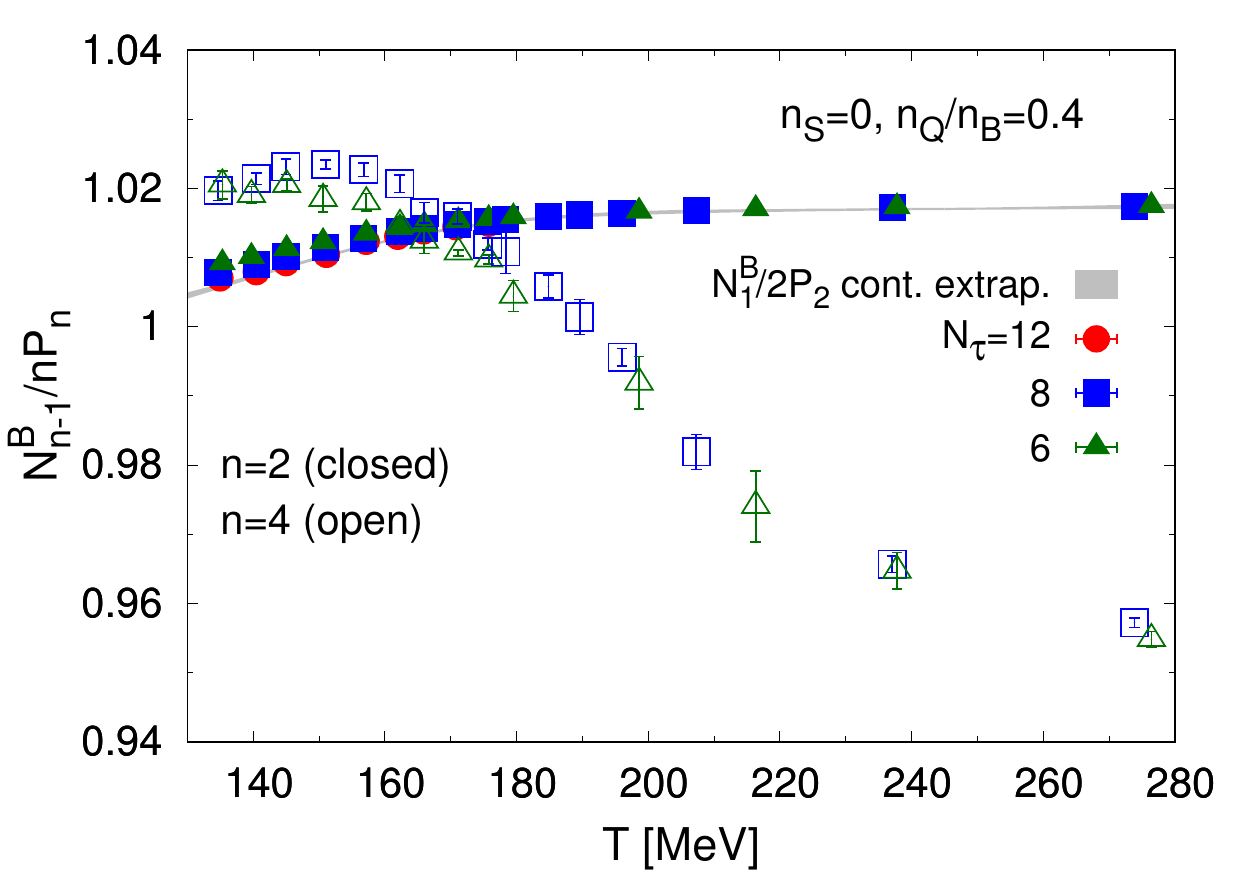}
\caption{Expansion coefficients of the pressure (top, and bottom left) 
and the ratio of net baryon-number density and pressure expansion 
coefficients (bottom, right) in strangeness neutral systems with $r=0.4$.
Broad bands show continuum extrapolations as discussed in Section~\ref{sec:outline}. The darker lines in the center of the error bands of these extrapolations
show the interpolating fits discussed in Subsection~\ref{sec:fits}. 
At low temperature lines for HRG model calculations based on hadron resonances
listed by the Particle Data Group is shown. 
}
\label{fig:NB}
\end{center}
\end{figure}

Using the constraints specified in Eq.~\ref{neutral} and the definition
of the pressure in terms of generalized susceptibilities, $\chi_{ijk}^{BQS}$,
the expansion coefficients $P_{2n}$ can easily be determined.
Here it advantageous to use the relation between the Taylor expansion 
coefficients of the pressure, $P_{2n}$, and number densities, $N_{2n-1}^X$, 
given in Eq.~\ref{PNX}, which simplifies considerably for strangeness neutral
systems. It now involves only the net baryon-number density
coefficients,
\begin{eqnarray}
P_2 &=& \frac{1}{2}\left[ N_1^B +r q_1 N_1^B \right] \; ,
\label{PN2} \\
P_4 &=&\frac{1}{4} \left[ N_3^B +r \left(q_1 N_3^B +3 q_3 N_1^B\right) \right] \; ,
\label{PN4} \\
P_6 &=&\frac{1}{6}\left[ N_5^B + r\left( q_1 N_5^B+ 3 q_3 N_3^B + 5 q_5 N_1^B   
\right) \right] \; .
\label{PN6}
\end{eqnarray} 
Explicit expressions for all $N_{n-1}^B$ and $q_{n-1}$, for $n=2,\ 4,\ 6$, are 
given in Appendix~\ref{app:constraint}. The resulting expansion 
coefficients for the pressure are shown in Fig.~\ref{fig:NB}.
Also shown in the bottom-right panel of this figure is the ratio of the 
expansion coefficients for the net baryon-number density, $N_{n-1}^B$ and the
appropriately rescaled expansion coefficients of the pressure, $nP_n$. In electric 
charge neutral systems, $r=0$ as well as in the isospin symmetric limit $r=1/2$, 
for which the expansion coefficients $q_i=0$ vanish for
all $i$, this ratio is unity. 
In both cases the simple relation given in
Eq.~\ref{Pncoef} holds. Also for other values of $r$ the contribution from 
terms proportional to $r$ are small. In Fig.~\ref{fig:NB}~(bottom, right)
we show the ratio $N_{2n-1}^B/nP_n$ 
for the case $r=0.4$ and $n=2$ and $4$, respectively.
At ${\cal O}(\hmu_B^2)$ differences
between $N_1^B$ and $2 P_2$ never exceed 2\% and at ${\cal O}(\hmu_B^4)$
the difference between $N_3^B$ and $4 P_4$ varies between 3\% at low 
temperature and -6\% at high temperature. In the infinite temperature 
ideal gas limit the ratios become
$N^B_{1}/2P_2=1.018$ and $N^B_{3}/4P_4= 0.927$, respectively.

\begin{figure}[t]
\begin{center}
\includegraphics[width=88mm]{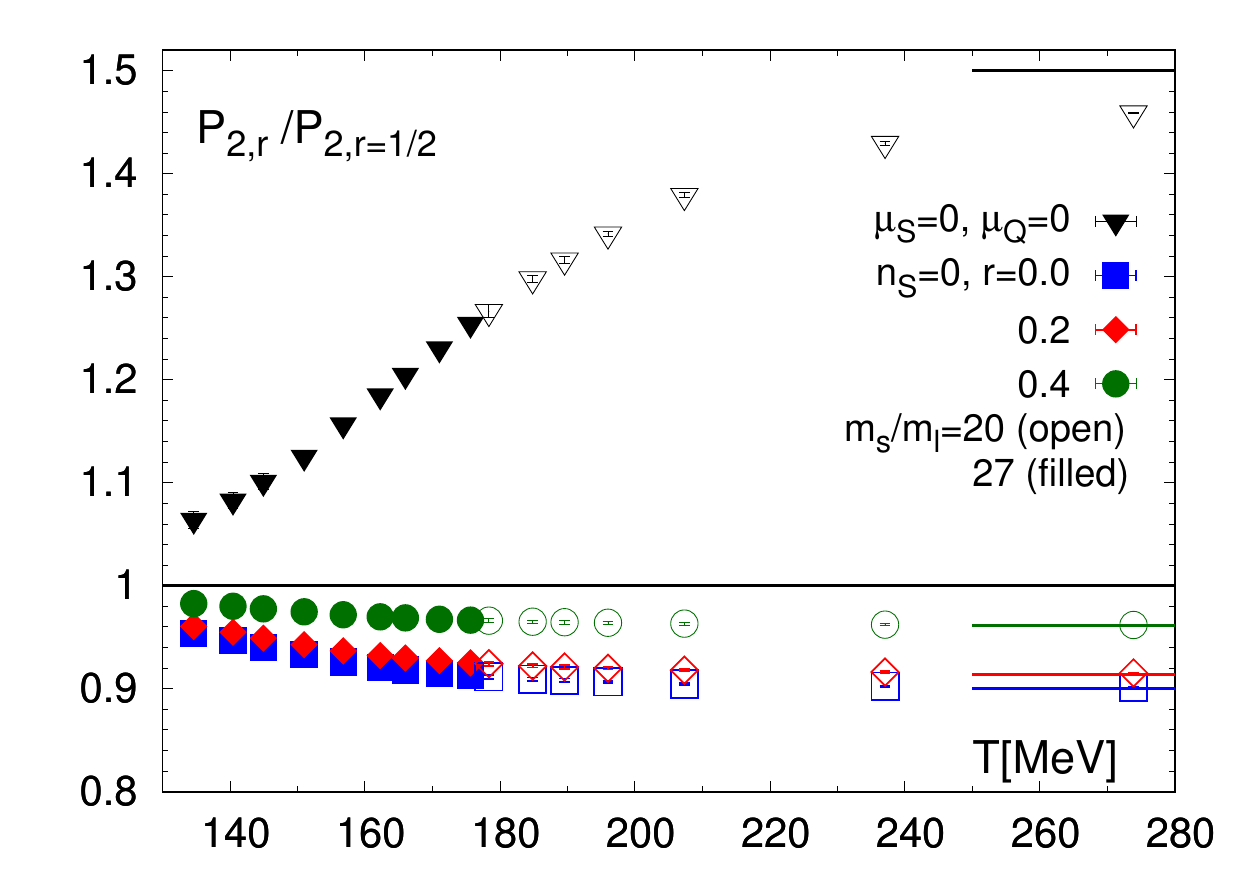}
\caption{Ratio of ${\cal O}(\hmu_B^2)$ expansion coefficients of the pressure
in systems with electric charge to net baryon-number ratio $r=n_Q/n_B$
relative to that of strangeness neutral, isospin symmetric systems ($r=1/2$).
Triangles show the ratio of the pressure in systems with vanishing
electric charge and strangeness chemical potential and the
strangeness neutral, isospin symmetric system. Horizontal lines at high
temperature show the corresponding free quark gas values. All data points
shown are from calculations on lattices with temporal extent $N_\tau=8$.
}
\label{fig:P2-r-ratio}
\end{center}
\end{figure}

In general one finds that the dependence of bulk thermodynamic observables
on the net electric charge to net baryon number-ratio is weak. 
The ${\cal O}(\hmu_B^2)$ expansion coefficient of the pressure in strangeness 
neutral systems differs by at most 10\% in electric charge neutral 
($r=0$) and isospin symmetric systems ($r=1/2$), respectively.
The expansion coefficient $P_2$ evaluated for different values of $r$ 
is shown in Fig.~\ref{fig:P2-r-ratio}.
For chemical potentials $\hmu \le 2$  this amounts to differences less 
than 1.5\% of the total pressure. 
On the other hand, strangeness neutral 
systems differ substantially from systems with vanishing strangeness
chemical potential. In this case the ${\cal O}(\hmu_B^2)$ expansion coefficients
differ by almost 50\% in the high temperature limit.  For $T< 150$~MeV
this difference is only about 10\% reflecting that the different treatment
of the strangeness sector becomes less important for the thermodynamics at low
temperature.
This is also shown in Fig.~\ref{fig:P2-r-ratio}.

\begin{figure}[htb]
\begin{center}
\includegraphics[width=78mm]{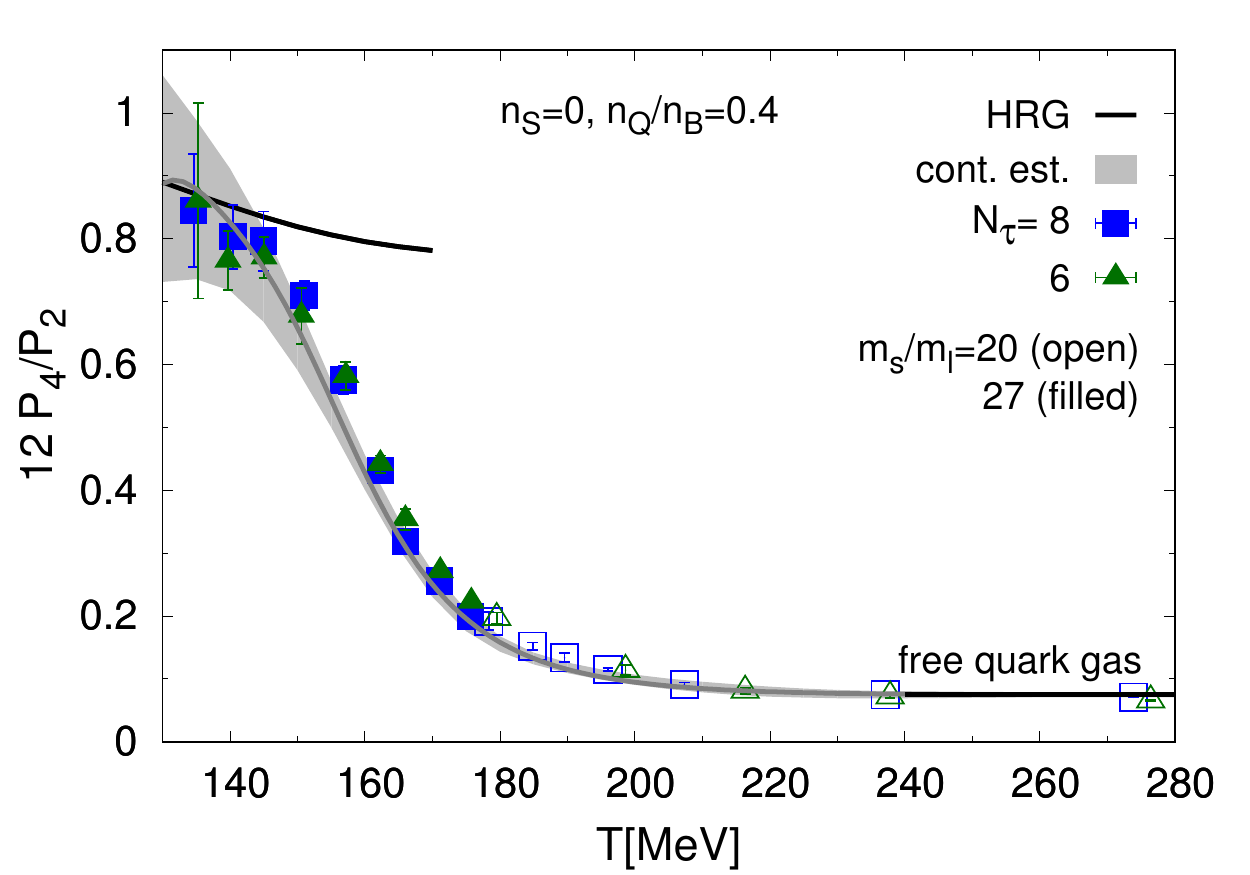}
\includegraphics[width=78mm]{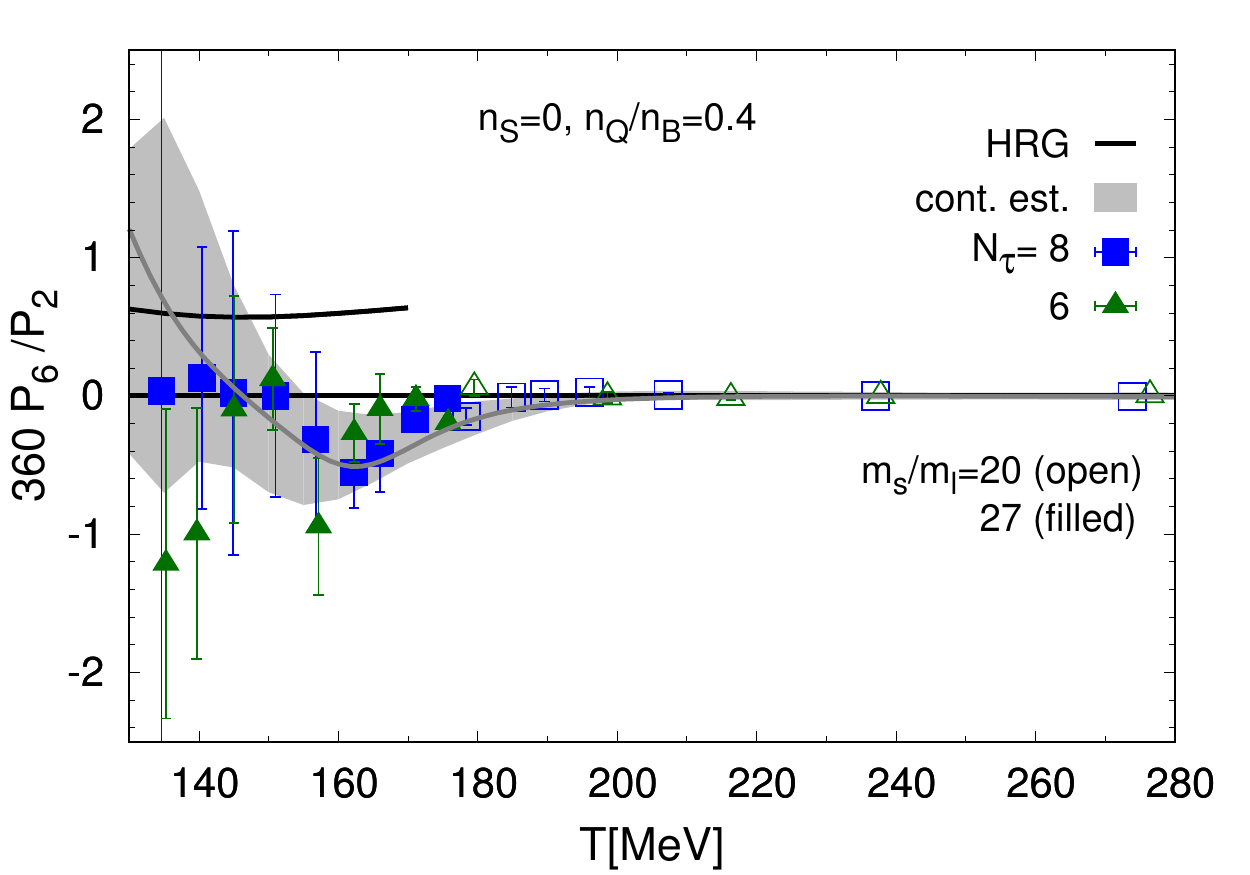}
\caption{Ratio of expansion coefficients of the pressure in
strangeness neutral systems with $r=0.4$.
The darker lines in the center of the error bands of these extrapolations
show results obtained with the parametrization discussed in 
Subsection~\ref{sec:fits}.
}
\label{fig:Pn}
\end{center}
\end{figure}

\begin{figure}[t]
\hspace{-0.9cm}\includegraphics[scale=0.81]{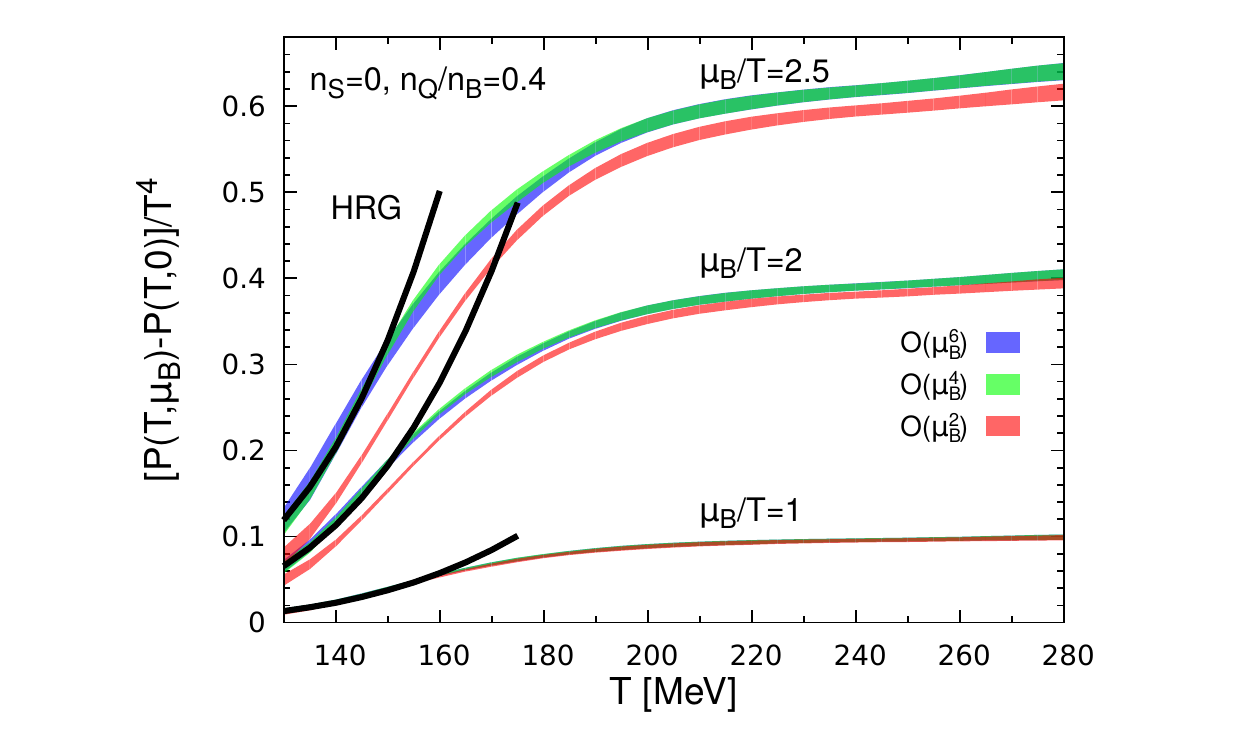}\hspace{-1.9cm}
\includegraphics[scale=0.81]{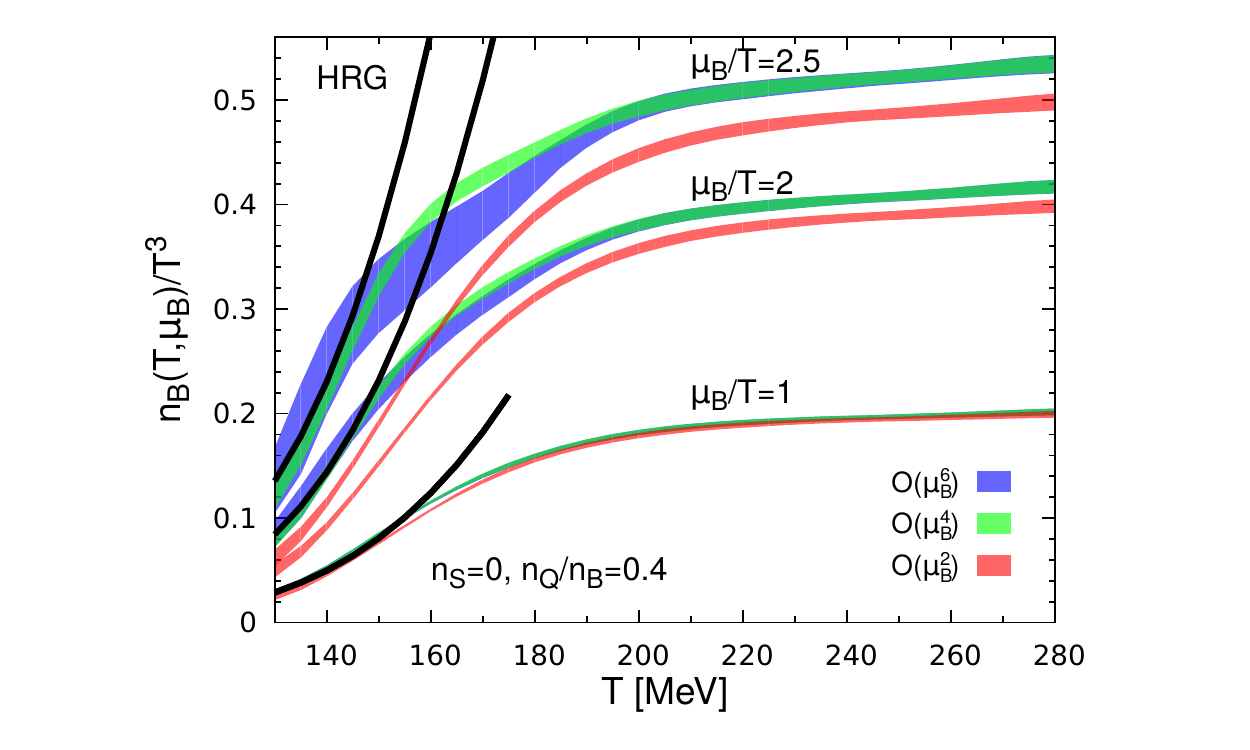}

\hspace{-0.9cm}\includegraphics[scale=0.81]{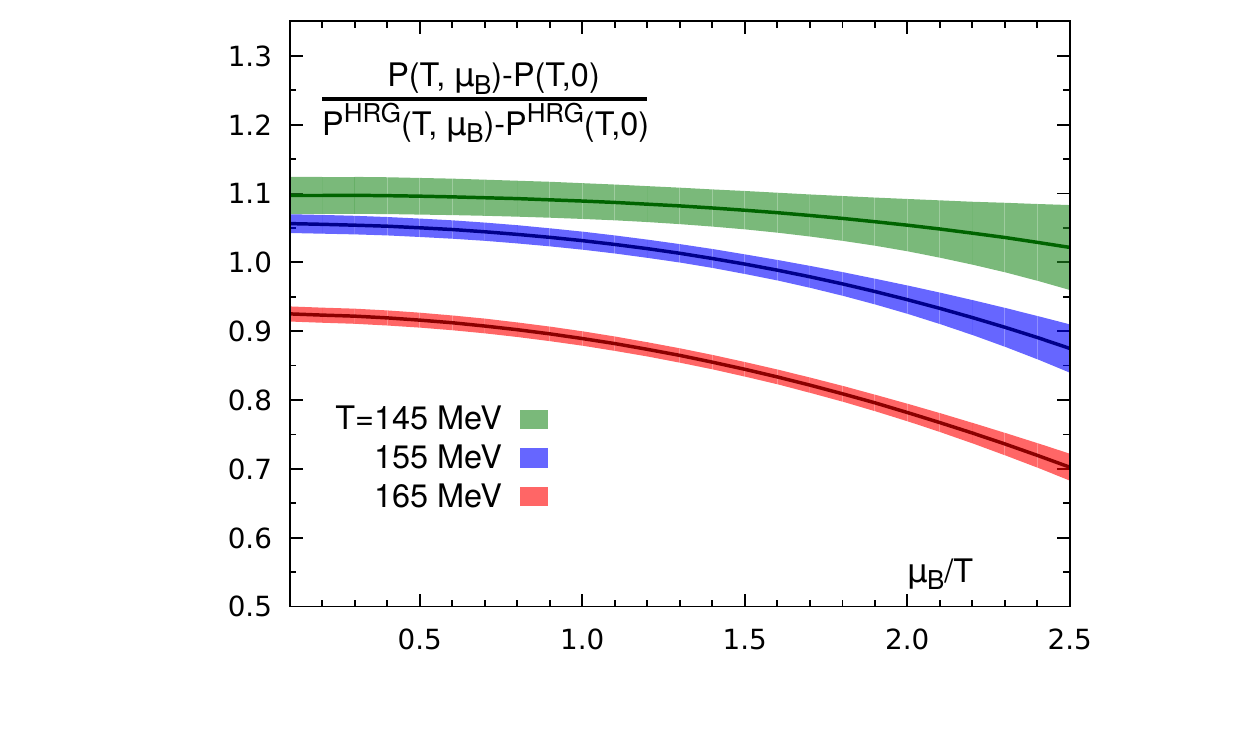}\hspace{-1.9cm}
\includegraphics[scale=0.81]{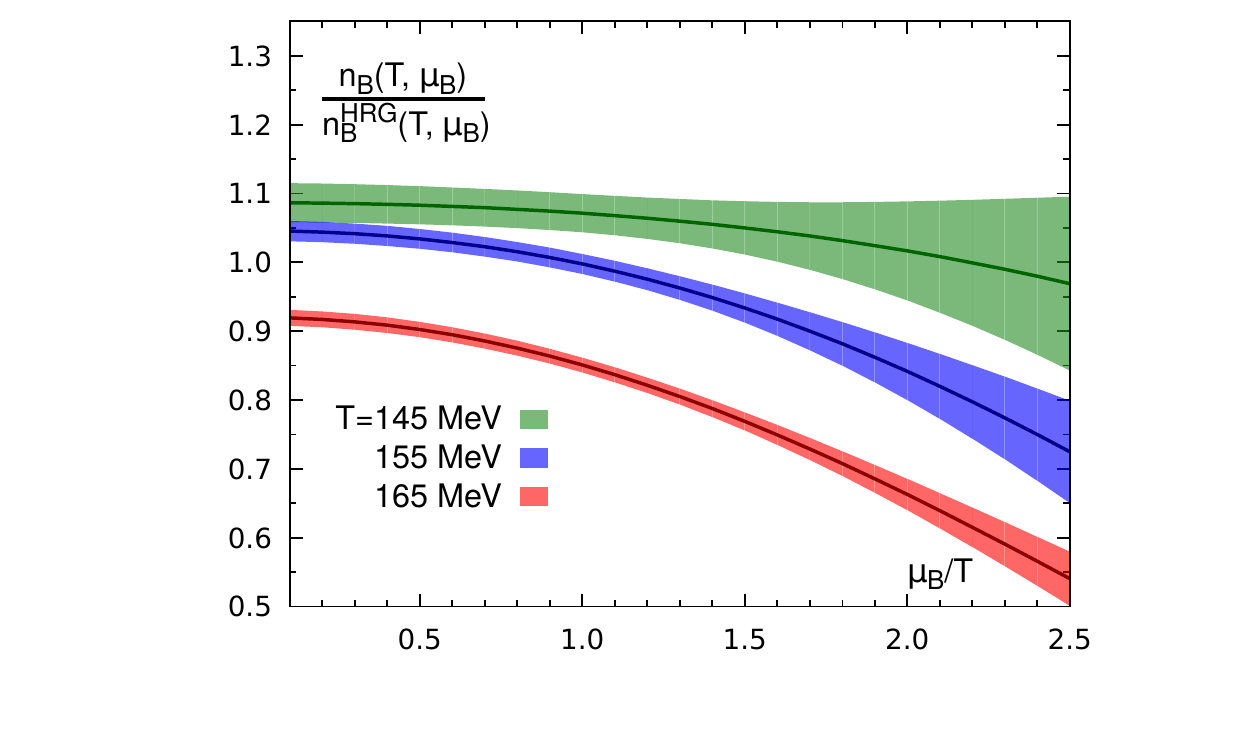}
\caption{The $\mu_B$ dependent contribution to
the pressure (top, left) and the baryon-number density (top, right)
for several values of the baryon chemical potential
in units of temperature.
The lower two panels show these quantities normalized to the corresponding
HRG model values, obtained from a calculation with all baryon resonances,
up to mass $m_H=2.5$~GeV, listed in the PDG tables, as function of $\mu_B/T$ 
for three values of the temperature.
}
\label{fig:pressureSN}
\end{figure}

Compared to the leading ${\cal O}(\hmu_B^2)$ contributions to
bulk thermodynamic observables
the ${\cal O}(\hmu_B^4)$ and ${\cal O}(\hmu_B^6)$
corrections are smaller in the strangeness neutral case than in the 
case $\mu_Q=\mu_S=0$, which we have discussed in the previous section. 
This is evident from
Fig.~\ref{fig:Pn}, where we show the ratios $12 P_4/P_2$ and $360 P_6/P_2$.
These combinations are unity in a HRG with $\mu_S=\mu_Q=0$ but smaller than
unity in the strangeness neutral case. Higher order corrections in Taylor
series for strangeness neutral systems thus are of less importance than 
in the case $\mu_S=0$. This also means that the errors,
which are large on e.g. sixth order expansion coefficients, are of less 
importance for the overall error budget of Taylor expansions in strangeness
neutral systems.
This is indeed reflected in the $\mu_B$-dependence
of $(P(T,\mu_B)-P(T,0))/T^4$ and $n_B(T,\mu_B)/T^3$ shown in the upper
panels of Fig.~\ref{fig:pressureSN} for the case $r=0.4$. 
As can be seen in these two figures, at low temperatures the $\mu_B$-dependent 
part of the pressure as well as the net baryon-number density agree quite well with 
HRG model calculations that describe the thermodynamics of a gas of non-interacting, point-like 
hadron resonances. This agreement, however, gets
worse at larger values of $\mu_B$. Not unexpectedly, at higher 
temperatures deviations from HRG model calculations become large already
at small values of $\mu_B$. This is apparent from the lower two panels of 
Fig.~\ref{fig:pressureSN}, where we show the ratio of the
$\mu_B$-dependent part of the pressure and the corresponding HRG model result 
(left) and the net baryon-number density divided by the corresponding HRG
model result (right).  In the HRG model
calculation $(P(T,\mu_B)-P(T,0))/T^4$ as well as $n_B(T,\mu_B)/T^3$ only 
depend on the baryon sector of the hadron spectrum. The results shown
in Fig.~\ref{fig:pressureSN} thus strongly suggest that HRG model 
calculations using resonance spectra in model calculations for
non-interacting, point-like hadron gases may be appropriate (within $\sim 10$\%
accuracy) to describe
the physics in the crossover region of strongly interacting matter at
vanishing or small values of the baryon chemical potential, but 
fail\footnote{It has been
pointed out that the point-like particle approximation is appropriate
in the meson sector but not in the baryon sector at high density. 
Introducing a non-zero size of hadron resonances
\cite{Hagedorn:1980kb,Dixit:1980zj} may,
for some observables, improve the comparison with QCD thermodynamics
\cite{Andronic:2012ut,Vovchenko:2016rkn}. 
However, it seems that
the introduction of several additional parameters will be needed to achieve 
overall good agreement with the many observables calculated now in QCD 
in the temperature range of interest, i.e. in the crossover region
from a hadron gas to strongly interacting quark-gluon matter.
} to do so at large $\mu_B/T$ and/or $T\gsim 160$~MeV. At $T=165$~MeV QCD and
HRG model results for the net baryon-number density differ by 40\% at
$\mu_B/T=2$.
This has consequences
for the determination of freeze-out conditions in heavy ion collisions.
We will come back to this discussion in Section~\ref{sec:LCP}. 

\begin{figure}[t]
\includegraphics[scale=0.7]{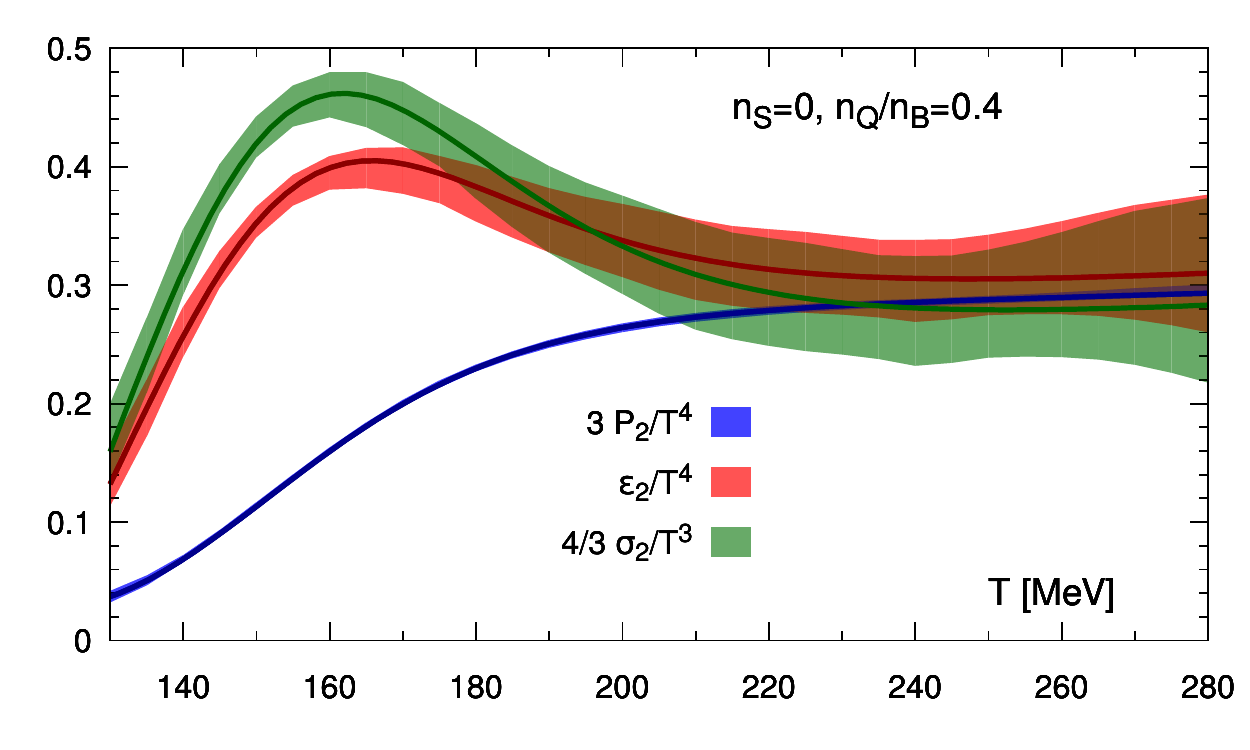}
\includegraphics[scale=0.7]{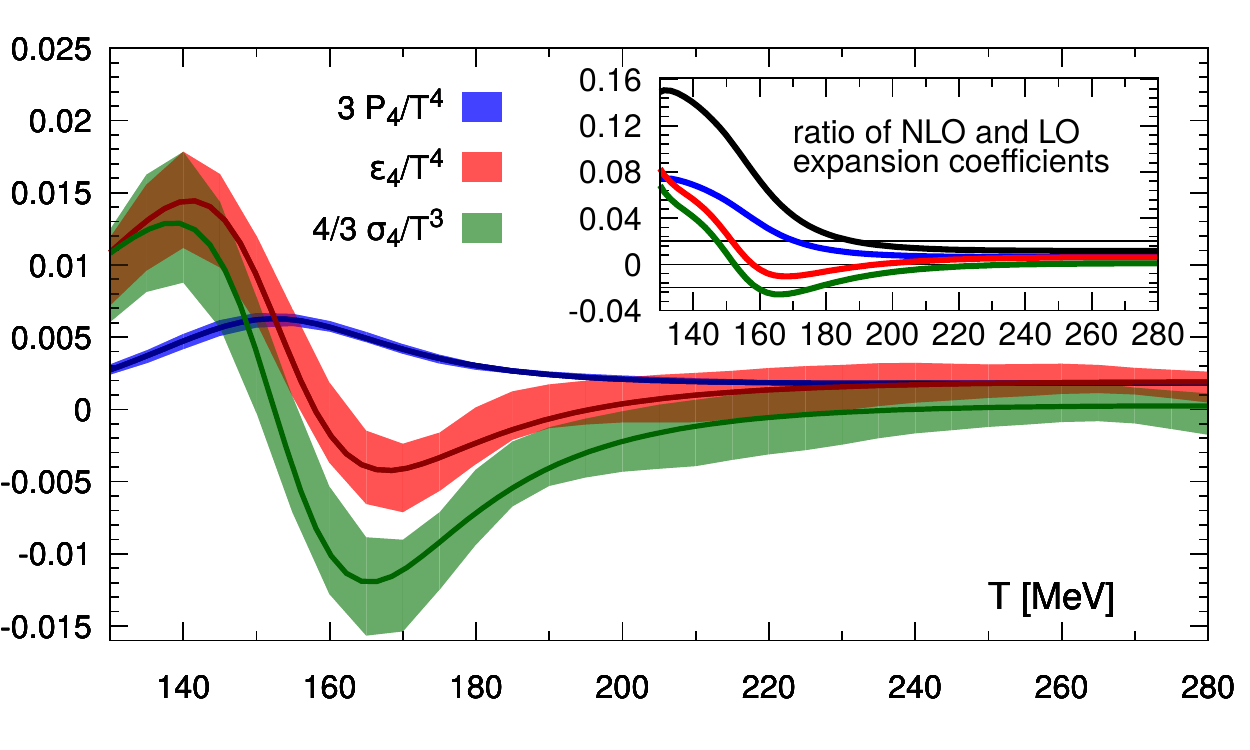}
\caption{Leading order (left) and next-to-leading order (right)
expansion coefficients for the $\mu_B$-dependent part of pressure,
the energy and entropy densities in the strangeness neutral
case with fixed electric charge to net baryon-number density, $n_Q/n_B=0.4$.
The darker lines in the center of the error bands of these extrapolations
show the interpolating fits discussed in Subsection~\ref{sec:fits}.
The insert in the right hand figure shows the ratios of NLO and LO
expansion coefficients $N^B_3/N^B_1$, $P_4/P_2$, $\epsilon_4/\epsilon_2$
and $\sigma_4/\sigma_2$. The influence of a non-vanishing electric
charge chemical potential, which formally gives rise to deviations from the
result in the isospin symmetric limit ($N^B_1=2 P_2$, $N^B_3=4 P_4$),
are negligible at ${\cal O}(\hmu_B^2)$ and ${\cal O}(\hmu_B^4)$. For that reason
we do not show results for $N_1^B$ and $N_3^B$. However, we show in the
insertion in the left hand figure the ratio $N_3^B/N_1^B$ (black line) which 
clearly shows that NLO corrections are a factor two larger in the Taylor series
for the number density then in the pressure series.
}
\label{fig:ecumulants_neutral}
\end{figure}

\begin{figure}[htb]
\hspace{-0.9cm}\includegraphics[scale=0.81]{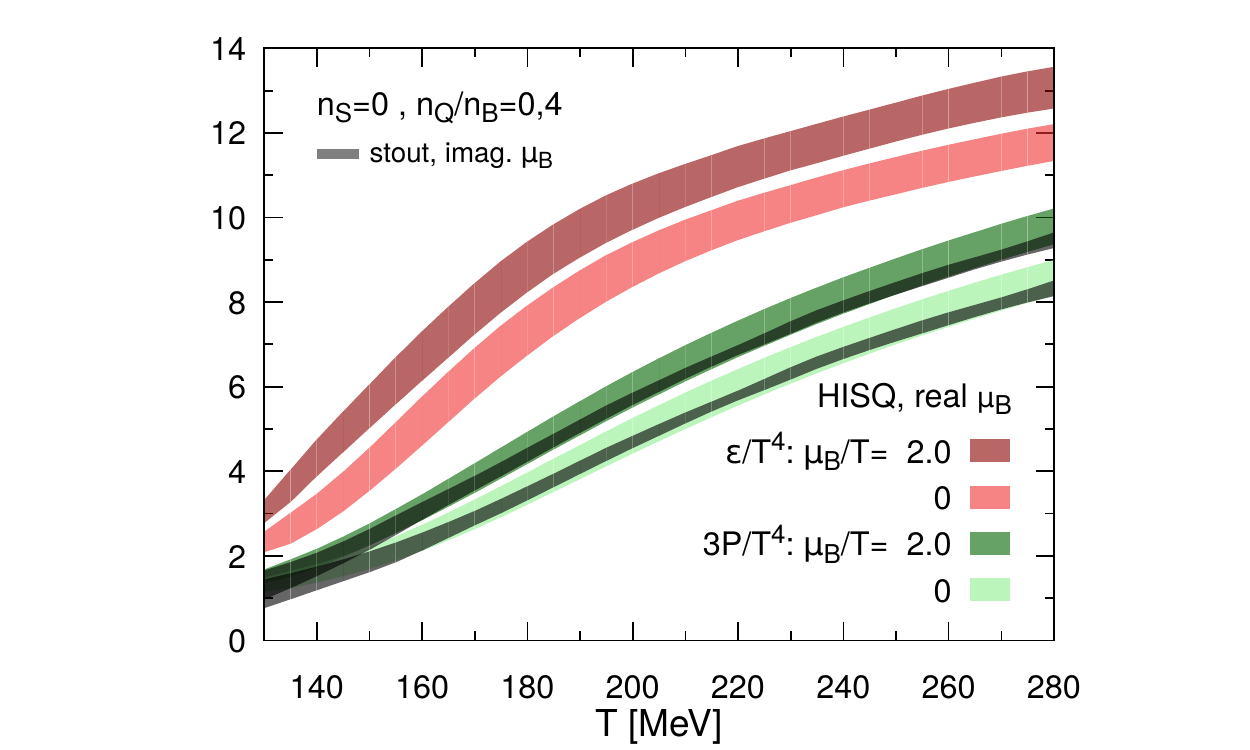}\hspace{-1.9cm}
\caption{
The total energy density (upper two curves) of 
(2+1)-flavor QCD
for $\mu_B/T=0$ and $2$, respectively. The lower two curves show
corresponding results for three times the pressure.
The dark lines show the results obtained with the stout action
from analytic continuation with sixth order polynomials in
$\hmu_B$ \cite{Gunther:2016vcp}.
}
\label{fig:etotal}
\end{figure}

The $\mu_B$-dependent contributions to the energy and entropy densities have
been defined in Eqs.~\ref{energyc} and \ref{entropyc}. In strangeness neutral
systems the expansion coefficients simplify considerably,
\begin{eqnarray}
\epsilon_{2n}(T) &=&3 P_{2n}(T)+T P_{2n}'(T) - r \sum_{k=1}^n T q'_{2k-1} 
N_{2n-2k+1}^B
\label{ec} \\
\sigma_{2n}(T) &=&4 P_{2n}(T)+T P_{2n}'(T) - N_{2n-1}^B-
r \sum_{k=1}^n (q_{2k-1}+T q'_{2k-1}) N_{2n-2k+1}^B
\label{sc}
\end{eqnarray}
Results for the ${\cal O}(\mu_B^2)$ and ${\cal O}(\mu_B^4)$ expansion 
coefficients are shown in 
Fig.~\ref{fig:ecumulants_neutral} together with the corresponding
expansion coefficients for the pressure and net baryon-number 
density. Results for the total energy density as well as the 
total pressure for $\mu_B/T=0$ and $2$ are shown in Fig.~\ref{fig:etotal}.
As discussed in the previous section also here it is evident that
current errors on the total pressure and energy density are dominated
by errors on these observables at $\mu_B=0$.

In Fig.~\ref{fig:etotal} we also show results for the total pressure
obtained within the stout discretization scheme. The result for $\hmu_B=0$ 
is taken from \cite{{Borsanyi:2013bia}}. The $\hmu_B$-dependent contribution
is based on calculations with an imaginary chemical potential 
\cite{Gunther:2016vcp}. These results have been analytically continued
to real values of $\hmu_B$ using a $6^{th}$ order polynomial in $\hmu_B$.
As can be seen the total pressure agrees quite well with the results
obtained with a sixth order Taylor expansion, although the results
obtained the analytic continuation within the stout discretization
scheme  tend to stay systematically below the central values obtained from 
the analysis of Taylor series expansions in the HISQ discretization scheme. 

\subsection{Parametrization of the equation of state}
\label{sec:fits}

At $\mu_B=0$ the HotQCD Collaboration presented a parametrization of the 
pressure,
obtained as interpolating curves for the continuum extrapolated fit, 
that also provided an adequate description of all the other basic thermodynamic 
quantities, i.e. the energy and entropy densities as well as the
specific heat and the velocity of sound \cite{Bazavov:2014pvz}. 
Here we want to extend this 
parametrization to the case $\hmu_B>0$. 
Similar to what has been done at $\mu_B=0$ it turns out that a 
ratio of fourth order polynomials in the inverse temperature
is flexible enough to describe the temperature dependence of
all required Taylor expansion coefficients in the temperature range
$T\in [130~{\rm MeV},280~{\rm MeV}]$. We use such an ansatz for the
three expansion coefficients of the net baryon-number density 
($N_1^B,\ N_3^B,\ N_5^B$) and the three electric charge chemical 
potentials ($q_1,\ q_3,\ q_5$). This suffices to calculate all 
thermodynamic observables in strangeness neutral systems.

We use a ratio of fourth order polynomials in $1/T$ as an ansatz
for the expansion coefficients of the net baryon-number density, 
\begin{eqnarray}
N^B_k(T) &=&  
\frac{N^B_{k,0n}+ N^B_{k,1n} \bar{t}+N^B_{k,2n} \bar{t}^2+
N^B_{k,3n} \bar{t}^3+N^B_{k,4n} \bar{t}^4}{1+N^B_{k,1d} \bar{t}+
N^B_{k,2d} \bar{t}^2+
N^B_{k,3d} \bar{t}^3+N^B_{k,4d} \bar{t}^4} \;\; , \;\; k=1,\ 3,\ 5 \;\; .
\label{NBfit}
\end{eqnarray}
Here $\bar{t} = T_c/T$ and the QCD transition temperature
$T_c=154$~MeV is used as a convenient normalization.  Similarly we
define the parametrization of the expansion coefficients for the
electric charge chemical potential,
\begin{eqnarray}
q_k(T) &=&  
\frac{q_{k,0n}+ q_{k,1n} \bar{t}+q_{k,2n} \bar{t}^2+
q_{k,3n} \bar{t}^3+q_{k,4n} \bar{t}^4}{1+q_{k,1d} \bar{t}+q_{k,2d} \bar{t}^2+
q_{k,3d} \bar{t}^3+q_{k,4d} \bar{t}^4} \;\; , \;\; k=1,\ 3,\ 5 \;\; .
\label{qnfit}
\end{eqnarray}
The parameters for these interpolating curves are summarized in 
Table~\ref{tab:pfit}.

The expansion coefficients of the pressure are then obtained by using
Eqs.~\ref{PN2}-\ref{PN6}. The resulting interpolating curves
for $P_k$ are shown as darker curves in Fig.~\ref{fig:NB}. All other
interpolating curves shown as darker curves in other figures have been
obtained by using the above interpolations. In particular, 
interpolating curves for
the energy and entropy densities are obtained by 
using Eqs.~\ref{ec} and \ref{sc} and calculating
analytically temperature derivatives of the parametrizations
of $P_n$ and $q_n$ given in Eqs.~\ref{NBfit} and \ref{qnfit}.
The resulting interpolating curves for the second and fourth
order Taylor expansion coefficients are shown in 
Fig.~\ref{fig:ecumulants_neutral}.

We also used a ratio of fourth order polynomials to interpolate results
for the pressure at $\mu_B=0$. We write the pressure as
\begin{eqnarray}
\frac{P(T,\mu_B=0)}{T^4} &=&  
\frac{p_{0n}+ p_{1n} \bar{t}+p_{2n} \bar{t}^2+
p_{3n} \bar{t}^3+p_{4n} \bar{t}^4}{1+p_{1d} \bar{t}+p_{2d} \bar{t}^2+
p_{3d} \bar{t}^3+p_{4d} \bar{t}^4} \;\; .
\label{pnfit}
\end{eqnarray}
The coefficients $p_{in}$ and $p_{id}$ are also given in Table~\ref{tab:pfit}.

\begin{table}[t]
\begin{center}
\begin{tabular}{|c||c|c|c|c|c|c|c|c|c|}
\hline
$k$&$N^B_{k,0n}$&$N^B_{k,1n}$&$N^B_{k,2n}$&$N^B_{k,3n}$&$N^B_{k,4n}$ &
$N^B_{k,1d}$&$N^B_{k,2d}$&$N^B_{k,3d}$&$N^B_{k,4d}$ \\
\hline
1&0.302182&-0.929305& 1.230560&-0.798724&0.204722 &-2.011836& 1.190147& 0.003869 &-0.076244 \\
3&0.000446650&0.00983742&-0.0315076&0.0323632&-0.0107642 &-1.327047&0.0472047 &0.0&0.323696  \\
5&0.0000104211&-0.000327321&0.00122751&-0.00158725&0.000672708&-1.467875&-0.264770 &0.796010 &-0.044968  \\
\hline
\hline
$k$&$q_{k,0n}$&$q_{k,1n}$&$q_{k,2n}$&$q_{k,3n}$&$q_{k,4n}$ &
$q_{k,1d}$&$q_{k,2d}$&$q_{k,3d}$&$q_{k,4d}$ \\
\hline
1&-0.114472&-0.631833&2.102001&-2.165174&0.739905 &16.565265&-35.328733&19.940335& 0.384797 \\
3&0.0505332 &-0.312052 &0.700958 &-0.662171&0.219351 &-23.224117&82.688725&-89.160400&31.381036 \\
5&0.0000842&-0.0005250&0.00113467&-0.00103897 &0.00034414&-2.095094&0.987940&0.146830&-0.0210650 \\
\hline
\hline
~&$p_{0n}$&$p_{1n}$&$p_{2n}$&$p_{3n}$&$p_{4n}$ &
$p_{1d}$&$p_{2d}$&$p_{3d}$&$p_{4d}$ \\
\hline
0&0.00556035&128.702341&-293.064074&228.763685&-58.084225&12.713331&0.0&-31.330957&26.524394 \\
\hline
\end{tabular}
\end{center}
\caption{Parameters used in the ansatz given in Eq.~\ref{NBfit} for 
the interpolation of the expansion coefficients of the net baryon-number 
density of (2+1)-flavor QCD with vanishing net strangeness and a fixed 
ratio of electric charge and net baryon-number density, $n_Q/n_B=0.4$.  
These interpolations have been determined for the temperature 
interval $T\in [130~{\rm MeV}, 280~{\rm MeV}]$.
Also given are parameters needed for the interpolation of the expansion
coefficients for the electric charge chemical potential (Eq.~\ref{qnfit})
\label{tab:pfit} and the coefficients for the parametrization of the pressure 
at $\mu_B=0$ given in Eq.~\ref{pnfit}.
}
\label{tab:pfit}
\end{table}

\section{Lines of constant physics to \boldmath${\cal O}(\mu_B^4)$}
\label{sec:LCP}

We want to use here the Taylor series for bulk thermodynamic
observables, i.e. the pressure, energy and entropy densities, 
to discuss contour lines in the $T$-$\mu_B$ plane on which these
observables stay constant.
It has been argued quite successfully that the thermal conditions
at the time of chemical freeze-out in heavy ion collisions can be
characterized by lines in the $T$-$\mu_B$ plane on which certain thermodynamic
observables or ratios thereof stay constant 
\cite{Cleymans:1999st,Cleymans:2005xv}, although the freeze-out
mechanism in the rapidly expanding fireball created in a heavy
ion collision is of dynamical origin and will in detail be more
complicated (see for instance \cite{Tomasik:2002qt}).
While lines of constant physics (LCPs) involving total baryon-number
densities, as used in \cite{Cleymans:1999st,Cleymans:2005xv}, are not
appropriate for calculations within the framework of quantum field
theories, other criteria like lines of constant entropy density in units
of $T^3$
\cite{Cleymans:2004hj} or
constant pressure \cite{Rafelski:2009jr,Petran:2013qla,Rafelski:2015hta} 
have been suggested to characterize freeze-out parameters
$(T_f,\mu_B^f)$ corresponding to heavy ion collisions at different 
values of the beam energy ($\sqrt{s_{NN}}$). Generally such criteria have
been established by comparing experimental data with model calculations
based on some version of a HRG model. We will determine here LCPs from
the lattice QCD calculations of pressure, energy and entropy densities
and confront them with freeze-out parameters that have been obtained by 
comparing particle yields, measured at different values of $\sqrt{s_{NN}}$,
to HRG model calculations.

We consider an observable $f(T,\mu_B)$,
i.e.\ the pressure, energy density or entropy density which are even
functions of $\mu_B$. We parametrize a 'line of constant $f$' by,
\begin{equation}
T_f(\mu_B) = T_0 \left(1-\kappa_2^f 
\left( \frac{\mu_B}{T_0}\right)^2- \kappa_4^f \left( \frac{\mu_B}{T_0}
\right)^4\right) \; .
\label{Tf}
\end{equation}
In order to determine the expansion coefficients $\kappa_2^f$ and
$\kappa_4^f$ we need to expand the function $f(T,\mu_B)$ up to $4^{th}$
order in $\mu_B$ and up to second order in $T$ around some point $(T_0,0)$,
\begin{eqnarray}
f(T,\mu_B) = && f(T_0,0) + \left. 
\frac{\partial f(T,\mu_B)}{\partial T}\right|_{(T_0,0)} (T-T_0)
+ \frac{1}{2}   \left. 
\frac{\partial^2 f(T,\mu_B)}{\partial \mu_B^2}\right|_{(T_0,0)}   \mu_B^2 
\label{Taylorfa} \\
&+& \frac{1}{2}\left. \frac{\partial^2 f(T,\mu_B)}{\partial T^2}\right|_{(T_0,0)} (T-T_0)^2 + \frac{1}{2} \left. \frac{\partial}{\partial T} \frac{\partial^2 f(T,\mu_B)}{\partial \mu_B^2}\right|_{(T_0,0)} (T-T_0) \mu_B^2 + \frac{1}{4!}   \left. 
\frac{\partial^4 f(T,\mu_B)}{\partial \mu_B^4}\right|_{(T_0,0)}   \mu_B^4
\; .
\nonumber
\end{eqnarray}
Note that we expand here in terms of $\mu_B$ rather than in 
$\hmu_B\equiv \mu_B/T$.
Replacing the temperature $T$ in Eq.~\ref{Taylorfa} by the ansatz for a 
line of constant $f$, Eq.~\ref{Tf}, and keeping terms up to 
${\cal O}(\mu_B^4)$
gives
\begin{eqnarray}
f(T(\mu_B),\mu_B) =&& f(T_0,0) + 
\left( - \left. \kappa_2^f \frac{\partial f(T,\mu_B)}{\partial T}\right|_{(T_0,0)} \frac{1}{T_0} + \frac{1}{2} 
\left. \frac{\partial^2 f(T,\mu_B)}{\partial \mu_B^2}\right|_{(T_0,0)}  \right)
\mu_B^2 
\nonumber \\
&+&\left(- \left. \kappa_4^f \frac{\partial f(T,\mu_B)}{\partial T}\right|_{(T_0,0)} \frac{1}{T_0^3}  
+ \left. \frac{1}{2} \frac{\partial^2 f(T,\mu_B)}{\partial T^2}\right|_{(T_0,0)} (\kappa_2^f)^2\frac{1}{T_0^2} -
\frac{1}{2} \left. \frac{\partial}{\partial T} \frac{\partial^2 f(T,\mu_B)}{\partial \mu_B^2}\right|_{(T_0,0)} \kappa_2^f \frac{1}{T_0} \right.
\nonumber \\
&&\hspace{0.5cm}+ \frac{1}{4!}   \left. \left. 
\frac{\partial^4 f(T,\mu_B)}{\partial \mu_B^4}\right|_{(T_0,0)} \right)  \mu_B^4
\; .
\nonumber
\end{eqnarray}
We then can determine $\kappa_2^f$ and $\kappa_4^f$ by demanding that the 
expansion coefficients at ${\cal O}(\mu_B^2)$ and ${\cal O}(\mu_B^4)$ vanish,
i.e.
\begin{eqnarray}
\kappa_2^f &=& \frac{T_0}{2}  \frac{
\left. \frac{\partial^2 f(T,\mu_B)}{\partial \mu_B^2}\right|_{(T_0,0)} }{
\left. \frac{\partial f(T,\mu_B)}{\partial T}\right|_{(T_0,0)} }
\; ,\label{k2f}  \\
\kappa_4^f &=& \frac{\frac{1}{2}T_0^2\left. \frac{\partial^2 f(T,\mu_B)}{\partial T^2}\right|_{(T_0,0)} (\kappa_2^f)^2 -
\frac{1}{2} \left. T_0^3 \frac{\partial}{\partial T} \frac{\partial^2 f(T,\mu_B)}{\partial \mu_B^2}\right|_{(T_0,0)} \kappa_2^f  + \frac{1}{4!} T_0^4  \left. 
\frac{\partial^4 f(T,\mu_B)}{\partial \mu_B^4}\right|_{(T_0,0)}}{\left. 
T_0 \frac{\partial f(T,\mu_B)}{\partial T}\right|_{(T_0,0)} }
\; .
\label{k4f}
\end{eqnarray}
As we will deal with observables that are given as a Taylor series in
$\hmu_B$ at fixed $T$, i.e. 
$f(T,\mu_B)= \sum_{k=0}^{\infty} f_{2k}(T)\hmu_B^{2k}$, the derivatives
with respect to $\mu_B$ appearing in Eqs.~\ref{k2f} and \ref{k4f} can
be replaced by suitable Taylor expansion coefficients of $f(T,\mu_B)$,
\begin{eqnarray}
\kappa_2^f &=&  \frac{ f_2(T_0)}{
\left. T_0 \frac{\partial f_0(T)}{\partial T}\right|_{(T_0,0)} }
\label{k2fb}  \\
\kappa_4^f &=& \frac{\frac{1}{2}\left. T_0^2\frac{\partial^2 f_0(T)}{\partial T^2}\right|_{(T_0,0)} (\kappa_2^f)^2 -
\left. 
\left( T_0 \frac{\partial f_2(T)}{\partial T} \right|_{(T_0,0)} -2 f_2(T_0) \right)
\kappa_2^f  + 
f_4(T_0)}{\left. T_0 
\frac{\partial f_0(T)}{\partial T}\right|_{(T_0,0)} }
\label{k4fb}
\end{eqnarray}
We will in the following work out detailed expressions for the 
quadratic correction coefficient, $\kappa_2^f$, for lines of constant
pressure ($f\equiv P$), energy density ($f\equiv \epsilon$) and 
entropy density ($f\equiv s$) in strangeness neutral systems with
electric charge to net baryon-number ratio $r=0.4$. 
Details for the quartic coefficient,
$\kappa_4^f$, are given in Appendix~\ref{app:curvature}. 

\begin{itemize}
\item[]
\underline{pressure $f\equiv P$:}\\
The function $f(T,\mu_B)$ is given by $P=T^4\sum_n P_n (\mu_B/T)^n$,
with $P_0=P(T,0)/T^4$ denoting the pressure in units of $T^4$ 
at vanishing baryon chemical potential and $P_n(T)$, $n>0$, denoting the
expansion coefficients of $P(T,\mu_B)/T^4$ as introduced in Eq.~\ref{Pn}.
In the denominator of Eq.~\ref{k2fb} we use the thermodynamic relation between
pressure and entropy density $ s=(\partial P/\partial T)_{\mu_B}$.
The numerator is given by $f_2(T)=T^4 P_2(T)$.
This gives
\begin{equation}
\kappa_2^P = \frac{P_2}{s/T^3} \; ,
\end{equation}
where $s/T^3$ is evaluated at $\hmu_B=0$.
\item[]
\underline{energy density $f\equiv \epsilon$:}\\
The function $f(T,\mu_B)$ is given by $\epsilon=T^4\sum_n \epsilon_n (\mu_B/T)^n$,
with $\epsilon_0=\epsilon(T,0)/T^4$ denoting the energy density in units of $T^4$
at vanishing baryon chemical potential .
In the denominator of Eq.~\ref{k2fb} we use the thermodynamic relation between
energy density and specific heat
$C_V= (\partial \epsilon/\partial T)_{\mu_B}$
In the numerator we have $f_2(T)=T^4 \epsilon_2(T)$. This gives
\begin{equation}
\kappa_2^\epsilon = \frac{\epsilon_2}{C_V/T^3} \; ,
\end{equation}
where $C_V/T^3$ is evaluated at $\hmu_B=0$.
\item[]
\underline{entropy density $f\equiv s$:}\\
The function $f(T,\mu_B)$ is given by 
$s=(\epsilon +P-\mu_B n_B-\mu_Q n_Q)/T = 
(\epsilon +P-\mu_B n_B (1+r \mu_Q/\mu_B))/T$.
As $n_B$ is of ${\cal O}(\mu_B)$ we need for the ratio of electric
charge and strangeness chemical potentials only the leading order
relation $\mu_Q/\mu_B=q_1$ defined in Eq.~\ref{qs}. 
In the denominator we use,
\begin{equation}
\frac{\partial s}{\partial T} =
\frac{\partial (\epsilon+P)/T}{\partial T} = -\frac{s}{T} 
+\frac{1}{T} \frac{\partial (\epsilon+P)}{\partial T} = \frac{C_V}{T}
\; .
\end{equation}
In the numerator we have 
$f_2(T)=T^3(\epsilon_2+P_2-N_1^B  (1+r q_1))$. With this we get,
\begin{equation}
\kappa_2^s = T^3 \frac{\epsilon_2+P_2-N_1^B  (1+r q_1)}{C_V} =\frac{\epsilon_2 -P_2}{C_V/T^3}
\; .
\end{equation}
where we have used Eq.~\ref{PN2} to replace $N_1^B$ in favor of $P_2$.
\end{itemize}
We note that $\kappa_2^\epsilon > \kappa_2^s$, i.e. with increasing
$\mu_B$ the entropy density  decreases on lines of constant energy 
density.

\begin{figure}[t]
\begin{center}
\includegraphics[width=78mm]{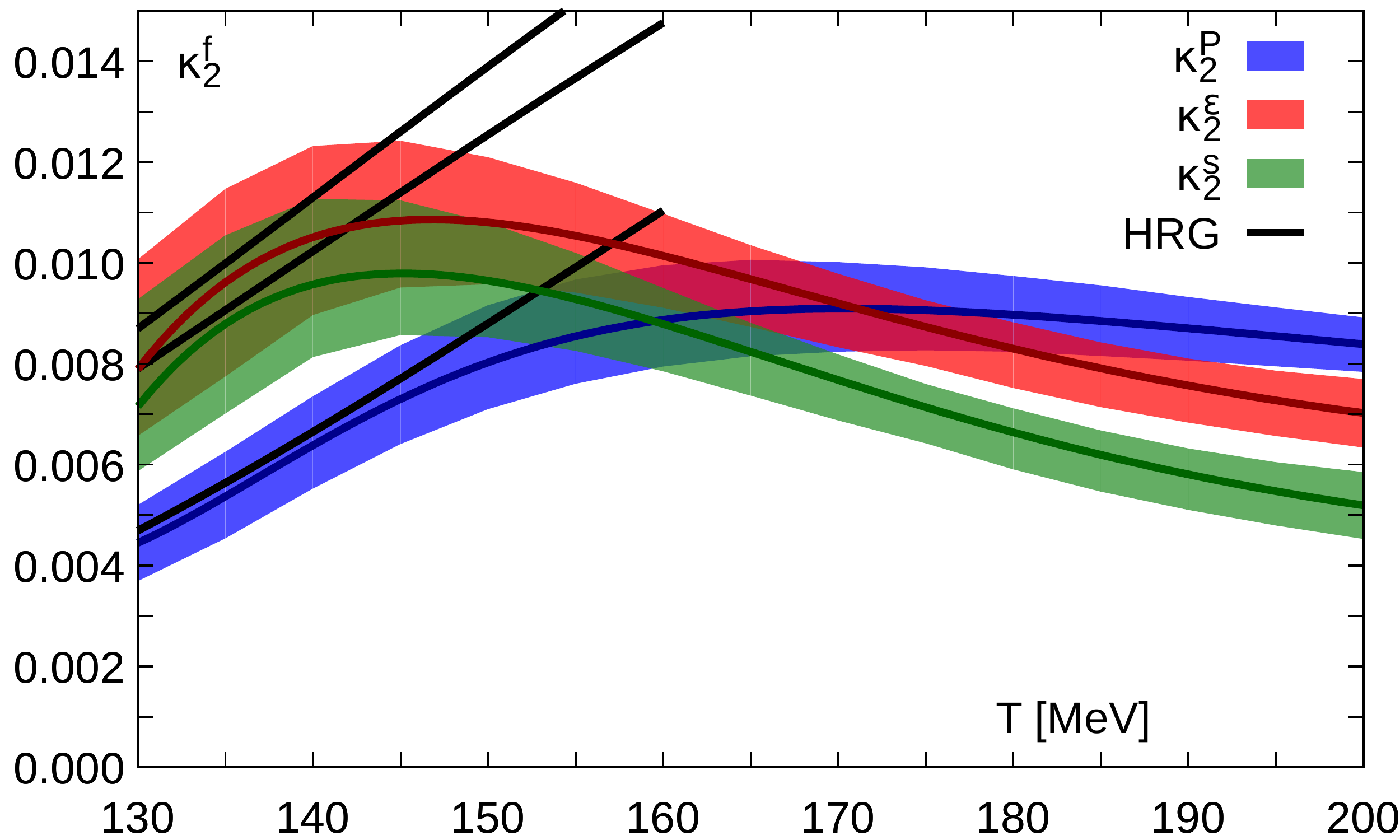}
\includegraphics[width=78mm]{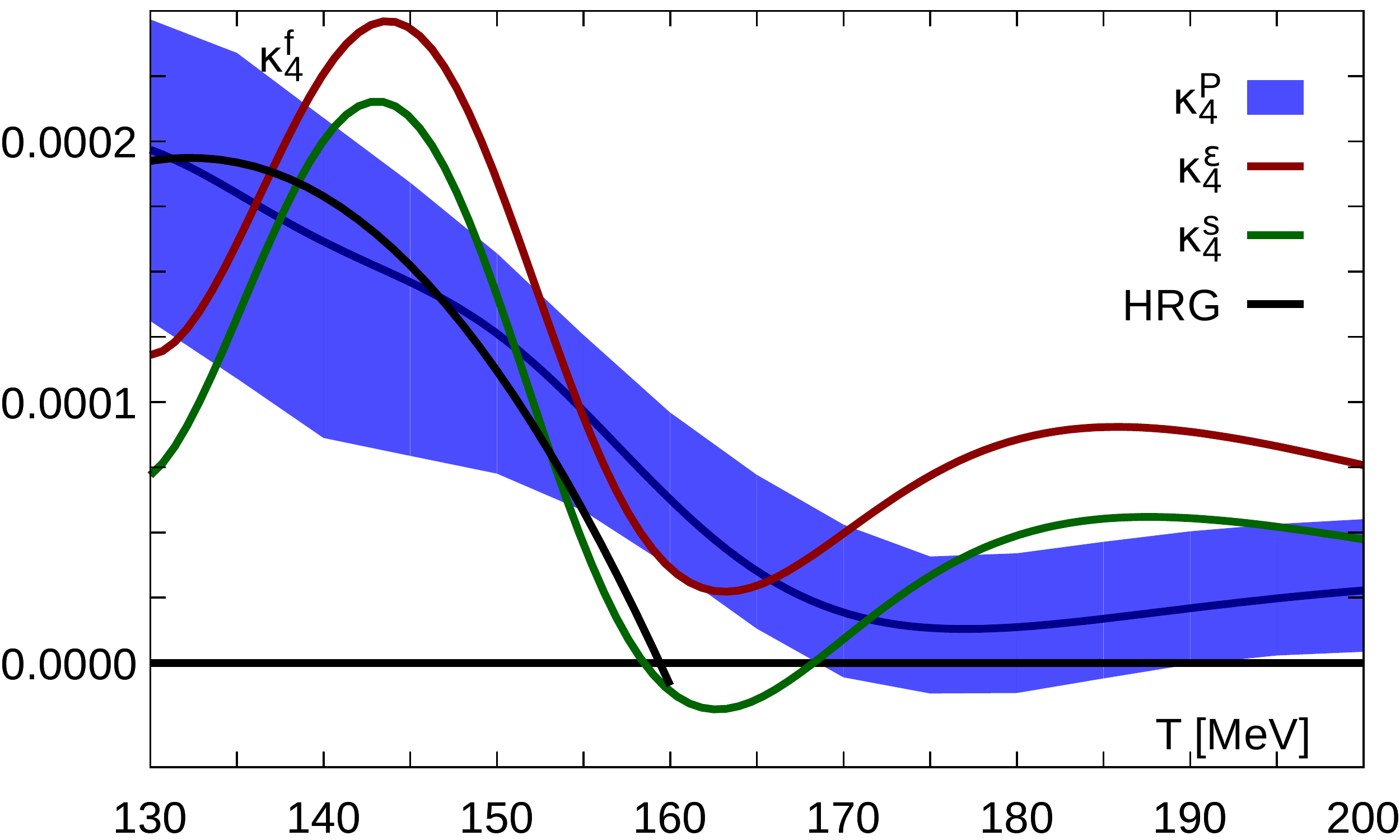}
\caption{{\it Left:} Second order curvature coefficients of lines of
constant pressure,
energy density and entropy density versus temperature in (2+1)-flavor
QCD (bands) and in a HRG model (lines). {\it Right:}
same as on the left, but for fourth order coefficients.
The darker lines in the center of the error bands 
show the interpolating fits discussed in subsection~\ref{sec:fits}.
For $\kappa_4^\epsilon$ and $\kappa_4^s$ only these interpolating curves
are shown.
}
\label{fig:kappa}
\end{center}
\end{figure}

The second order coefficients for the lines of constant physics thus
can directly be calculated using the continuum extrapolated results
for the pressure and energy density obtained at vanishing chemical
potential in \cite{Bazavov:2014pvz} and the leading order expansion coefficient
of the pressure shown in Fig.~\ref{fig:Pn}. Similarly we obtain
the quartic coefficients from the fourth order expansion of the
pressure using the relations given in Appendix~\ref{app:curvature}.
We show results for $\kappa_2^f$ and $\kappa_4^f$ in Fig.~\ref{fig:kappa}.

In the interval around $T_c$, i.e. 
$T\in [145~{\rm MeV},165~{\rm MeV}]$ we find,
\begin{equation}
0.0064 \le \kappa_2^P  \le 0.0101 \;\; ,\;\; 
0.0087 \le \kappa_2^\epsilon  \le 0.012  \;\; ,\;\;
0.0074 \le \kappa_2^s  \le 0.011 \; .
\label{kappa2}
\end{equation}
Apparently, at ${\cal O}(\mu_B^2)$, lines of constant pressure and constant energy or entropy densities agree
quite well and they also agree, within currently large errors, with the curvature
of the transition line in (2+1)-flavor QCD. 
The coefficient of the quartic correction for the contour lines turns out to 
be about two orders of magnitude smaller than the
leading order coefficients. This, of course, reflects the small contribution
of the NLO corrections to the $\mu_B$-dependent part of pressure and energy 
density. For all fourth order coefficients we find $|\kappa_4^f|\le 0.00024$
in the temperature interval around $T_c$.
For $\mu_B/T\le 2$ the contribution arising from $\kappa_4^f$ only
leads to modifications of $T_f(\mu_B)$ that stays within the error band
arising from the uncertainty in $\kappa_2^f$. 

\begin{figure}[t]
\begin{center}
\includegraphics[width=81mm]{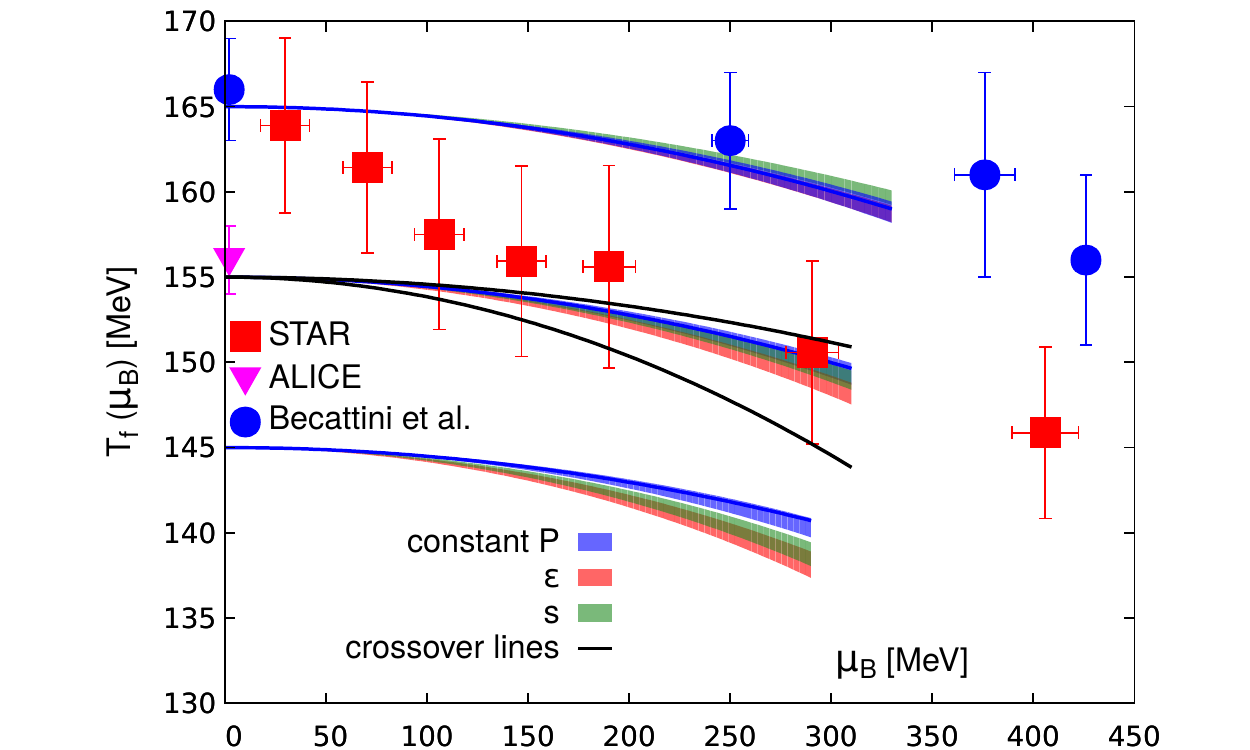}
\includegraphics[width=81mm]{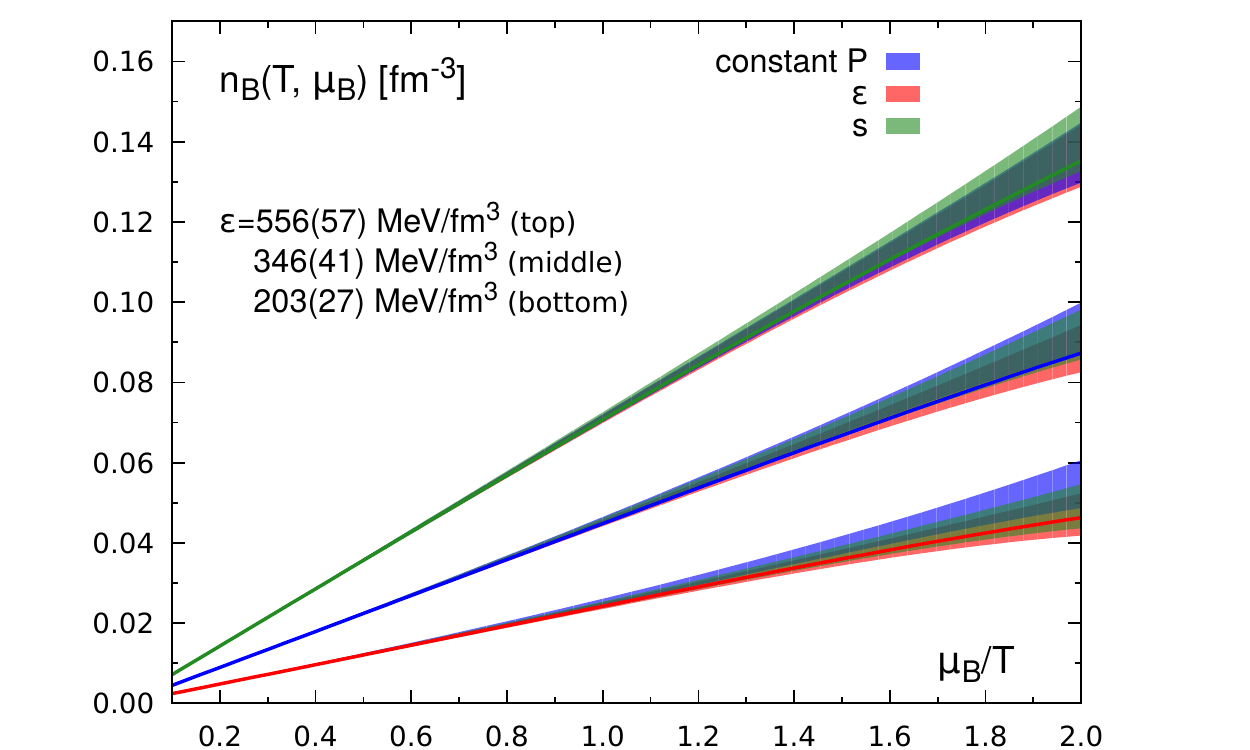}
\caption{{\it Left:} Lines of constant pressure,
energy density and entropy density versus temperature in (2+1)-flavor
QCD for three different initial sets of values fixed at $\mu_B=0$ 
and $T_0=145$~MeV, $155$~MeV and $165$~MeV, respectively 
(see Table~\ref{tab:LCP}). 
Data points show freeze-out temperatures determined by the
STAR Collaboration in the BES at RHIC (squares) \cite{Das:2014qca}
and the ALICE Collaboration at the LHC (triangle)
\cite{Floris:2014pta}. The circles denote hadronization temperatures
obtained by comparing experimental data on particle yields with a
hadronization model calculation \cite{Becattini:2016xct}. 
Also shown are two lines representing the current spread in determinations
of the $\mu_B$-dependence of the QCD crossover transition line (see text).
{\it Right:} Net baryon-number density on the lines of constant physics
for three values of the energy density at $\mu_B=0$.
Other thermodynamic parameters characterizing these lines
are summarized in Table~\ref{tab:LCP}.
}
\label{fig:LCP}
\end{center}
\end{figure}

\begin{table}[t]
\centering
\begin{tabular}{|c|ccc||lcc|}
\hline
~&\multicolumn{3}{|c||}{at $\mu_B=0$}&\multicolumn{3}{|c|}{on LCP} \\
\hline
$T_0~{\rm [MeV]}$  &   $p/T_0^4$   &   $\epsilon/T_0^4$  & $s/T_0^3$ & $~p~{\rm [GeV/fm}^3]$ &$\epsilon~{\rm [GeV/fm}^3]$&$s~[{\rm fm}^{-3}]$ \\
\hline
145 &   0.586(80)  &3.52(47) & 4.11(53) & ~0.0337(46)  &  0.203(27)  &   1.63(21) \\
155 &   0.726(95)  &4.61(55) & 5.34(63) & ~0.0546(71)  &  0.346(41)  &   2.59(30)\\
165 &   0.898(110) &5.76(59) & 6.66(69) & ~0.0868(106) &  0.556(57)  &   3.90(40) \\
\hline
\end{tabular}
\caption{Pressure, energy density and entropy density, characterizing
lines of constant physics which correspond to the conditions met for
$\mu_B=0$ at $T_0=145$~MeV, $155$~MeV and $165$~MeV. Columns 2-4 give
results in appropriate units of temperature, while columns 5-7 give 
the same results expressed in units of $GeV$ and $fm$. 
}
\label{tab:LCP}
\end{table}

The resulting lines of constant physics in the $T$-$\mu_B$ plane are shown
in Fig.~\ref{fig:LCP}~(left) for three values of the temperature, $T=145$~MeV,
$155$~MeV and $165$~MeV. These correspond to constant energy densities
$\epsilon = 0.203(27)$~GeV/fm$^3$, $0.346(41)$~GeV/fm$^3$ and 
$0.556(57)$~GeV/fm$^3$, which roughly correspond to the energy density
of cold nuclear matter, a hard sphere gas of nucleons at dense packing
and the interior of a nucleus, respectively. 
Values of other bulk thermodynamic observables characterizing these LCPs
are summarized in Table~\ref{tab:LCP}.
 The corresponding
net baryon-number densities on these LCPs are shown in 
Fig.~\ref{fig:LCP}~(right). It is apparent from Fig.~\ref{fig:LCP}~(left)
that LCPs for constant pressure, energy or entropy density agree well
with each other up to baryon chemical potentials $\mu_B/T = 2$, where the 
difference in temperature on different LCPs is at most $2$~MeV. We also note 
that the temperature
on a LCP varies by about $7$~MeV or, equivalently, 5\% between $\hmu_B=0$ 
and $\hmu_B=2$. 
Thus on a line of constant pressure, the entropy in units 
of $T^3$ changes by about 15\%. I.e. constant $P$ or constant $s/T^3$, 
which both have been suggested as phenomenological descriptions for 
freeze-out conditions in heavy ion collisions, can not hold simultaneously, 
although a change of 15\% of one of these observables may phenomenologically
not be of much relevance. We also stress that at large values of $\hmu_B$
the comparison of experimental data with HRG model calculations, e.g.
the use of single particle Boltzmann distributions used to extract
freeze-out temperatures and chemical potentials, becomes questionable.
As shown in Fig.~\ref{fig:pressureSN} net baryon-number densities extracted
from HRG and QCD calculations differ substantially at $\mu_B/T \simeq 2$. 

Also shown in Fig.~\ref{fig:LCP}~(left) are
results on freeze-out parameters and
hadronization temperatures extracted from particle yields measured in 
heavy ion experiments \cite{Das:2014qca,Floris:2014pta,Becattini:2016xct}
by comparing data with model calculations based on the hadron resonance gas 
models. 
The region $\mu_B/T\le 2$ corresponds to  beam energies
$\sqrt{s_{NN}}\ge 11.4$~GeV in the RHIC beam energy scan.
Obviously, the freeze-out parameters extracted from the beam
energy scan data \cite{Das:2014qca} do not follow any of the LCPs. However,
the discrepancy between the freeze-out parameters determined at the LHC
\cite{Floris:2014pta} and the highest beam energy at RHIC \cite{Das:2014qca}
suggests that also these determinations are not consistent among each other.

Finally we note that the lines of constant physics discussed above 
compare also well with the crossover line for the QCD transition.
At non-zero values of the baryon chemical potential the change of
the (pseudo)-critical temperature has been determined, using various
approaches at real \cite{Kaczmarek:2011zz,Endrodi:2011gv} and imaginary
\cite{Bonati:2015bha,Bellwied:2015rza,Cea:2015cya} values of the chemical potential.
To leading order one obtains,
\begin{equation}
T_c (\mu_B) = T_c(0) \left( 1 - \kappa^c_2 \left( \frac{\mu_B}{T_c(0)}\right)^2\right)
\label{TcmuB}
\end{equation}
with $\kappa^c_2$ ranging from $0.0066(7)$ \cite{Kaczmarek:2011zz,Endrodi:2011gv}
to $0.0135(20)$ \cite{Bonati:2015bha}, $0.0149(21)$ \cite{Bellwied:2015rza} and $0.020(4)$ \cite{Cea:2015cya}. Lines
that cover this spread in curvature parameters are also shown in 
Fig.~\ref{fig:LCP}~(left) for $T_c(0)=155$~MeV. While a small curvature
for the crossover line would suggest that the crossover transition happens
under more or less identical bulk thermodynamic conditions a large
curvature obviously would indicate that the crossover transition
happens already at significantly smaller values of pressure and energy
density as $\mu_B/T$ increases. 

\section{Radius of convergence and the critical point}
\label{sec:radius}
As discussed in the previous sections we generally find that the 
Taylor series for all basic thermodynamic quantities converge well
for values of baryon chemical potentials $\mu_B\le 2 T$.
Even in the low temperature regime the relative contribution of higher
order expansion coefficients are generally smaller than in corresponding
HRG model calculations.
This, of course, also has consequences for our current understanding
of the location of a possible critical point in the QCD phase diagram.

The results on the expansion coefficients of the Taylor series for e.g. 
the pressure can be cast into 
estimates for the location of a possible critical point in the QCD 
phase diagram. In general the radius of convergence can be obtained
from ratios of subsequent expansion coefficients in the Taylor series 
for the pressure. Equally well one may use one of the derivatives of the
pressure series. As one has to rely on estimates of the radius of 
convergence that generally are based on a rather short series, it may
indeed be of advantage to use as a starting point the series for the 
net baryon-number susceptibility \cite{Gavai:2004sd}, which diverges at the 
critical point, but still contains information from all expansion coefficients
of the pressure series. 
The radius of convergence of this series is identical to that of the 
pressure. Model calculations also suggest that the estimators obtained
from the susceptibility
series converge faster to the true radius of convergence \cite{Karsch:2011yq}.
For $\mu_Q=\mu_S=0$ the expansion coefficients of the Taylor series  for the
net baryon-number susceptibility are again simply related to that of the 
pressure,
\begin{equation}
\chi_2^B(T,\mu_B) = \sum_{n=0}^\infty \frac{1}{(2n)!} \chi_{2n+2}^B \hmu_B^{2n}
\; .
\label{chi2Bseries}
\end{equation}
From this one obtains estimators for the radius of convergence of the pressure
and susceptibility series,
\begin{eqnarray}
r_{2n}^P= \left| \frac{(2n+2)(2n+1) \chi_{2n}^B}{\chi_{2n+2}^B}\right|^{1/2} \; &,& \;
r_{2n}^{\chi} = \left| \frac{2n (2n-1)\chi_{2n}^B}{\chi_{2n+2}^B} \right|^{1/2} \; .
\label{rnchi}
\end{eqnarray}
Both estimators converge
to the true radius of convergence in the limit $n\rightarrow\infty$.
In order for this to correspond to a singularity at real values of $\hmu_B$, 
all expansion coefficients should asymptotically stay positive.
 
Obviously, the estimators $r_{2n}^P$ and $r_{2n}^{\chi}$ are proportional to 
each other, $r_{2n}^P =\sqrt{(2n+2)(2n+1)/[2n (2n-1)]} r_{2n}^{\chi}$.
The difference between these to estimators may be taken as a systematic
error for any estimate of the radius of convergence obtained from a
truncated Taylor series.
In the hadron resonance gas limit one finds for estimators involving
sixth order cumulants, $r_{4}^P = 1.58 r_4^\chi$. In the following
we restrict our discussion to an analysis of $r_{2n}^\chi$, which
at finite $n$ leads to the smaller estimator for the radius of convergence.
This seems to be appropriate
in the present situation where we only can construct two independent
estimators from ratios of three distinct susceptibilities. We thus may
hope to identify regions in the QCD phase diagram at small values of 
$\hmu_B$ which are unlikely locations for a possible critical point.

\begin{figure}[t]
\begin{center}
\includegraphics[width=81mm]{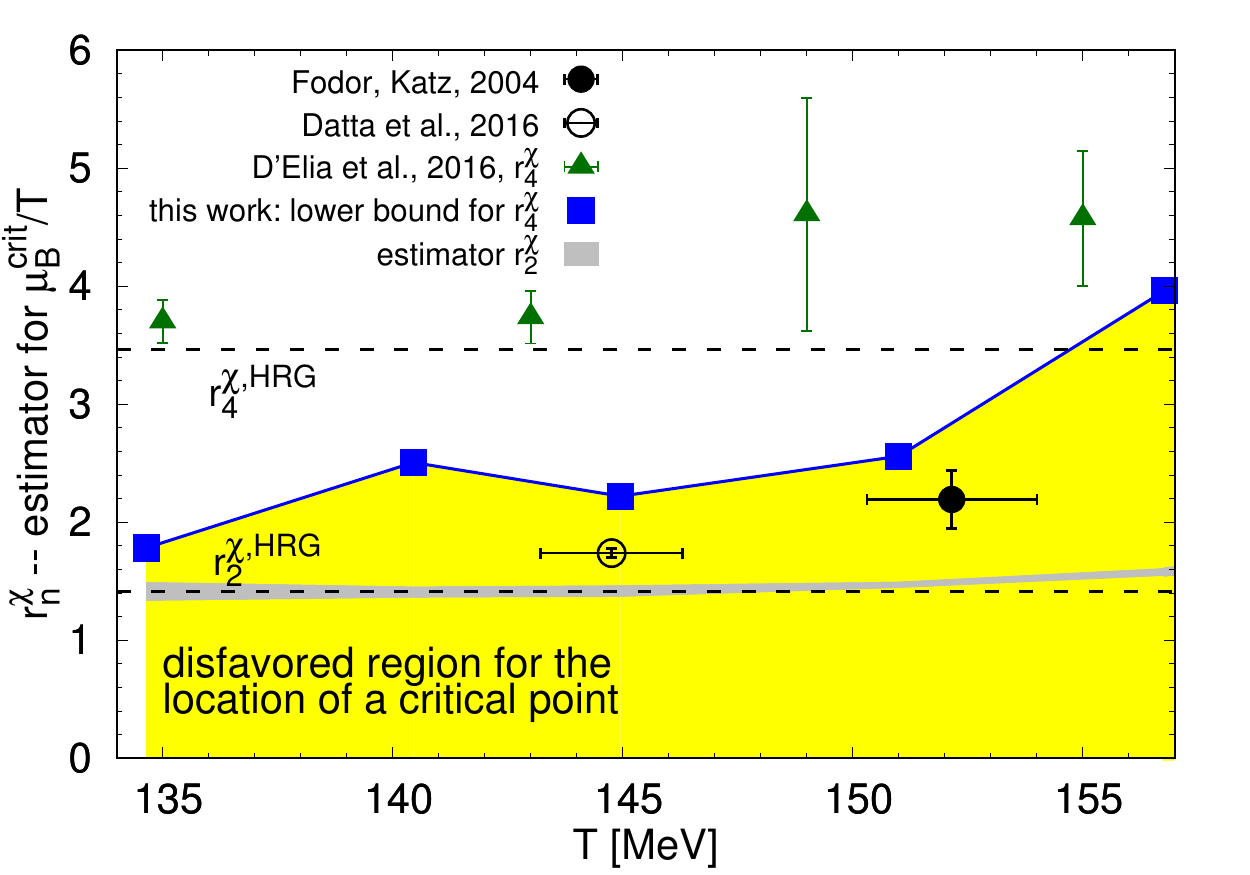}
\caption{Estimators for the radius of convergence of the Taylor
series for net baryon-number fluctuations, $\chi_2^B(T,\mu_B)$, in the
case of vanishing electric charge and strangeness chemical potentials
obtained on lattices with temporal extent $N_\tau =8$. Shown are lower
bounds for the estimator $r_4^{\chi}$ obtained in this work (squares)
and results for this estimator obtained from calculations with an
imaginary chemical potential (triangles) \cite{DElia:2016jqh}.
Also shown are estimates for the location of the critical point
obtained from calculations with unimproved staggered fermions using
a reweighting technique \cite{Fodor:2004nz} and Taylor expansions \cite{Datta:2016ukp}.
In both cases results have been rescaled using $T_c=154$~MeV.
}
\label{fig:radius}
\end{center}
\end{figure}

An immediate consequence of the definitions given in Eq.~\ref{rnchi} 
is that the ratios of generalized susceptibilities
need to grow asymptotically like $|\chi_{n+2}^B/\chi_{n}^B|\sim n^2$ in order
to arrive in the limit $n\rightarrow \infty$
at a finite value for the radius of convergence. At least
for large values of $n$ one thus needs to find large deviations from
the hadron resonance gas results $|\chi_{n+2}^B/\chi_{n}^B|^{HRG}=1$.
As is obvious from the results presented in the previous sections,
in particular from Fig.~\ref{fig:chi6B},
the analysis of up to sixth order Taylor expansion coefficients does 
not provide any hints for such large deviations. The ratio 
$\chi_4^B/\chi_2^B$ turns out to be less than unity in the entire 
temperature range explored so far, i.e. for $T\ge 135$~MeV or $T/T_c> 0.87(6)$.
Below the crossover temperature, $T\sim 155$~MeV, the sixth order expansion 
coefficients also are consistent with HRG model results. They still have large 
errors. However, using the upper value of the error for $\chi_6^B/\chi_4^B$ 
provides a lower limit for the value of the estimator $r_4^\chi$. For temperatures 
in the interval $135~{\rm MeV}\le T\le 155~{\rm MeV}$ 
(or equivalently $0.87(5)\le T/T_c\le 1$) we currently obtain a lower
limit on $r_4^\chi$ from the estimate
$\chi_6^B/\chi_4^B\simeq \chi_6^B/\chi_2^B < 3$. This converts into the bound
$r_4^{\chi} \ge 2$, which
is consistent with our observation
that the Taylor series of all thermodynamic observables discussed in the
previous sections is well behaved up to $\mu_B= 2T$.
A more detailed analysis, using the current errors on $\chi_6^B/\chi_4^B$ at
five temperature values below and in the crossover region of the
transition at $\mu_B=0$, is shown in 
Fig.~\ref{fig:radius}. This shows that the bound arising from $r_4^\chi$ is actually 
more stringent at temperatures closer to $T_c$, where $\chi_6$ starts to become small 
and eventually tends to become negative. 

These findings are consistent with recent results for susceptibility ratios
obtained from calculations with an imaginary chemical potential \cite{DElia:2016jqh}.
Also in that case all susceptibility ratios are consistent with HRG model results.
At present one thus cannot rule out that the radius of convergence may actually
be infinite. Results for  $r_4^{\chi}$ obtained in Ref.~\cite{DElia:2016jqh} 
lead to even larger estimators for the radius
of convergence than our current lower bound. This is also shown in Fig.~\ref{fig:radius}.

The observations and conclusions discussed above 
are in contrast to estimates for the location of a 
critical point obtained from a calculation based on a 
reweighting technique \cite{Fodor:2004nz} as well as 
from Taylor series expansion in 2-flavor QCD \cite{Datta:2014zqa,Datta:2016ukp}.
Both these calculations have been performed with unimproved staggered
fermion discretization schemes and thus may suffer from large cut-off
effects. Moreover, the latter calculation also suffers from large
statistical errors on higher order susceptibilities. 
Results from Ref.~\cite{Fodor:2004nz} and Ref.~\cite{Datta:2016ukp}
are also shown in Fig.~\ref{fig:radius}.

We thus conclude from our current analysis that a critical point at
chemical potentials smaller than $\mu_B= 2T$ is strongly disfavored
in the temperature range $135~{\rm MeV}\le T\le 155$~MeV and its location
at higher values of temperature seems to be ruled out.
Our results suggest that the radius of convergence in that temperature 
interval will turn out
to be significantly larger than the current bound once the 
statistics on $6^{th}$ order cumulants gets improved and higher order
cumulants become available. 

\section{Conclusions}
\label{sec:conclusion}

We have presented results on the equation of state of strong-interaction
matter obtained from a sixth order Taylor-expansion of the pressure of
(2+1)-flavor QCD with physical light and strange quark masses.
We discussed expansions at vanishing strangeness chemical potential
$\mu_S=0$ as well as for strangeness neutral systems $n_S=0$. We
have discussed in detail the latter case for a fixed electric charge to
net baryon-number ratio, $n_Q/n_B=0.4$, which is appropriate for situations
met in heavy ion collisions. The results, however, can easily be extended
to arbitrary ratios of $n_Q/n_B$. We find that the  dependence 
of basic thermodynamic observables on $n_Q/n_B$ is small for 
$0\le n_Q/n_B \le 1/2$. This may be 
of interest for applications in heavy ion collisions where strong external 
magnetic fields and non-trivial topology in QCD can lead to charge asymmetries 
in different regions of phase space. 

We have presented a parametrization of basic thermodynamic observables
in terms of ratios of fourth order polynomials in the inverse temperature
which is appropriate in the temperature range studied here, i.e. 
$T\in [130~{\rm MeV},330~{\rm MeV}]$.

We presented results for lines of constant pressure, energy and entropy
density in the $T$-$\mu_B$ plane and showed that corrections of 
${\cal O}(\hmu_B^4)$ are negligible for $\hmu_B<2$. For all three 
observables the curvature term at ${\cal O}(\hmu_B^2)$ is smaller 
than $\kappa^{\rm max}=0.012$. This suggest that, e.g. energy density
and pressure, would drop on the crossover line for the chiral transition,
if the corresponding curvature coefficient turns out to be larger than 
$\kappa^{\rm max}$.

The Taylor series for pressure and net baryon-number density as well
as energy density and entropy density determined for $\mu_S=0$ as well
as $n_S=0$ have expansion coefficients that are close to HRG model results
at low temperature. In general ratios of subsequent expansion coefficients
approach the corresponding HRG model values from below when lowering
the temperature. As a consequence, in the entire temperature range 
explored so far, the expansions are ''better behaved''
than the HRG model series, which have an infinite radius of convergence.
Assuming that the current results obtained with expansion coefficients
up to $6^{th}$ order are indicative for the behavior of higher order
expansion coefficients and taking into account the current errors on
$6^{th}$ order expansion coefficients we concluded that at temperatures
$T> 135$~MeV the presence of a critical point in the QCD phase diagram 
for $\mu_B\le 2 T$ is unlikely.

\section*{Acknowledgments}
\label{sec:acknowledge}

This work was supported in part through Contract No. DE-SC001270 with the 
U.S. Department of Energy, through the Scientific Discovery through Advanced 
Computing (SciDAC) program funded by the U.S. Department of Energy, Office of 
Science, Advanced Scientific Computing Research and Nuclear Physics, 
the DOE funded BEST topical collaboration, 
the NERSC Exascale Application Program (NESAP),
the grant 05P12PBCTA of the German Bundesministerium f\"ur Bildung und 
Forschung, the grant 56268409 of the German Academic Exchange Service (DAAD), 
grant 283286 of the European Union, the National Natural Science Foundation 
of China under grant numbers 11535012 and 11521064 and the Early Career Research 
Award of the Science and Engineering Research Board of the Government of India.
Numerical calculations have been made possible through an
INCITE grant of USQCD, ALCC grants in 2015 and 2016, and PRACE grants 
at CINECA, Italy, and the John von Neumann-Institute for Computing (NIC) in 
Germany. These grants provided access to resources on Titan
at ORNL, BlueGene/Q at ALCF and NIC, Cori I and II at NERSC and Marconi at 
CINECA.
Additional numerical calculations have been performed on USQCD GPU and KNL
clusters at JLab and Fermilab, as well as GPU clusters at
Bielefeld University, Paderborn University, and Indiana University.
We furthermore acknowledge the support of NVIDIA through the CUDA Research 
Center at Bielefeld University. 

\clearpage
\appendix
\section{Details on simulation parameters and data sets}
\label{app:statistics}

Our main data sets have been generated on lattices of size
$N_\sigma^3\times N_\tau$, with $N_\sigma/N_\tau =4$ and $N_\tau =6,\ 8$
and $12$.
We performed calculations with two different light to strange quark mass
ratios, $m_l/m_s=1/20$ and $1/27$, respectively. The simulation 
parameters are summarized in Table~\ref{tab:statistics20} and
Table~\ref{tab:statistics27}.

\begin{table}[htb]
\centering
\begin{tabular}{|cccr||cccr|}
\hline
\hline
\multicolumn{4}{|c||}{ $N_\tau=6$}&\multicolumn{4}{c|}{ $N_\tau=8$} \\
\hline
$\beta$ & $m_l$ & T[MeV] & \#conf. & $\beta$ & $m_l$ & T[MeV] & \#conf. \\
\hline
6.245 & 0.00415 & 179.52 & 14521  & 6.515 & 0.00302 & 178.36 & 16933 \\
6.341 & 0.00370 & 198.61 &  3745  & 6.550 & 0.00291 & 184.84 & 15853 \\
6.423 & 0.00335 & 216.33 &  1481  & 6.575 & 0.00282 & 189.58 & 11853 \\
6.515 & 0.00302 & 237.81 &  1408  & 6.608 & 0.00271 & 196.01 & 16760 \\
6.664 & 0.00257 & 276.43 &  1364  & 6.664 & 0.00257 & 207.32 &  8358 \\
~&~&~& ~&                           6.800 & 0.00224 & 237.07 &  5816 \\
~&~&~& ~&                           6.950 & 0.00193 & 273.88 &  9550 \\
~&~&~& ~&                           7.150 & 0.00160 & 330.23 &  9184 \\
\hline
\hline
\end{tabular}
\caption{Ensemble parameters for calculations with light to
strange quark mass ratio $m_l/m_s=1/20$ on lattices
of size $N_\sigma^3 N_\tau$ with $N_\tau=6,\ 8$ and $N_\sigma =4 N_\tau$.
Columns 4 and 8 give the number of gauge field configurations,
separated by 10 RHMC steps, that contributed to the analysis 
of up to sixth order generalized susceptibilities $\chi_{ijk}^{BQS}$.} 
\label{tab:statistics20}
\end{table}

\begin{table}[htb]
\centering
\begin{tabular}{|cccr||cccr||cccr|}
\hline
\hline
\multicolumn{4}{|c||}{ $N_\tau=6$}&\multicolumn{4}{c|}{ $N_\tau=8$}&\multicolumn{4}{c|}{ $N_\tau=12$} \\
\hline
$\beta$ & $m_l$ & T[MeV] & \#conf. & $\beta$ & $m_l$ & T[MeV] & \#conf. & $\beta$ & $m_l$ & T[MeV] & \#conf. \\
\hline
5.980 & 0.00435 & 135.29 &  81200  &6.245 & 0.00307 & 134.64 & 180320 &6.640 & 0.00196 & 134.94 &  5834 \\
6.010 & 0.00416 & 139.71 &  120790 &6.285 & 0.00293 & 140.45 & 172110 &6.680 & 0.00187 & 140.44 &  5833 \\
6.045 & 0.00397 & 145.05 &  120770 &6.315 & 0.00281 & 144.95 & 138150 &6.712 & 0.00181 & 144.97 & 13846 \\
6.080 & 0.00387 & 150.59 &  79390  &6.354 & 0.00270 & 151.00 & 107510 &6.754 & 0.00173 & 151.10 & 14200 \\
6.120 & 0.00359 & 157.17 &  66180  &6.390 & 0.00257 & 156.78 & 135730 &6.794 & 0.00167 & 157.13 & 15476 \\
6.150 & 0.00345 & 162.28 &  79660  &6.423 & 0.00248 & 162.25 & 115850 &6.825 & 0.00161 & 161.94 & 16772 \\
6.170 & 0.00336 & 165.98 &  49760  &6.445 & 0.00241 & 165.98 & 120270 &6.850 & 0.00157 & 165.91 & 19542 \\
6.200 & 0.00324 & 171.15 &  122700 &6.474 & 0.00234 & 171.02 & 139980 &6.880 & 0.00153 & 170.77 & 21220 \\
6.225 & 0.00314 & 175.76 &  122730 &6.500 & 0.00228 & 175.64 & 133070 &6.910 & 0.00148 & 175.76 & 12303 \\
\hline
\hline
\end{tabular}
\caption{Same as Table~\protect\ref{tab:statistics20} but
for the light to strange quark mass ratio $m_l/m_s=1/27$ and including results
for $N_\tau=12$.}
\label{tab:statistics27}
\end{table}

\section{Constraints on chemical potential for strangeness neutral systems
with fixed electric charge to baryon-number ratio}
\label{app:constraint}
We are interested in expansion coefficients for strangeness neutral systems
in which the net electric-charge is proportional 
to the net baryon-number. {\it I.e.} we introduce the constraint given in 
Eq.~\ref{neutral}.
These constraints can be fulfilled order by order in the Taylor expansion
of the number densities by choosing the expansion coefficients of the series
for $\hmu_Q$ and $\hmu_S$, given in Eq.~\ref{qs}, appropriately, 
i.e. the coefficients $s_n$ and $q_n$ can be determined order by order. 
We start with the Taylor series for the number densities introduced 
in Eq.~\ref{nX} and define the expansion coefficients as

\begin{eqnarray}
N^B_{n}&=&s_{n}\chi_{11}^{BS}+q_{n}\chi_{11}^{BQ}+m^B_{n}
\label{eq:nB}\\
N^Q_{n}&=&s_{n}\chi_{11}^{QS}+q_{n}\chi_{2}^{Q}+m^Q_{n}
\label{eq:nQ}\\
N^S_{n}&=&s_{n}\chi_{2}^{S}+q_{n}\chi_{11}^{QS}+m^S_{n}
\label{eq:nS}
\end{eqnarray}
for $n=1,\ 3,\ 5$.
At each order in the expansion we then have to solve a set of two linear 
equations, which always have the same structure. We find as solutions
\begin{equation}
s_n=-\frac{q_n\chi_{11}^{QS}+m^S_n}{\chi_{2}^{S}}\;,
\end{equation}
and
\begin{equation}
q_n=\frac{-m^B_n r \chi_{2}^{S}+ m^Q_n \chi_{2}^{S}+ m^S_n 
(r \chi_{11}^{BS}- \chi_{11}^{QS})}
{(\chi_{11}^{QS})^2-r \chi_{11}^{BS} \chi_{11}^{QS}+r \chi_{2}^{S} \chi_{11}^{BQ}-\chi_{2}^{S} \chi_{2}^{Q}}\;.
\end{equation}

At leading order one finds for the terms $m_1^X$,

\begin{equation}
m^B_1=\chi_{2}^{B}\;\; ,\;\; m^Q_1=\chi_{11}^{BQ} \;\; ,\;\;
m^S_1=\chi_{11}^{BS} \;\; ,
\end{equation}
and the contributions to the next-to-leading order expansion terms, $m_3^X$,
are given by

\begin{eqnarray}
m^B_3=\frac{1}{6} \big(&&
3 q_1^2 s_1 \chi_{121}^{BQS}
+3 q_1 s_1^2 \chi_{112}^{BQS}
+6 q_1 s_1 \chi_{211}^{BQS}
+q_1^3 \chi_{13}^{BQ}
+3 q_1^2 \chi_{22}^{BQ}
\nonumber \\
&&+3 q_1 \chi_{31}^{BQ}
+s_1^3 \chi_{13}^{BS}
+3 s_1^2 \chi_{22}^{BS}
+3 s_1 \chi_{31}^{BS}
+\chi_{4}^{B}\big)
\nonumber \\
m^Q_3=\frac{1}{6} \big(&&3 q_1^2 s_1 \chi_{31}^{QS}
+3 q_1 s_1^2 \chi_{22}^{QS}
+6 q_1 s_1 \chi_{121}^{BQS}
+q_1^3 \chi_{4}^{Q}
+3 q_1^2 \chi_{13}^{BQ}
\nonumber \\
&&+3 q_1 \chi_{22}^{BQ}
+s_1^3 \chi_{13}^{QS}
+3 s_1^2 \chi_{112}^{BQS}
+3 s_1 \chi_{211}^{BQS}
+\chi_{31}^{BQ}\big)
\nonumber \\
m^S_3=\frac{1}{6} \big(&&3 q_1^2 s_1 \chi_{22}^{QS}
+3 q_1 s_1^2 \chi_{13}^{QS}
+6 q_1 s_1 \chi_{112}^{BQS}
+q_1^3 \chi_{31}^{QS}
+3 q_1^2 \chi_{121}^{BQS}
\nonumber \\
&&+3 q_1 \chi_{211}^{BQS}
+s_1^3 \chi_{4}^{S}
+3 s_1^2 \chi_{13}^{BS}
+3 s_1 \chi_{22}^{BS}
+\chi_{31}^{BS}\big)
\end{eqnarray}

Finally the contributions to the next-to-next-to-leading order expansion 
terms, $m_5^X$, are given by
\begin{eqnarray}
m^B_5=\frac{1}{120} \big(&&5 q_1^4 s_1 \chi_{141}^{BQS}
+10 q_1^3 s_1^2 \chi_{132}^{BQS}
+20 q_1^3 s_1 \chi_{231}^{BQS}
+60 q_1^2 s_3 \chi_{121}^{BQS}
+10 q_1^2 s_1^3 \chi_{123}^{BQS}
+30 q_1^2 s_1^2 \chi_{222}^{BQS}
+30 q_1^2 s_1 \chi_{321}^{BQS}
\nonumber \\
&&+120 q_1 s_1 s_3 \chi_{112}^{BQS}
+5 q_1 s_1^4 \chi_{114}^{BQS}
+120 q_3 q_1 s_1 \chi_{121}^{BQS}
+120 q_1 s_3 \chi_{211}^{BQS}
+20 q_1 s_1^3 \chi_{213}^{BQS}
+30 q_1 s_1^2 \chi_{312}^{BQS}
\nonumber \\
&&+20 q_1 s_1 \chi_{411}^{BQS}
+60 q_3 s_1^2 \chi_{112}^{BQS}
+120 q_3 s_1 \chi_{211}^{BQS}
+q_1^5 \chi_{15}^{BQ}
+5 q_1^4 \chi_{24}^{BQ}
+10 q_1^3 \chi_{33}^{BQ}
+60 q_3 q_1^2 \chi_{13}^{BQ}
\nonumber \\
&&+10 q_1^2 \chi_{42}^{BQ}
+120 q_3 q_1 \chi_{22}^{BQ}
+5 q_1 \chi_{51}^{BQ}
+60 q_3 \chi_{31}^{BQ}
+60 s_1^2 s_3 \chi_{13}^{BS}
+s_1^5 \chi_{15}^{BS}
+120 s_1 s_3 \chi_{22}^{BS}
\nonumber \\
&&+5 s_1^4 \chi_{24}^{BS}
+60 s_3 \chi_{31}^{BS}
+10 s_1^3 \chi_{33}^{BS}
+10 s_1^2 \chi_{42}^{BS}
+5 s_1 \chi_{51}^{BS}
+\chi_{6}^{B}\big)
\nonumber \\
m^Q_5=\frac{1}{120} \big(&&5 q_1^4 s_1 \chi_{51}^{QS}
+10 q_1^3 s_1^2 \chi_{42}^{QS}
+20 q_1^3 s_1 \chi_{141}^{BQS}
+60 q_1^2 s_3 \chi_{31}^{QS}
+10 q_1^2 s_1^3 \chi_{33}^{QS}
+30 q_1^2 s_1^2 \chi_{132}^{BQS}
+30 q_1^2 s_1 \chi_{231}^{BQS}
\nonumber \\
&&+120 q_1 s_1 s_3 \chi_{22}^{QS}
+5 q_1 s_1^4 \chi_{24}^{QS}
+120 q_3 q_1 s_1 \chi_{31}^{QS}
+120 q_1 s_3 \chi_{121}^{BQS}
+20 q_1 s_1^3 \chi_{123}^{BQS}
+30 q_1 s_1^2 \chi_{222}^{BQS}
\nonumber \\
&&+20 q_1 s_1 \chi_{321}^{BQS}
+60 q_3 s_1^2 \chi_{22}^{QS}
+120 q_3 s_1 \chi_{121}^{BQS}
+q_1^5 \chi_{6}^{Q}
+5 q_1^4 \chi_{15}^{BQ}
+10 q_1^3 \chi_{24}^{BQ}
+60 q_3 q_1^2 \chi_{4}^{Q}
\nonumber \\
&&+10 q_1^2 \chi_{33}^{BQ}
+120 q_3 q_1 \chi_{13}^{BQ}
+5 q_1 \chi_{42}^{BQ}
+60 q_3 \chi_{22}^{BQ}
+60 s_1^2 s_3 \chi_{13}^{QS}
+s_1^5 \chi_{15}^{QS}
+120 s_1 s_3 \chi_{112}^{BQS}
\nonumber \\
&&+5 s_1^4 \chi_{114}^{BQS}
+60 s_3 \chi_{211}^{BQS}
+10 s_1^3 \chi_{213}^{BQS}
+10 s_1^2 \chi_{312}^{BQS}
+5 s_1 \chi_{411}^{BQS}
+\chi_{51}^{BQ}\big)
\nonumber \\
m^S_5=\frac{1}{120} \big(&&5 q_1^4 s_1 \chi_{42}^{QS}
+10 q_1^3 s_1^2 \chi_{33}^{QS}
+20 q_1^3 s_1 \chi_{132}^{BQS}
+60 q_1^2 s_3 \chi_{22}^{QS}
+10 q_1^2 s_1^3 \chi_{24}^{QS}
+30 q_1^2 s_1^2 \chi_{123}^{BQS}
+30 q_1^2 s_1 \chi_{222}^{BQS}
\nonumber \\
&&+120 q_1 s_1 s_3 \chi_{13}^{QS}
+5 q_1 s_1^4 \chi_{15}^{QS}
+120 q_3 q_1 s_1 \chi_{22}^{QS}
+120 q_1 s_3 \chi_{112}^{BQS}
+20 q_1 s_1^3 \chi_{114}^{BQS}
+30 q_1 s_1^2 \chi_{213}^{BQS}
\nonumber \\
&&+20 q_1 s_1\chi_{312}^{BQS}
+60 q_3 s_1^2 \chi_{13}^{QS}
+120 q_3 s_1 \chi_{112}^{BQS}
+q_1^5 \chi_{51}^{QS}
+5 q_1^4 \chi_{141}^{BQS}
+10 q_1^3 \chi_{231}^{BQS}
+60 q_3 q_1^2 \chi_{31}^{QS}
\nonumber \\
&&+10 q_1^2 \chi_{321}^{BQS}
+120 q_3 q_1 \chi_{121}^{BQS}
+5 q_1 \chi_{411}^{BQS}
+60 q_3 \chi_{211}^{BQS}
+60 s_1^2 s_3 \chi_{4}^{S}
+s_1^5 \chi_{6}^{S}
+120 s_1 s_3 \chi_{13}^{BS}
\nonumber \\
&&+5 s_1^4 \chi_{15}^{BS}
+60 s_3 \chi_{22}^{BS}
+10 s_1^3 \chi_{24}^{BS}
+10 s_1^2\chi_{33}^{BS}
+5 s_1 \chi_{42}^{BS}
+\chi_{51}^{BS}\big)
\end{eqnarray}

\begin{figure}[t]
\includegraphics[width=0.50\textwidth]{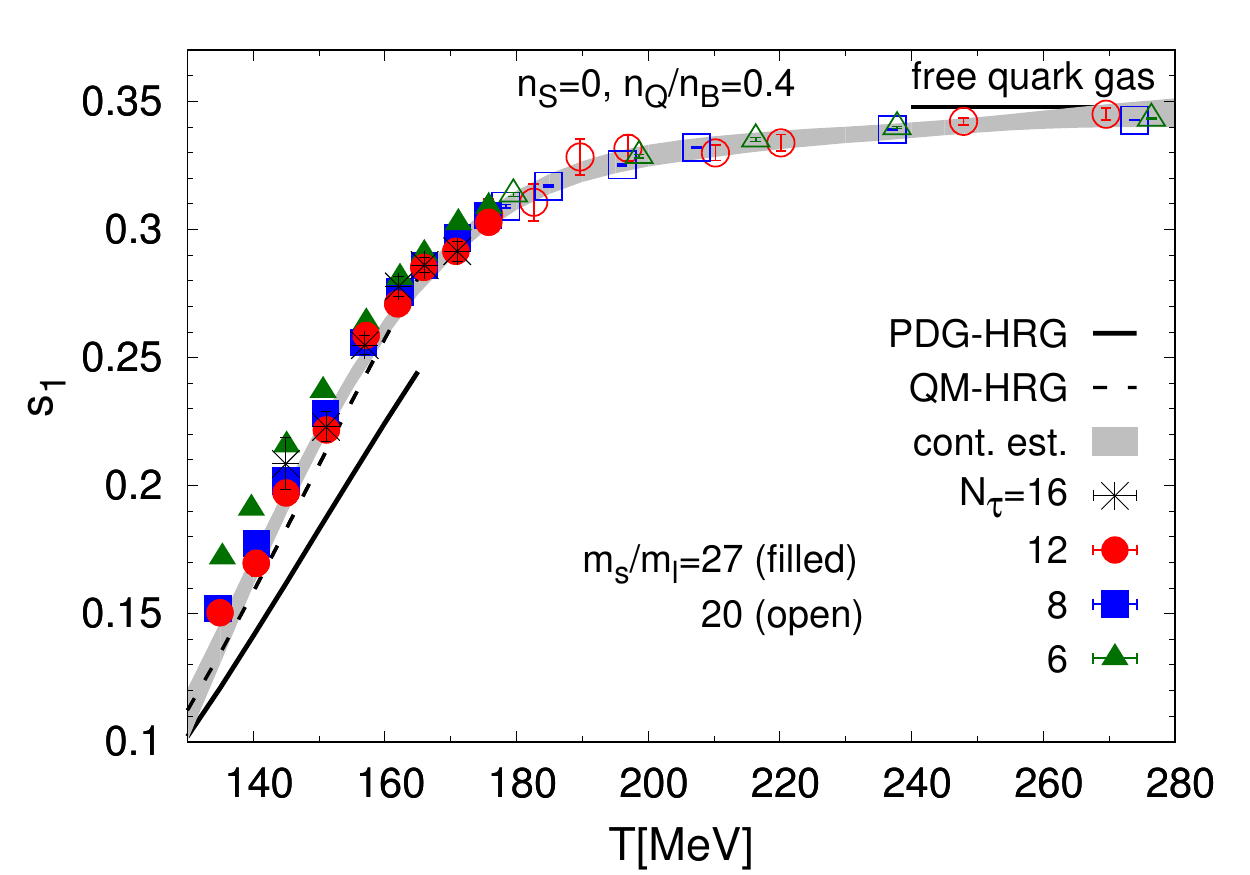}\hspace*{-0.1cm}
\includegraphics[width=0.50\textwidth]{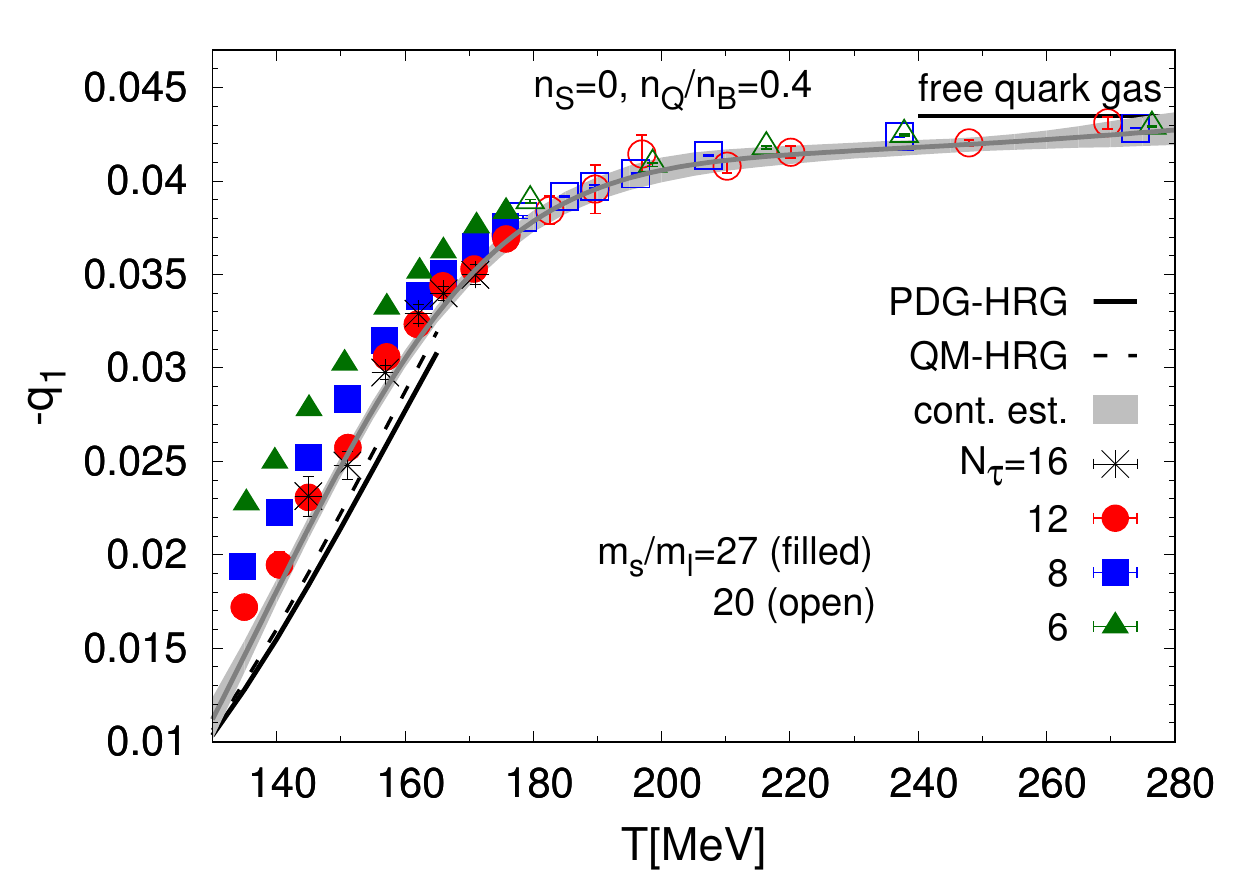}%

\includegraphics[width=0.50\textwidth]{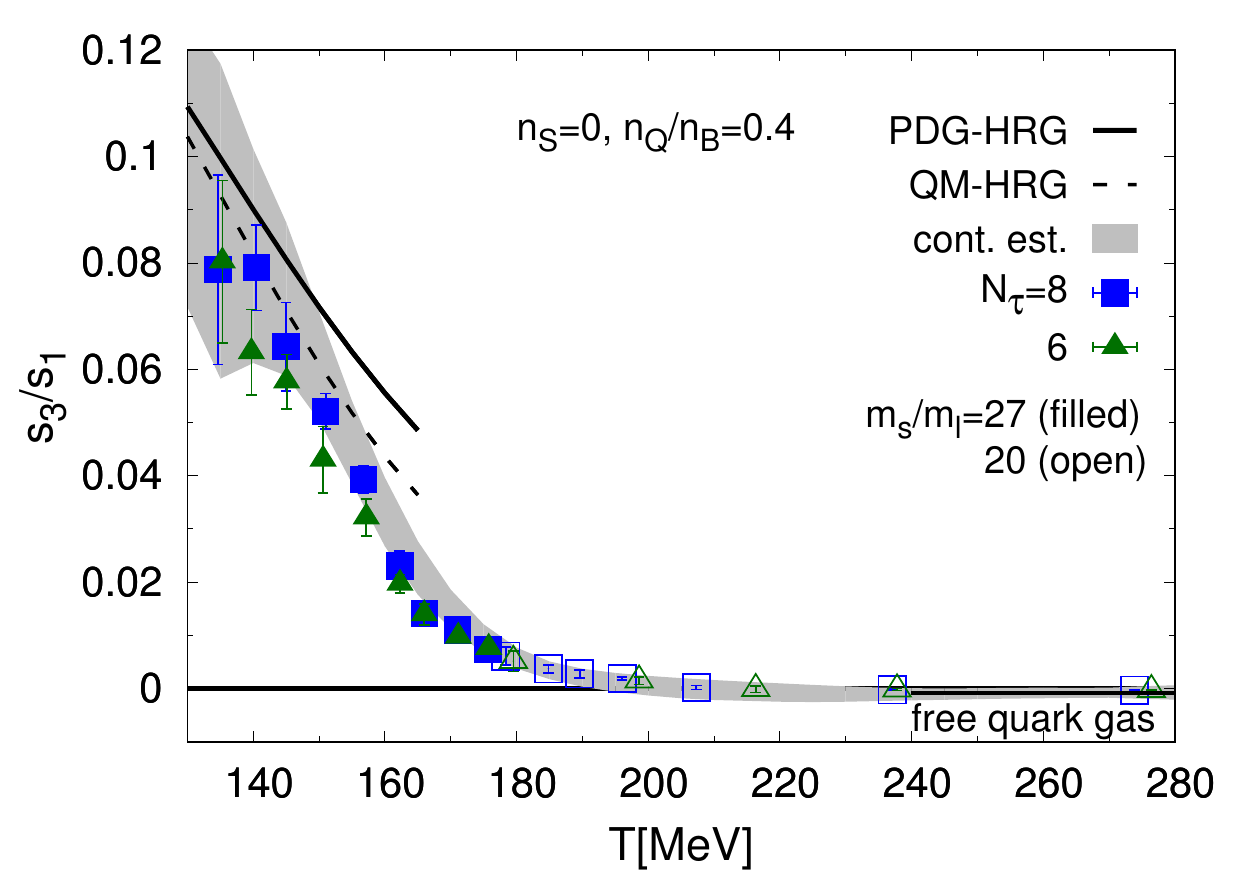}\hspace*{-0.1cm}
\includegraphics[width=0.50\textwidth]{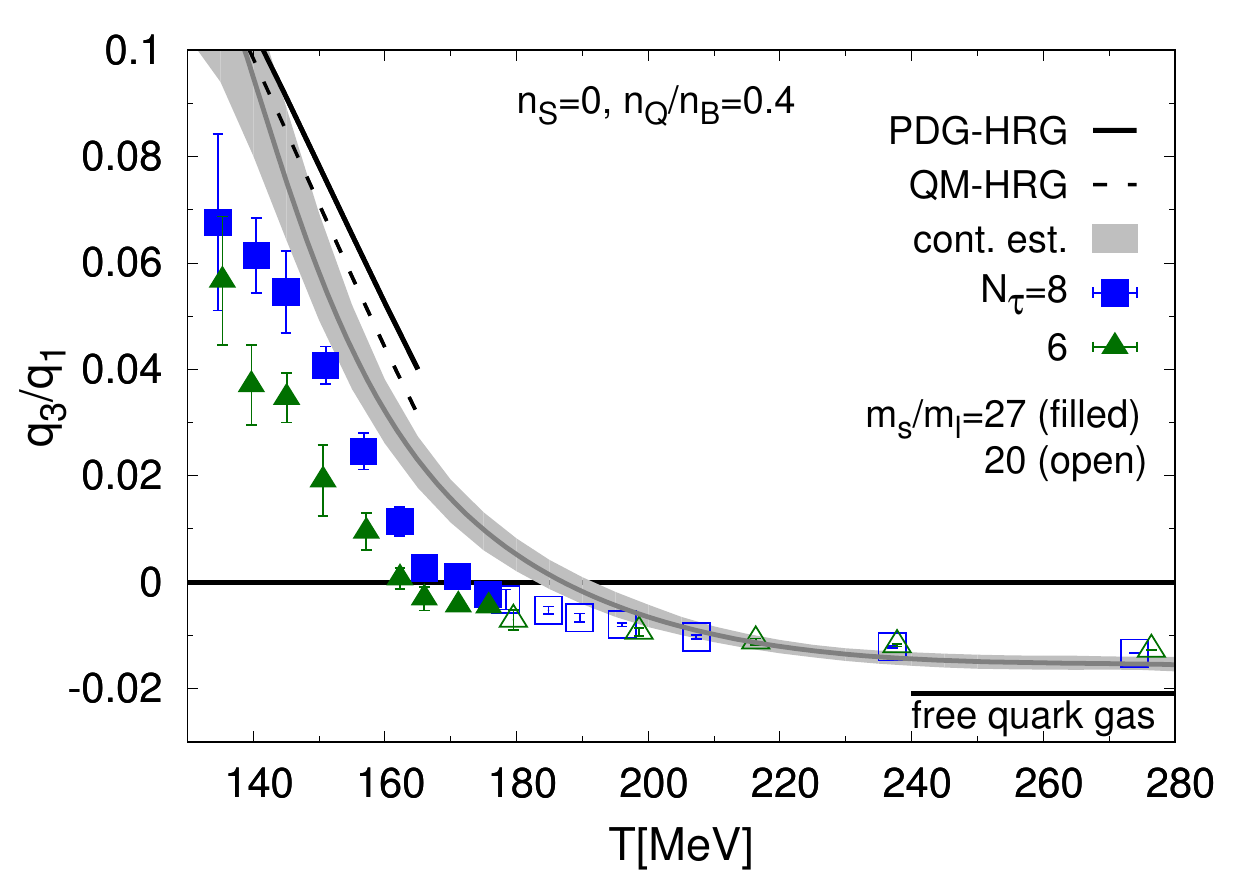}
\caption{The LO Taylor expansion coefficients $s_1$ (top, left) and $q_1$
(top, right) of the
expansions of $\hmu_S$ and $\hmu_Q$ w.r.t. $\hmu_B$.
The bottom set of figures show the ratios of NLO and LO expansion
coefficients. The broad bands give the continuum extrapolated
results. The curves inside these bands show results obtained with
the interpolating curves introduced in Eq.~\ref{qnfit}.
 Also shown are the PDG-HRG and QM-HRG results (see text).
The solid black lines labeled `free quark gas' denote the $T\to\infty$
non-interacting massless quark gas result.}
\label{fig:qisi}
\end{figure}

In (2+1)-flavor QCD calculations the light ($u,\ d$) quark masses are
taken to be degenerate. A consequence
of this degeneracy is that not all generalized susceptibilities
$\chi_{ijk}^{BQS}$ that enter the above expressions are independent. 
In a given order $n\equiv 2l \equiv i+j+k$ this
results in a set of relations among the expansion coefficients. In general,
at order $n=2l$, there are $l(l+1)$ constraints, i.e.
for $l=1$ this gives rise to two relations,
\cite{Bazavov:2012vg}
\begin{eqnarray}
0 &=& \chi_2^B - 2\chi_{11}^{BQ} + \chi_{11}^{BS}   
\nonumber \\
0 &=& \chi_2^S - 2\chi_{11}^{QS} + \chi_{11}^{BS} \; ,
\label{k2}
\end{eqnarray}
for $l=2$ there are six constraints,
\begin{eqnarray}
0 &=& \chi_4^B-2 \chi_{31}^{BQ}+\chi_{31}^{BS} 
\nonumber \\
0 &=& \chi_4^S-2 \chi_{13}^{QS}+\chi_{13}^{BS}
\nonumber \\
0 &=& \chi_{22}^{BS} + \chi_{13}^{BS} - 2 \chi_{112}^{BQS} 
\nonumber \\
0 &=& \chi_{31}^{BS} + \chi_{22}^{BS} - 2 \chi_{211}^{BQS}
\nonumber \\
0 &=& \chi_4^B-6 \chi_{31}^{BQ} +12 \chi_{22}^{BQ} -8 \chi_{13}^{BQ} 
+3 \chi_{31}^{BS} + 3 \chi_{22}^{BS} 
 + \chi_{13}^{BS} -12 \chi_{211}^{BQS} + 12 \chi_{121}^{BQS} 
-6 \chi_{112}^{BQS} 
\nonumber \\
0 &=& \chi_4^S 
+ \chi_{31}^{BS}
+3  \chi_{22}^{BS}
+ 3 \chi_{13}^{BS} 
- 8 \chi_{31}^{QS}
+12 \chi_{22}^{QS} 
  -6 \chi_{13}^{QS}
-6 \chi_{211}^{BQS} +12 \chi_{121}^{BQS} -12 \chi_{112}^{BQS}
\label{k4}
\end{eqnarray}
and for $l=3$ there are twelve constraints,
\begin{eqnarray}
0 &=& \chi_6^B- 2\chi_{51}^{BQ} + \chi_{51}^{BS},\nonumber \\
0 &=& \chi_{15}^{BS}- 2\chi_{15}^{QS} + \chi_6^S,\nonumber \\
0 &=& \chi_{42}^{BS}- 2\chi_{312}^{BQS} + \chi_{33}^{BS},\nonumber \\
0 &=& \chi_{33}^{BS}- 2\chi_{213}^{BQS} + \chi_{24}^{BS},\nonumber \\
0 &=& \chi_{51}^{BS}- 2\chi_{411}^{BQS} + \chi_{42}^{BS},\nonumber \\
0 &=& \chi_{24}^{BS}- 2\chi_{114}^{BQS} + \chi_{15}^{BS},\nonumber \\
0 &=& \chi_6^B - 6\chi_{51}^{BQ} + 12\chi_{42}^{BQ} - 8\chi_{33}^{BQ} + 3\chi_{51}^{BS}
-12\chi_{411}^{BQS} + 12\chi_{321}^{BQS} + 3\chi_{42}^{BS} - 6\chi_{312}^{BQS} + \chi_{33}^{BS},\nonumber \\
0 &=& \chi_{33}^{BS} - 6\chi_{213}^{BQS} + 12\chi_{123}^{BQS} - 8\chi_{33}^{QS} + 3\chi_{24}^{BS}
-12\chi_{114}^{BQS} + 12\chi_{24}^{QS} + 3\chi_{15}^{BS} - 6\chi_{15}^{QS} + \chi_6^S,\nonumber \\
0 &=& \chi_{42}^{BS} - 6\chi_{312}^{BQS} + 12\chi_{222}^{BQS} - 8\chi_{132}^{BQS} + 3\chi_{33}^{BS}
-12\chi_{213}^{BQS} + 12\chi_{123}^{BQS} + 3\chi_{24}^{BS} - 6\chi_{114}^{BQS} + \chi_{15}^{BS},\nonumber \\
0 &=& \chi_{51}^{BS} - 6\chi_{411}^{BQS} + 12\chi_{321}^{BQS} - 8\chi_{231}^{BQS} + 3\chi_{42}^{BS}
-12\chi_{312}^{BQS} + 12\chi_{222}^{BQS} + 3\chi_{33}^{BS} - 6\chi_{213}^{BQS} + \chi_{24}^{BS},\nonumber \\
0 &=& \chi_6^B - 10\chi_{51}^{BQ} + 40\chi_{42}^{BQ} - 80\chi_{33}^{BQ} + 80\chi_{24}^{BQ}
- 32\chi_{15}^{BQ} +  5\chi_{51}^{BS} - 40\chi_{411}^{BQS} +120\chi_{321}^{BQS} \nonumber \\
   &&      -160\chi_{231}^{BQS} + 80\chi_{141}^{BQS} + 10\chi_{42}^{BS} - 60\chi_{312}^{BQS}
+120\chi_{222}^{BQS} - 80\chi_{132}^{BQS} + 10\chi_{33}^{BS} - 40\chi_{213}^{BQS} \nonumber \\
   &&      + 40\chi_{123}^{BQS} +  5\chi_{24}^{BS} - 10\chi_{114}^{BQS} +      \chi_{15}^{BS},\nonumber \\
0 &=& \chi_{51}^{BS} - 10\chi_{411}^{BQS} + 40\chi_{321}^{BQS} - 80\chi_{231}^{BQS} + 80\chi_{141}^{BQS}
- 32\chi_{51}^{QS} +  5\chi_{42}^{BS} - 40\chi_{312}^{BQS} +120\chi_{222}^{BQS} \nonumber \\
   &&      -160\chi_{132}^{BQS} + 80\chi_{42}^{QS} + 10\chi_{33}^{BS} - 60\chi_{213}^{BQS}
+120\chi_{123}^{BQS} - 80\chi_{33}^{QS} + 10\chi_{24}^{BS} - 40\chi_{114}^{BQS} \nonumber \\
   &&      + 40\chi_{24}^{QS} +  5\chi_{15}^{BS} - 10\chi_{15}^{QS} +      \chi_6^S.
\label{k6}
\end{eqnarray}
Using these constraints it is tedious, but straightforward, to show
that in the isospin symmetric case, $r=1/2$, indeed all expansion
coefficients for the electric charge chemical potential vanish, i.e.
$\hmu_Q=0$ to all orders in $\mu_B$.

We show results for the LO expansion coefficients $s_1$ and $q_1$ and 
the ratios of the NLO and LO expansion coefficients, $s_3/s_1$ and
$q_3/q_1$ in Fig.~\ref{fig:qisi}. As can be seen the NLO coefficients
are already negligible for $T\gsim 170$~MeV. The absolute value of the 
NNLO expansion coefficients $s_5$ and $q_5$ never is larger than 1\% of the 
corresponding LO coefficients.

In Fig.~\ref{fig:qisi},  we also show results from hadron resonance gas (HRG)
model calculations. The black curves are the predictions of the usual HRG
model which consists of all the resonances listed in the Particle Data Group
Tables up to 2.5 GeV (PDG-HRG). The PDG-HRG results for $s_1$ are
substantially smaller than the continuum extrapolated lattice QCD results.
It has been argued in \cite{Bazavov:2014xya} that this can
be caused by contributions from additional, experimentally not yet observed,
strange hadron resonances which are predicted in quark model calculations.
A HRG model
calculation based on such an extended resonance spectrum (QM-HRG) is also
shown in  Fig.~\ref{fig:qisi}. At finite values of the lattice cut-off
we observe significant differences between lattice QCD calculations and
both versions of the HRG models. This is in particular the case for the
expansion coefficients of the electric charge chemical potentials. One
thus may wonder whether these deviations can be understood in terms of
taste violations in the staggered fermion formulation which result
in a modification of the resonance spectrum and affect most strongly the light
pseudo-scalar (pion) sector.

\section{The coefficient \boldmath$\kappa_4^f$ of lines of constant physics at \boldmath ${\cal O}(\mu_B^4)$}
\label{app:curvature}
We will present here results for the expansion coefficient $\kappa_4^f$ of lines of constant physics defined in Eq.~\eqref{k4f},
\begin{eqnarray}
\kappa_4^f &=& \frac{\frac{1}{2}\left. \frac{\partial^2 f(T,\mu_B)}{\partial T^2}\right|_{(T_0,0)} (\kappa_2^f)^2\frac{1}{T_0^2} -
\frac{1}{2} \left. \frac{\partial}{\partial T} \frac{\partial^2 f(T,\mu_B)}{\partial \mu_B^2}\right|_{(T_0,0)} \kappa_2^f \frac{1}{T_0} + \frac{1}{4!}   \left. 
\frac{\partial^4 f(T,\mu_B)}{\partial \mu_B^4}\right|_{(T_0,0)}}{\left. 
\frac{\partial f(T,\mu_B)}{\partial T}\right|_{(T_0,0)} \frac{1}{T_0^3} }
\nonumber \\
&=& \frac{\frac{1}{2}\left. T_0^2\frac{\partial^2 f_0(T)}{\partial T^2}\right|_{(T_0,0)} (\kappa_2^f)^2 - \left. 
\left( T_0 \frac{\partial f_2(T)}{\partial T} \right|_{(T_0,0)} -2 f_2(T_0) \right)
\kappa_2^f  + f_4(T_0)}{\left. T_0 \frac{\partial f_0(T)}{\partial T}\right|_{(T_0,0)} }
\label{k4fx}
\end{eqnarray}
The coefficients $f_{2k}$ are defined by 
\begin{equation}
f(T,\mu_B) = \sum_{k=0}^\infty f_{2k}\hat{\mu}_B^{2k}.
\label{eq:f}
\end{equation}
In particular, we will give explicit expressions for the case of constant pressure ($f\equiv P$), constant energy density ($f\equiv \epsilon$) and constant entropy density($f\equiv s$). For the pressure we had the earlier expression (Eq.~\eqref{Pn})
\begin{equation}
\frac{P(T,\mu_B) - P(T,0)}{T^4} = \sum_{n=1}^\infty P_{2n}\hat{\mu}_B^{2n}.
\label{eq:P}
\end{equation}
Comparing Eqs.~\eqref{eq:P} and \eqref{eq:f} we have, $f_0 = P(T,0) \equiv T^4 P_0$, $f_2 = T^4P_2$ and $f_4 = T^4P_4$. Thus,
\begin{subequations}
\begin{align} 
\left.\frac{\partial f_0}{\partial T}\right|_{\mu_B} &= \left.\frac{\partial P_0T^4}{\partial T}\right|_{\mu_B}
= T^3\left(TP_0' + 4P_0\right) \equiv s, \\
\left.\frac{\partial^2 f_0}{\partial T^2}\right|_{\mu_B} &= \left.\frac{\partial^2 P_0T^4}{\partial T^2}\right|_{\mu_B}
= T^2\left(T^2P_0'' + 8TP_0' + 12P_0\right) \equiv \frac{C_V}{T}.
\end{align}
\label{eq:P0_derivatives}
\end{subequations}
Here $s$ and $C_V$ are the entropy density and specific heat per unit volume 
at vanishing chemical potential. Similarly, 
\begin{equation}
\left.\frac{\partial f_2}{\partial T}\right|_{\mu_B} = \left.\frac{\partial P_2T^4}{\partial T}\right|_{\mu_B}
= T^3\left(TP_2' + 4P_2\right).
\label{eq:df2}
\end{equation}
Putting everything together we get, for the pressure:
\begin{align}
\kappa_4^P &= \frac{1}{TP_0' + 4P_0}\left[P_4 - \kappa_2^P\left(TP_2' + 2P_2\right) + 
\frac{1}{2}\left(\kappa_2^P\right)^2\left(T^2P_0'' + 8TP_0' + 12P_0\right)\right] 
\nonumber \\
&=\frac{T^3}{s} \left[ 
P_4(T) - \kappa_2^P \sigma_2(T) +
\frac{1}{2}\left(\kappa_2^P\right)^2 \frac{C_V}{T^3} 
\right] \; ,
\label{eq:k4P}
\end{align}
where $\sigma_2$ denotes the ${\cal O}(\hmu_B^2)$ expansion coefficient of the 
entropy density as introduced in Eq.~\ref{entropyc}.

Next we consider $\kappa_4^\epsilon$. Since the energy density is also of 
dimension four, we only need to replace $P_{2n}$ with $\epsilon_{2n}$ in the 
first line of Eq.~\eqref{eq:k4P}. With this we obtain,
\begin{equation}
\kappa_4^\epsilon = \frac{1}{T\epsilon_0' + 4\epsilon_0}\left[
\epsilon_4 - \kappa_2^\epsilon\left(T\epsilon_2' + 2\epsilon_2\right) + 
\frac{1}{2}\left(\kappa_2^\epsilon\right)^2\left(T^2\epsilon_0'' + 8T\epsilon_0' + 12\epsilon_0\right)
\right].
\label{eq:k4e}
\end{equation}
Since $C_V \equiv \left(\partial \epsilon_0 /\partial T\right)_{\mu_B}$, the above may be written as
\begin{equation}
\kappa_4^\epsilon = \frac{T^3}{C_V}\left[
\epsilon_4 - \kappa_2^\epsilon\left(T\epsilon_2' + 2\epsilon_2\right) + 
\frac{1}{2}\left(\kappa_2^\epsilon\right)^2\frac{1}{T^2}\frac{\partial C_V}{\partial T}
\right].
\label{eq:k4e-2}
\end{equation}

Finally we consider $\kappa_4^s$. Since the entropy density is of dimension three, Eqs.~\eqref{eq:P0_derivatives} become
\begin{align} && &&
\left.\frac{\partial(sT^3)}{\partial T}\right|_{\mu_B} = T^2\left(Ts' + 3s\right), && 
\left.\frac{\partial^2(sT^3)}{\partial T^2}\right|_{\mu_B} = T\left(T^2s'' + 6Ts' + 6s\right), && &&
\label{eq:S0_derivatives}
\end{align}
and therefore
\begin{equation}
\kappa_4^\sigma = \frac{1}{Ts' + 3s}\left[
\sigma_4 - \kappa_2^\sigma\left(T\sigma_2' + \sigma_2\right) + 
\frac{1}{2}\left(\kappa_2^\sigma\right)^2\left(T^2s'' + 6Ts' + 6s\right)
\right].
\label{eq:k4s}
\end{equation}
To zeroth order, the specific heat is also given by $C_V = \left(\partial(Ts)/\partial T\right)_{\mu_B}$. Thus,
\begin{equation}
\kappa_4^\sigma = \frac{T^3}{C_V}\left[
\sigma_4 - \kappa_2^\sigma\left(T\sigma_2' + \sigma_2\right) + 
\frac{1}{2}\left(\kappa_2^\sigma\right)^2\frac{1}{T^2}\frac{\partial C_V}{\partial T}
\right].
\label{eq:k4s-2}
\end{equation}


\end{document}